\documentclass[journal=jctcce,manuscript=article,layout=traditional]{achemso}
\pdfoutput=1

\pdftrailerid{}
\usepackage[T1]{fontenc} 
\usepackage{amssymb}
\usepackage{mathtools}
\usepackage{graphicx}
\usepackage{multirow}
\usepackage{siunitx}
\usepackage{hyperref}
\usepackage{pdfpages}

\usepackage{mdframed}
\mdfsetup{linewidth=1pt, linecolor=red}

\SectionNumbersOn
\setkeys{acs}{doi = true}
\newcommand{\doi}[1]{\href{http://dx.doi.org/#1}{\nolinkurl{#1}}}
\DeclareSIUnit\atm{atm}

\title{
    STable AutoCorrelation Integral Estimator (STACIE):
    Robust and accurate transport properties
    from molecular dynamics simulations
}

\author{G\"{o}zdenur Toraman}
\affiliation[Soete]{Soete Laboratory, Ghent University, Technologiepark-Zwijnaarde 46, 9052 Ghent, Belgium}
\author{Dieter Fauconnier}
\affiliation[Soete]{Soete Laboratory, Ghent University, Technologiepark-Zwijnaarde 46, 9052 Ghent, Belgium}
\alsoaffiliation[FlandersMake]{FlandersMake@UGent, Core Lab MIRO, 3001 Leuven, Belgium}
\author{Toon Verstraelen}
\affiliation[Ghent University]{Center for Molecular Modeling (CMM), Ghent University, Technologiepark-Zwijnaarde 46, 9052 Ghent, Belgium}
\email{toon.verstraelen@ugent.be}

\begin{document}


\newpage

\begin{abstract}
    STACIE (STable AutoCorrelation Integral Estimator) is a novel algorithm and Python package that delivers robust, uncertainty-aware estimates of autocorrelation integrals from time-correlated data.
    While its primary application is deriving transport properties from equilibrium molecular dynamics simulations,
    STACIE is equally applicable to time-correlated data in other scientific fields.
    A key feature of STACIE is its ability to provide robust and accurate estimates without requiring manual adjustment of hyperparameters.
    Additionally, one can follow a simple protocol to prepare sufficient simulation data to achieve a desired relative error of the transport property.
    We demonstrate its application by estimating the ionic electrical conductivity of a NaCl-water electrolyte solution.
    We also present a massive synthetic benchmark dataset to rigorously validate STACIE,
    comprising 15360 sets of time-correlated inputs generated with diverse covariance kernels with known autocorrelation integrals.
    STACIE is open source and available on GitHub and PyPI, with comprehensive documentation and examples.
\end{abstract}

\begin{tocentry}
    \includegraphics{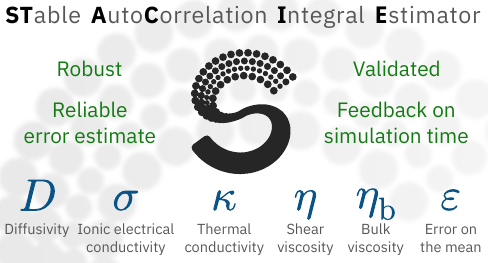}
\end{tocentry}

\newpage

\section{Introduction}\label{sec:introduction}
In molecular dynamics (MD) simulations, the autocorrelation function (ACF) plays an essential role in the computation of transport properties such as diffusivity, viscosity, and electrical or thermal conductivity.
The Green--Kubo (GK) linear response theory expresses these properties as time integrals of ACFs.~\cite{Green1954, Kubo1957}
Their work established the foundations for the computation of transport properties from equilibrium MD (EMD) simulations,~\cite{Frenkel2002, Allen2017} with noteworthy applications including the viscosity of simple fluids,~\cite{Levesque1973, Reese2012} the thermal conductivity of liquids~\cite{Boone2019, Manjunatha2021} and solids,~\cite{Reese2012, Knoop2023} or the viscosity and electrical conductivity of molten salts~\cite{Wang2020} and ionic liquids.~\cite{Zhang2015}
Its applicability, however, extends far beyond the scope of MD simulations.
For example, in high-energy physics, the GK framework has been employed to estimate the shear viscosity of quark-gluon plasmas produced in heavy-ion collisions.~\cite{Muronga2004}
In plasma physics, it aids in studying transport properties in dusty plasmas, offering insights into the behavior of strongly coupled charged particles.~\cite{Huang2021}
Beyond transport properties, the ACF and its integral are employed in various fields,
most notably to estimate the uncertainty of the average of time-correlated data.~\cite{Friedberg1970, Goodman2010, Allen2017, ForemanMackey2017}
Robust and accurate algorithms to estimate the integral of an ACF and its uncertainty are thus critical for the analysis of MD simulations and are also relevant to other fields.

For the calculation of transport properties from EMD simulations, an ACF of the following form is considered:
\begin{align}
  c(\Delta_t) = \Bigl\langle
    \hat{x}(t_0)\, \hat{x}(t_0 + \Delta_t)
  \Bigr\rangle
\end{align}
where $\hat{x}(t)$ is the time-correlated data, $t_0$ is the time origin, $\Delta_t$ is the time lag, and $\left\langle \cdot \right\rangle$ denotes an ensemble average.
In this work, all stochastic quantities are denoted with a hat to distinguish them from deterministic ones, and the adjective ``sampling'' is used to distinguish them from expectation values in the text when needed for clarity.
Practically, the average is taken over multiple time origins and independent sequences, e.g.\ different Cartesian components and/or different simulation runs.
Examples of time-correlated inputs include particle velocities (used for diffusivity) and stress components (used for shear viscosity).
This expression for the ACF assumes that $\langle \hat{x} \rangle = 0$, which is often the case for transport properties.
An exception is the computation of the bulk viscosity, where $\hat{x}(t)$ corresponds to the isotropic instantaneous pressure as a function of time,
which fluctuates around its mean value. In this case, the mean pressure must be subtracted from $\hat{x}(t)$ before computing the ACF.~\cite{Hafner2022}

In general, a transport property can be written as an integral of an ACF of the following form:
\begin{align}
  \mathcal{I} = F\, \frac{1}{2} \int_{-\infty}^{\infty} c(\Delta_t)\, \mathrm{d}\Delta_t
  \label{eq:acint_cont}
\end{align}
where $F$ is a property-dependent factor.
For example, for diffusivity, $F = 1$, and for viscosity, $F = V / k_\mathrm{B} T$, where $V$ is the volume of the simulation cell, $T$ is the temperature, and $k_\mathrm{B}$ is the Boltzmann constant.~\cite{Hansen2013, Frenkel2002}
Note that while it is more common to integrate from $0$ to $\infty$, the integral is symmetric, allowing the limits to be extended to $-\infty$ to $\infty$ by including the factor $\frac{1}{2}$.
This notation simplifies the discussion of STACIE in later sections.
The integration of the ACF also yields the integrated correlation time:
\begin{align}
  \tau_\text{int} = \frac{\mathcal{I}}{F c(0)}
  \label{eq:tauint_cont}
\end{align}
It is primarily used to quantify the number of independent samples in a time series of length $T$ as $T / (2\tau_\text{int})$.~\cite{Friedberg1970, Goodman2010, Allen2017, ForemanMackey2017}

There are several sources of error when numerically integrating the ACF.
In the context of MD simulations these include approximations of the function $\hat{x}(t)$, due to force fields, finite-size effects, Ewald convergence, etc.
Even if these are well controlled, $\hat{x}(t)$ is a stochastic quantity (due to thermal fluctuations) and can only be sampled at a finite number of time steps (due to limited computational resources).
As a result, a direct computation of $\mathcal{I}$ via numerical quadrature of Eq.~\eqref{eq:acint_cont} involves two key approximations: (i) estimating the ACF from sampled data and (ii) truncating the integration bounds to a finite domain $[-\Delta_\text{cut},\Delta_\text{cut}]$.
While collecting more data can systematically reduce the variance of the ACF estimate, addressing the second issue is more challenging.~\cite{Toraman2023}
Although extending the integration domain reduces bias due to truncation, it also increases variance for two reasons.
First, integrating over a larger domain introduces more uncertain terms in the quadrature.
Second, the variance of the integrand, the sampling ACF, grows at larger time lags.~\cite{Bartlett1980, Flyvbjerg1989, Francq2009}
To achieve a favorable trade-off between bias and variance, numerous algorithms have been proposed, some of which reformulate the integral to analyze alternative quantities, such as the mean squared displacement (MSD).\cite{Helfand1960}
Existing algorithms can be broadly classified into the four categories shown in Table~\ref{tab:overview}, roughly in chronological order: (i) methods that directly analyze the ACF, (ii) methods that analyze the MSD, (iii) methods that apply cepstral analysis and (iv) spectral methods that directly analyze the power spectral density (PSD) of $\hat{x}(t)$.
Section S1 of the Supporting Information lists known software implementations in each category.
We will summarize each category below with equations for continuous and infinite time.
This simplifies the notation as these quantities are not affected by sampling limitations.
However, we will also discuss the limitations when the algorithms are applied to a finite amount of data.

\begin{table}
    \centering
    \renewcommand{\arraystretch}{2.0}
    \begin{tabular}{ll}
      \textbf{Category} &
      \textbf{Key Quantity}
      \\ \hline
      Green--Kubo (GK) &
      ACF $=c(\Delta_t)=\displaystyle\Bigl\langle
        \hat{x}(t_0)\, \hat{x}(t_0 + \Delta_t)
      \Bigr\rangle$
      \\
      Einstein--Helfand (EH) &
      MSD $=\displaystyle\Bigl\langle
        \bigl|\hat{y}(t_0 + \Delta_t) - \hat{y}(t_0)\bigr|^2
      \Bigr\rangle$
      \\
      Cepstral &
      Cepstrum $=\mathcal{C}(\Delta_t)=\displaystyle\int_{-\infty}^{\infty} \log C(f)\, e^{2\pi i f \Delta_t} \,\mathrm{d}f$
      \\
      Spectral &
      PSD $=C(f) = \displaystyle\int_{-\infty}^{\infty} c(\Delta_t)\, e^{-2\pi i f \Delta_t} \,\mathrm{d}\Delta_t$
      \\ \hline
    \end{tabular}
    \caption{
      Classification of algorithms for estimating the integral of an ACF and the key quantity used in each case.
      $\hat{x}(t)$ denotes the time-correlated data, $\hat{y}(t)$ is its antiderivative, and $C(f)$ is its PSD.
      The cepstrum is the inverse Fourier transform of the logarithm of the PSD.
    }
    \label{tab:overview}
\end{table}

A first class of algorithms directly analyzes the ACF or its numerical quadrature.
These are often referred to as Green--Kubo (GK) algorithms or methods, although all algorithms discussed here ultimately rely on the same underlying linear response theory.
To manage the inherent bias-variance trade-off in estimating the autocorrelation integral,
these methods typically either truncate the numerical quadrature at a carefully chosen time lag~\cite{Guo2001, Chen2009, Danel2012},
or fit a model to the sampling ACF (or its antiderivative) and extrapolate it to infinite time~\cite{Hess2002, Guo2002, vanderSpoel1998, Zhang2015}.
Model-based methods offer the advantage of reduced variance by fitting to multiple data points and reduced bias by extrapolating beyond the cutoff.
However, they are also susceptible to bias introduced by the choice of model, fitting range, and regression weights.
In the case of  the time-decomposition method (TDM),~\cite{Zhang2015} it was shown that such bias can be significant by sampling a range of reasonable settings in the regression and investigating its impact on the outcome.~\cite{Toraman2023}
A further challenge is that these methods often apply (non)linear regression without accounting for the correlated uncertainties in the sampling ACF at different time lags~\cite{Bartlett1980, Francq2009},
which can result in underestimating the true uncertainty of the estimated transport property.
Model-free approaches also assume uncorrelated errors, such as the method by Liu \textit{et al},~\cite{Liu2024} where a model for the mean squared error of the viscosity is minimized to determine the optimal $\Delta_\text{cut}$.

A second class of Einstein--Helfand (EH) algorithms reformulates the integral in terms of the MSD:
\begin{align}
    \mathcal{I} =
        F\,\frac{1}{2} \lim_{\Delta_t \rightarrow \infty} \frac{\mathrm{d}}{\mathrm{d}\Delta_t}
        \Bigl\langle
            \bigl|\hat{y}(t_0 + \Delta_t) - \hat{y}(t_0)\bigr|^2
        \Bigr\rangle
\end{align}
where $\hat{y}(t)$ is the antiderivative of $\hat{x}(t)$.
(The subtraction of $\langle \hat{x}(t) \rangle$ is usually not necessary, as mentioned above.)
This relation was originally developed by Einstein for diffusion and was later generalized by Helfand to other transport properties.~\cite{Frenkel2002, Hansen2013, Allen2017, Helfand1960}
MD engines such as the Large-scale Atomic/Molecular Massively Parallel Simulator (LAMMPS)~\cite{Thompson2022, Gissinger2024} and GROMACS~\cite{Pll2020}, and several post-processing tools~\cite{Kneller1995, Gowers2016, deBuyl2018, Humbert2019, MichaudAgrawal2011, Smidstrup2019, Brehm2020, Baerends2025}, can efficiently compute the MSD.
A common approach is to visually identify a linear regime in a log-log plot of the MSD versus time.
In such a plot, a slope of 1 indicates a first-degree time dependence of the MSD, corresponding to normal diffusive or linear transport behavior.
The derivative of the MSD is then typically estimated by applying simple linear regression to the MSD for a fixed set of time lags associated with normal diffusion.~\cite{Jamali2019}

EH algorithms face limitations similar to those based on the ACF.
For instance, the choice of the fitting range is subjective and introduces bias into the final estimate.
In addition, simple linear regression to the sampling MSD assumes normally distributed errors~\cite{Pranami2015} and neglects the fact that these errors are correlated across time lags,
which complicates uncertainty quantification and can lead to underestimated errors.~\cite{Flyvbjerg1989, Crdoba2021, Moustafa2024}
A notable implementation of the EH algorithm is the On-the-fly Calculation of Transport Properties (OCTP) plug-in~\cite{Jamali2019} for LAMMPS,~\cite{Thompson2022} which uses the order-$n$ algorithm by Dubbeldam \textit{et al.} to efficiently compute the MSD at multiple time lags.~\cite{Dubbeldam2009}
While this approach significantly reduces the cost of storing and processing large data files, it still requires users to perform post-processing on the efficiently written data.

A third and novel class of algorithms, introduced by Baroni and coworkers, makes use of the cepstrum.~\cite{Ercole2017, Baroni2018, Bertossa2019, Ercole2022}
While it was originally referred to as a ``spectral'' method, we refer to it as the ``cepstral'' method to clearly distinguish it from the fourth class of algorithms discussed below.
To summarize their approach, we first review the Wiener--Khinchin theorem, which states that the PSD of a time-correlated signal is the Fourier transform of the ACF:~\cite{Ricker2003, Proakis2007}
\begin{align}
  C(f)
  = \int_{-\infty}^{\infty} c(\Delta_t)\, e^{-2\pi i f \Delta_t} \,\mathrm{d}\Delta_t
  = \lim_{T\rightarrow\infty} \frac{1}{2T} \left|\int_{-T}^{T} \hat{x}(t)\, e^{-2\pi i f t} \,\mathrm{d}t\right|^2
  \label{eq:wiener_khinchin_cont}
\end{align}
where $C(f)$ is the PSD.
The discrete form of this identity allows for an efficient computation of the sampling ACF using the Fast Fourier Transform (FFT): the PSD is proportional to the square of the FFT of the signal, and its inverse FFT equals the ACF.

The cepstrum is obtained by taking the inverse Fourier transform of the logarithm of the PSD as follows:
\begin{align}
  \mathcal{C}(\Delta_t)
  = \int_{-\infty}^{\infty} \log C(f)\, e^{2\pi i f \Delta_t} \,\mathrm{d}f
  \label{eq:cepstrum_cont}
\end{align}
When the cepstrum is computed from a finite amount of data, one can show that uncertainties of $\log(\hat{C}(f))$ are uncorrelated and that their statistical distribution is independent of $f$, except for $f=0$.
By invoking the law of large numbers, Ercole \textit{et al.}~\cite{Ercole2017} justify that the sampling cepstrum has practically uncorrelated errors, which greatly simplifies the uncertainty quantification.
Furthermore, the time integral of the cepstrum can be related to the autocorrelation integral. Using the notation of this work, that is:
\begin{align}
  \log\left(\frac{2\mathcal{I}}{F}\right) = \int_{-\infty}^{\infty} \mathcal{C}(\Delta_t)\, \mathrm{d}\Delta_t,
  \label{eq:cepstrum_integral}
\end{align}
The cepstral algorithm truncates the integral of the sampling cepstrum to a finite domain and properly accounts for statistical uncertainties.
It is implemented in the open-source software package SporTran and has been extensively tested with thermal conductivity calculations.~\cite{Ercole2022}

While the cepstral algorithm has clear conceptual advantages, it still inherits some limitations of the GK algorithm.
Notably, the truncation of the numerical quadrature of the cepstrum to a finite domain is difficult to automate.
The Akaike Information Criterion (AIC) has been proposed to fix the optimal cutoff, but it has been found to produce a suboptimal balance between bias and variance.~\cite{Ercole2017}
To address this, SporTran offers a graphical user interface (GUI) that allows users to manually inspect and adjust the cutoff.~\cite{Ercole2022}
In addition, SporTran introduces spectral filtering to smoothen the cepstrum and make the cutoff easier to identify visually.
While these interactive features can be helpful in practice, they also introduce potential subjectivity and user bias in the final results.

Recently, a fourth class of algorithms has been proposed that directly analyze the sampling PSD (or periodogram),~\cite{Drigo2024, Pegolo2025} and is currently being implemented in SporTran~\cite{Pegolo2025}.
This new algorithm was initially developed to estimate off-diagonal Onsager coefficients,~\cite{Drigo2024} with an application to the Seebeck coefficient of molten salts.
Later, Pegolo \textit{et al.}~\cite{Pegolo2025} generalized the approach to include also conventional ``diagonal'' transport properties.
Building on the Wiener--Khinchin theorem in Eq.~\eqref{eq:wiener_khinchin_cont}, the DC-component of the PSD is directly proportional to the autocorrelation integral:
\begin{align}
  \mathcal{I} = F\,\frac{1}{2}\, C(f=0)
  \label{eq:zerofreq}
\end{align}
By definition, obtaining an exact value for the integral would require an infinitely long time series, which is not feasible in practice.
Moreover, simply estimating $C(f=0)$ from a finite amount of data, even if possible, is of little use in practice due to its high variance.
A more practical approach is to obtain a low-variance estimate by fitting a model to the low-frequency part of a sampling PSD and extrapolating it to $f=0$, provided that the fit is sufficiently well-behaved.
The statistical properties of the sampling PSD are convenient (no correlated errors) and well understood,~\cite{Priestley1982, Fuller1995, Shumway2017, Ercole2017} allowing for a direct application of Bayesian regression and uncertainty quantification.
So far, spline models have been fitted to a fixed low-frequency part of the sampling PSD, with the number of knots optimized using the AIC.~\cite{Drigo2024, Pegolo2025}
In this work, we propose a new spectral algorithm that incorporates different models and automatically identifies the low-frequency range to be fitted.

In our overview of algorithms, we implicitly assumed that the inputs $\hat{x}(t)$ have been sampled with a sufficient number of steps to represent the relevant dynamics of a system.
However, this assumption is not trivial: the sampled sequences must be long enough to capture the slowest modes in the system, but their characteristic timescales are often unknown \textit{a priori}.
Ercole \textit{et al.}~\cite{Ercole2017} demonstrated that cepstral analysis can significantly reduce the sequence length required for accurate estimates.
However, if the time series is too short, even the most advanced algorithms will yield unreliable and biased estimates of the autocorrelation integral.
Despite the importance of this issue, only a few studies have provided recommendations for the minimum required simulation times, and most rely on heuristics.~\cite{Carlson2022}
Determining the minimum required simulation time remains an open problem, adding to the challenges of accurately calculating transport properties.~\cite{Maginn2020}

In summary, existing algorithms for estimating the integral of an ACF suffer from one or more of the following limitations:
\begin{enumerate}
    \item
    They often rely on \textit{ad hoc} tunable algorithmic hyperparameters that must be fixed by the user, which can bias the estimates.
    \item
    There is a limited understanding of the required simulation time to obtain a reliable estimate, which is especially important for systems exhibiting slow dynamics.
    \item
    Not all methods quantify the uncertainty of the integral $\mathcal{I}$, which is essential for assessing reliability and accuracy.
\end{enumerate}
In this work, we introduce a new algorithm and a Python package called STACIE (STable AutoCorrelation Integral Estimator) that aims to address these three challenges.
First, it eliminates the need for users to tune algorithmic hyperparameters, reducing the risk of bias.
Second, it provides clear feedback on the required simulation time to ensure converged estimates.
Third, it quantifies the uncertainty of the autocorrelation integral, offering a robust assessment of the accuracy and reliability of the results.

New algorithms for estimating the integral of an ACF have always been validated against a few physical test cases, such as the viscosity of argon or water.~\cite{Hess2002, Viscardy2007, Chen2009, Reese2012, Ercole2017}
However, this approach limits the scope of testing due to the computational cost and the logistics of running MD simulations.
Furthermore, in many physical systems, the true value of the transport property is not precisely known, making it difficult to assess the accuracy of a given method.
While comparisons to experimental data are valuable, they come with their own limitations, such as uncertainties arising from the force field models used in the simulations.

To enable more robust algorithm validation, we generated a massive synthetic benchmark dataset, called the AutoCorrelation Integral Drill (ACID).~\cite{ACID}
This dataset consists of 15360 sets of time-correlated sequences for which the autocorrelation integral is always exactly equal to one.
Because STACIE can be applied without manual judgment and is implemented as a Python library without a GUI, it can be efficiently validated with the ACID test on a compute node of a high-performance cluster.
This setup enables efficient, reproducible benchmarking and statistically meaningful validation of estimation accuracy across a wide range of input scenarios.

The remainder of this paper is organized as follows: Section~\ref{sec:algorithm} describes the STACIE algorithm.
In Section~\ref{sec:examples}, we present two illustrative use cases.
The first is a minimal, self-contained Python script including both the generation and analysis of time-correlated data.
The second example showcases a typical workflow for computing ionic conductivity of an electrolyte solution, highlighting how to address the challenge of an \textit{a priori} unknown required simulation length.
Section~\ref{sec:validation} introduces the ACID benchmark dataset and discusses the validation of STACIE across a wide range of test cases.
Finally, Section~\ref{sec:conclusions} summarizes our findings and outlines directions for future work.

\section{Algorithm}\label{sec:algorithm}
This section first describes the components of the STACIE algorithm.
In essence, STACIE computes the sampling PSD, fits a model to its low-frequency part, and evaluates the model at zero frequency to estimate the autocorrelation integral. (See Eq.~\eqref{eq:zerofreq}.)
We begin by summarizing well-known computational and statistical aspects of sampling PSDs of stationary stochastic time-correlated sequences.
STACIE employs locally weighted Bayesian regression to estimate parameters of a simple model for the low-frequency part of the spectrum.~\cite{Shao2019}
For numerical efficiency, it determines parameters that maximize the posterior probability and estimates uncertainties using the Laplace approximation.~\cite{MacKay2005}
A key challenge is the automatic identification of the cutoff frequency, which defines the range of frequencies for which the spectrum that can be adequately explained by the model. This identification is the final component of the algorithm.
The last two subsections are not part of the STACIE algorithm itself but provide guidance on preparing inputs for STACIE: how to generate sufficient data and how to store it efficiently.

\subsection{Statistical properties of sampling power spectra}\label{subsec:psd}

While we are, in principle, interested in the autocorrelation integral of continuous time-correlated physical quantities averaged over an entire thermodynamic ensemble, we must use a limited number of finite and discrete time series in practice.
These restrictions will be introduced step by step below, continuing with the notation from the introduction.

First, we discretize the time axis using a step size $h$ and an integer time lag $\Delta$.
Assuming infinitely long sequences, the autocorrelation integral in Eq.~\eqref{eq:acint_cont} must then be approximated using a quadrature rule:
\begin{align}
  \mathcal{I} \approx \frac{Fh}{2} \sum_{\Delta=-\infty}^{\infty} c_\Delta \qquad \text{with} \qquad c_\Delta  = \Bigl\langle \hat{x}(t_0)\, \hat{x}(t_0 + h \Delta) \Bigr\rangle
\end{align}
where $c_\Delta$ is the discrete ACF.

To rewrite the Wiener--Khinchin theorem in discrete time, we introduce a discretized form of $\hat{x}(t)$:
\begin{align}
  \hat{x}_\text{d}(t) = h \sum_{n=-\infty}^{\infty} \hat{x}(hn)\,\delta(t - hn)
\end{align}
where $\delta(t)$ is the Dirac delta function and the step size $h$ ensures that the units are consistent.
Then $\hat{x}_\text{d}(t)$ can be substituted into the Wiener--Khinchin theorem in Eq.~\eqref{eq:wiener_khinchin_cont} to obtain:
\begin{align}
  C_\text{d}(f)
  = h\sum_{\Delta=-\infty}^{\infty} c_\Delta\, e^{-2\pi i f h \Delta}
  = \lim_{N\rightarrow\infty} \frac{h}{2N+1} \left| \sum_{n=-N}^N \hat{x}(hn)\, e^{-2\pi i f h n} \right|^2
  \label{eq:wiener_khinchin_disc_time}
\end{align}
The spectrum of the discrete-time process, $C_\text{d}(f)$, is periodic with period $1/h$.~\cite{Oppenheim1999}
Hence, if the domain of the PSD of the original continuous process, $C(f)$, is contained within $[-1/2h, 1/2h]$, it will be well reproduced by the spectrum of the discrete process, $C_\text{d}(f)$, without so-called aliasing artifacts.~\cite{Oppenheim1999}
Since the time step in MD simulations must be small enough to ensure that the Nyquist frequency, $1/2h$, is significantly larger than the highest frequency of the molecular system, there is no risk for aliasing when using all the time steps as input.
In this case, we can safely use $\mathcal{I}\approx F C_\text{d}(0) / 2$.
However, writing results to disk at every time step for post-processing is not always possible due to storage limitations, for which a practical solution will be provided in Section~\ref{subsec:efficient-storage}.

In addition to discretization, time series are also finite in numerical applications.
Because Discrete Fourier Transforms (DFTs) are used to compute a sampling PSD, our treatment will only be exact for periodic sequences, whereas they are typically aperiodic.
Using aperiodic inputs introduces so-called leakage artifacts in the sampling PSD: the sampling PSD converges to the true PSD convolved with a sinc function, which results in the smearing of spectral amplitudes over neighboring bins~\cite{Oppenheim1999}.
This results in a loss of resolution on the scale of $1/(Nh)$.
In practice, this means that $N$ should be sufficiently large to ensure that $1/(Nh)$ is much smaller than the relevant features in the low-frequency part of the spectrum.

Consider $M$ discrete and finite time series of length $N$: $\{x^{(m)}_n\}_{m=1}^M$ with $\hat{x}^{(m)}_n=\hat{x}^{(m)}(nh)$.
Spectra derived from them will also be stochastic.
To reduce the statistical noise, we can average over $M$ independent spectra computed using the fully discrete equivalent of the Wiener--Khinchin theorem:~\cite{Oppenheim1999}
\begin{align}
  \hat{C}_k
  = \frac{h}{M} \sum_{m=1}^M \sum_{n=0}^{N-1} \hat{c}_\Delta\, e^{-2\pi i k \Delta / N}
  = \frac{h}{NM} \sum_{m=1}^M \left| \hat{X}^{(m)}_k \right|^2
  \label{eq:wiener_khinchin_fully_disc}
\end{align}
with
\begin{align}
  \hat{X}^{(m)}_k = \sum_{n=0}^{N-1} \hat{x}^{(m)}_n\, e^{-2\pi i k n / N}
\end{align}
and where $\hat{c}_\Delta$ is the sampling ACF:
\begin{align}
  \hat{c}_\Delta = \frac{1}{N} \sum_{n=0}^{N-1} \hat{x}^{(m)}_n\, \hat{x}^{(m)}_{n+\Delta}
\end{align}
The time step $h$ is included for consistency with the continuous case.
The integer frequency index $k\in\{0, 1, \ldots, N\}$ corresponds to a real frequency $f_k = k / (Nh)$.
In most applications, $M$ is greater than 1 by construction.
For example, diffusivity is computed using all Cartesian components of each particle's velocity.
In the case of diffusivity in 3D, this results in $M = 3N_\text{atom}$ and $F = 1$.

One may propose to approximate the autocorrelation integral simply as $\mathcal{I} \approx F \hat{C}_0 / 2$.
However, if the true PSD $\langle \hat{C}_k \rangle$ varies slowly with frequency index $k$, a smaller variance can be obtained by fitting a model to the low frequency part of the sampling PSD and evaluating this model at zero frequency.
For convenience, STACIE always stores the spectrum as $\hat{I}_k = F \hat{C}_k / 2$, and we will always use this rescaled spectrum, including the factor $F/2$, in the remainder of the text.
Before performing any analysis, one can already plot the low frequency part of this rescaled sampling PSD to get a first visual impression of the autocorrelation integral.

In order to quantify the uncertainty of the regression and extrapolation to zero frequency,
we need to characterize the statistical properties of the sampling PSD.
To this end, we first assume that the input sequences, $\hat{x}^{(m)}$, are samples of a discrete, periodic and stationary Gaussian process (GP) with zero mean.
Trajectories derived from EMD simulations do not fully satisfy this assumption,
i.e.\ they are obtained by numerical integration of coupled ODEs.
However, for the slowest modes in the system, close to $f = 0$, thus of relevance for the autocorrelation integral, this is a reasonable approximation.
These slow modes are embedded in a bath of high-frequency thermal noise due to many atomic collisions and can therefore be treated as stochastic oscillators.

The covariance of a discrete, periodic and stationary GP is a circulant matrix, which becomes diagonal in the Fourier domain.
Hence, the DFT of a sample from this GP is a complex vector, whose real and imaginary parts are independent and normally distributed with zero mean.
By taking the modulus squared of these DFTs and averaging over $M$ independent sequences, one can show that the components of the sampling PSD are also statistically independent and $\operatorname{Gamma}(\alpha,\theta)$-distributed:~\cite{Priestley1982, Fuller1995, Shumway2017, Ercole2017}
\begin{align}
  \hat{I}_k &= \frac{F h}{2 N M} \sum_{m=1}^{M} \left|\hat{X}^{(m)}_k\right|^2
  \sim \operatorname{Gamma}\left(\frac{\nu_k}{2}, \frac{2I_k}{\nu_k} \right)
\end{align}
with
\begin{align}
  \nu_k = \begin{cases}
    M & \text{if $k=0$} \\
    M & \text{if $k=N/2$ and $N$ is even} \\
    2M & \text{otherwise}
  \end{cases}
  \label{eq:nu_k}
\end{align}
and where we used the notation $I_k = \langle \hat{I}_k \rangle$.
Most spectrum amplitudes receive $2M$ contributions (degrees of freedom): $M$ real and $M$ imaginary components of $\hat{X}^{(m)}\,\forall m \in \{1, \ldots, M\}$.
There are only $M$ degrees of freedom when $\hat{X}^{(m)}$ is real by construction.

The importance of the statistical independence of the sampling PSD amplitudes cannot be overstated.
Because the amplitudes are independent, regression and other statistical analyses can treat the sampling PSD as a product of one-dimensional distributions, greatly simplifying the analysis compared to modeling a large, coupled multivariate distribution.
This well-defined mathematical structure reduces both computational cost and programming complexity when deriving properties and their uncertainties from the spectrum.
It represents a significant advantage over algorithms that post-process time-dependent quantities such as the ACF, its running integral, or the MSD, all of which exhibit correlated uncertainties.

\subsection{Power spectrum model}\label{subsec:model}

Simply reading the sampling PSD at zero frequency, $\hat{I}_0$, yields an unreliable estimate of the autocorrelation integral because the sampling PSD is inherently noisy.
Instead, STACIE fits a model to the low-frequency part of the sampling PSD and evaluates this model at zero frequency to obtain a more robust estimate.
STACIE offers several built-in models and is easily extended with user-defined models.
To avoid making strong assumptions about the shape of the spectrum, this work employs a general, smooth, and positive model of the following form:
\begin{align}
  I^\text{model}(f, \mathbf{b}) = \exp\left(\sum_{s\in S} b_s f^s\right)
  \label{eq:exppoly}
\end{align}
where $\mathbf{b}$ is the vector of model parameters, and $S$ is the set of polynomial degrees, which must include $0$.
In most cases, using $S=\{0, 1, 2\}$ is sufficient.
However, for spectra with large statistical uncertainties (when $M$ is small), it is advisable to use fewer terms.
For example, $S = \{0\}$ is suitable for modeling a white-noise spectrum, while $S = \{0,1\}$ can extract a meaningful trend from highly noisy spectra.
If the spectrum is expected to have a zero derivative at the origin, then $S = \{0,2\}$ can be used.
A demonstration of the model for different sets of polynomial degrees $S$ is shown in Section S2 of the Supplementary Information.
More elaborate models can be employed to extract additional properties from the spectrum, such as the exponential correlation time, which will be explored in a future paper.~\cite{Toraman2025viscosity}
Regardless of the model used, STACIE is designed to provide reliable error estimates.
Therefore, it is recommended to select the model that minimizes the predicted uncertainty.

\subsection{Parameter estimation}\label{subsec:parameters}

In this subsection, we assume that a suitable low-frequency part of the spectrum has already been identified.
We only describe how to optimize the spectrum model parameters for this choice.
The next subsection will explain how to select the low-frequency part of the spectrum automatically by scanning a range of cutoff frequencies.
Rather than imposing a hard cutoff that ignores all higher frequencies, we use a smooth switching function that gradually decreases from 1 to 0 for an increasing frequency $f$.
\begin{align}
  w(f|f_\text{cut}) = \frac{1}{1 + (f / f_\text{cut})^\beta}
  \label{eq:wfull}
\end{align}
The switching function is $1/2$ at the cutoff frequency $f_\text{cut}$ and the exponent $\beta$ controls its steepness.
As $\beta\rightarrow\infty$, the function approaches a hard cutoff.
Using a smooth cutoff makes the search for suitable cutoff frequencies less sensitive to noise in the rescaled sampling PSD $\hat{I}_k$, as described in the next subsection.
By default, STACIE uses $\beta=8$, for which the weight decreases from 0.9 to 0.1 over a frequency range of approximately $[0.76 f_\text{cut}, 1.31 f_\text{cut}]$.
STACIE's results are not very sensitive to the value of $\beta$.
Although other switching functions exist, this form is simple and has conceptual advantages that become clearer when rewritten in terms of the hyperbolic tangent function:
\begin{align}
  w(f|f_\text{cut}) = \frac{1}{2} \left( 1 - \tanh\left(\frac{\beta}{2}\ln\frac{f}{f_\text{cut}}\right) \right)
  \label{eq:wfull_tanh}
\end{align}
The dependence on $\ln(f/f_\text{cut})$ reveals that the switching function is scale-invariant and that $\beta$ controls the width of the transition from 1 to 0 on a logarithmic scale.

For a given $f_\text{cut}$ and $\beta$, model parameters are estimated with locally weighted Bayesian regression.~\cite{Shao2019}
For the sake of numerical efficiency, however, we use the maximum \textit{a posteriori} (MAP) estimate of the model parameters and derive the uncertainty of the parameters using the Laplace approximation.~\cite{MacKay2005}
In addition, a uniform prior is used for the model parameters, meaning that the practical implementation is equivalent to a local likelihood maximization.~\cite{Fan1998}
Local regression methods were mainly used for non-parametric smoothing of noisy data, but we use the same approach to fit a parametric model to a subset of the data.

The model parameters are determined by minimizing the negative log-likelihood of the observations under that model.
Building on the properties of the Gamma distribution, this cost function takes the following form:
\begin{align}
  \begin{split}
  \operatorname{cost}(\mathbf{b}|f_\text{cut})
  &= -\ln\mathcal{L}(\mathbf{b}|f_\text{cut})
  =-\sum_{k\in K} w(f_k| f_\text{cut})\, \ln\bigl( p_{\Gamma(\alpha_k,\theta_k(\mathbf{b}))}(\hat{I}_k)\bigr)
  \\
  &=\sum_{k\in K}
  w(f_k|f_\text{cut})\, \left[
    \ln \Gamma(\alpha_k)
    + \ln\bigl(\theta_k(\mathbf{b})\bigr)
    + (1 - \alpha_k)\ln\left(\frac{\hat{I}_k}{\theta_k(\mathbf{b})}\right)
    + \frac{\hat{I}_k}{\theta_k(\mathbf{b})}
  \right]
  \end{split}
  \label{eq:cost}
\end{align}
where
\begin{align*}
  \alpha_k &= \frac{\nu_k}{2} \quad \text{and} \quad \theta_k(\mathbf{b}) = \frac{2 I^\text{model}_k(\mathbf{b})}{\nu_k}
\end{align*}
Here, $\nu_k$ denotes the number of degrees of freedom for each component, see Eq.~\eqref{eq:nu_k}.
The frequency $f_k=k/Nh$ is the standard frequency grid in DFT analysis.
The set $K$ includes at most $\{0, 1, \ldots, \lfloor N/2 \rfloor\}$, but some elements can be excluded.
If the inputs $\hat{x}^{(m)}$ have a non-zero or biased mean, e.g.\ when $\langle \hat{x} \rangle$ is constrained, the DC component ($k=0$) should be omitted.
Furthermore, grid points with low weights can be discarded.
(In STACIE, all points with $w(f_k|f_\text{cut})$ below 0.001 are excluded.)
This form of weighted likelihood is also found in robust regression techniques that down-weight outliers.~\cite{Wang2005,Miller2018}

The initial guess of the parameters is obtained with multivariate linear regression.
For the model in Eq.~\eqref{eq:exppoly} used in work, they are found by fitting a polynomial to the logarithm of the sampling PSD.
The parameters are then refined further by numerically minimizing the cost function with the ``trust-constr'' algorithm from SciPy.~\cite{Virtanen2020}
Analytical first and second derivatives of the log-likelihood are implemented with vectorized NumPy~\cite{Harris2020} operations to make the optimization fast and robust.
These derivatives were validated against finite differences with the numdifftools library.~\cite{Brodtkorb2015}
After the optimization, the covariance matrix of the parameters in the Laplace approximations equals the inverse of the Hessian matrix of the cost function.

One can derive an estimate of the autocorrelation integral and its variance from the optimized parameters, $\hat{\mathbf{b}}$, and their estimated covariance, $\hat{\mathbf{C}}_\mathbf{b}$.
Because the parameters $\mathbf{b}$ appear in an exponential function in the model, the autocorrelation integral follows a log-normal distribution with its estimated mean and variance given by:
\begin{align}
  \begin{aligned}
    \hat{\mathcal{I}} &= \exp\left(\hat{b}_0 + \frac{\hat{\sigma}^2_{b_0}}{2}\right) \\
    \hat{\sigma}^2_\mathcal{I} &=
      \exp\Bigl(2\hat{b}_0 + \hat{\sigma}^2_{b_0}\Bigr)
      \Bigl( \exp(\hat{\sigma}^2_{b_0}) - 1 \Bigr)
  \end{aligned}
  \label{eq:lognormal}
\end{align}
For a fixed frequency cutoff, the variance quantifies the uncertainty of $\hat{\mathcal{I}}$ due to statistical fluctuations in the sampling PSD.
However, there are other sources of error that are not yet accounted for.
For example, if the cutoff frequency is too high, the spectrum below the cutoff may contain features that the model cannot explain, biasing the autocorrelation integral estimate.
Moreover, any manual or automated algorithm that selects the cutoff relies on noisy input data, meaning that the cutoff selection in the next subsection is also a source of uncertainty that must be quantified.

The use of weights has significant practical advantages: it enables the use of simpler models and reduces the computational cost by fitting only a subset of the sampling PSD.
However, it also introduces a conceptual difficulty, since the weights cannot be rigorously justified in a Bayesian framework.
A practical limitation is that the unit of the likelihood depends on the cutoff frequency.
(This is not related to the use of a smooth switching function, but rather a consequence of fitting to a variable number of data points.)
As a result, the likelihood and marginal likelihood cannot be used to compare models with different cutoff frequencies.
Instead, we use cross validation to find suitable cutoff frequencies, as described in the next subsection.

\subsection{Frequency cutoff}\label{subsec:cutoff}

\begin{figure}
  \centering
  \includegraphics{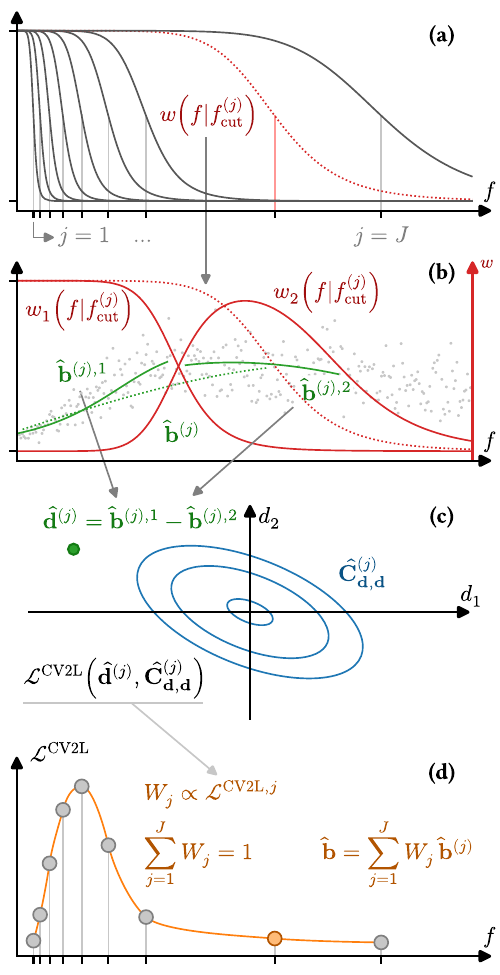}
  \caption{
    Schematic representation of the algorithm used to identify suitable cutoff frequencies.
    (a) Construction of a logarithmic cutoff grid with $J$ points and a switching function for every cutoff.
    (b) Local nonlinear regression for cutoff $j$ (dotted lines, red = switching function, green = fitted model with parameters $\hat{\mathbf{b}}^{(j)}$).
    Cross validation is implemented with two first-order corrections to the parameters, $\hat{\mathbf{b}}^{(j),1}$ and $\hat{\mathbf{b}}^{(j),2}$, for the first and second halves, using weight functions $w_1$ and $w_2$, respectively (solid lines).
    (c) Evaluation of the likelihood of the difference between the first-order corrections, $\hat{\mathbf{d}}^{(j)}$, accounting for the expected covariance $\hat{\mathbf{C}}^{(j)}_{\mathbf{d},\mathbf{d}}$.
    (d) Results are averaged over the cutoff frequency grid, where typically more than one cutoff contributes significantly.
  }
  \label{fig:cutoff-scan}
\end{figure}

A suitable cutoff frequency strikes a balance between two requirements.
If the cutoff is too high, the selected part of the spectrum becomes too complex for the model to explain, resulting in underfitting and biased parameters.
Conversely, if the cutoff is too low, useful data from the spectrum is discarded, resulting in a large variance of the parameters.
To find suitable cutoffs, we consider a grid of possible cutoff frequencies.
For each cutoff, we optimize model parameters and compute the likelihood that the same parameters are found when refitting them to the first and second halves of the data below the cutoff.
Rather than selecting ``the best'' cutoff (or optimizing it together with the model parameters), we marginalize the model parameters over the cutoff grid, as there are often multiple suitable cutoffs with slightly different predictions of $\hat{\mathcal{I}}$.
Only by considering all suitable cutoffs can the uncertainty of $\hat{\mathcal{I}}$ be reliably estimated.
This approach is similar to the marginalization over a hyperparameter in Bayesian regression,~\cite{Rasmussen2005} except that we cannot use the likelihood $\mathcal{L}(\mathbf{b}|f_\text{cut})$ introduced in the previous subsection, because its units depend on $f_\text{cut}$.
Instead, we use cross validation to construct a proxy for the likelihood, as a function of $f_\text{cut}$, whose units are independent of the amount of data used in the regression.
A visual summary of the cutoff scan is shown in Fig.~\ref{fig:cutoff-scan} and is explained in more detail below.

A logarithmic cutoff grid is used, defined by a minimal cutoff frequency $f_\text{cut,min}$ and the ratio between two neighboring grid points, $r$:
\begin{align}
  f_{\text{cut}, j} = f_{\text{cut,min}}\, r^j
\end{align}
By using a logarithmic grid, we assume a uniform prior for the logarithm of the cutoff frequency,
meaning all grid points are considered equally likely \textit{a priori}.
This step is depicted in Fig.~\ref{fig:cutoff-scan}(a).
The highest possible cutoff is determined by the Nyquist frequency ($1/2h$) but the search for suitable cutoffs can be stopped earlier in practice, as will be explained below.
The minimal cutoff frequency is determined by the relation between the effective number of points used in the regression and the cutoff frequency:
\begin{align}
    N_\text{eff} = \sum_{k\in K} w(f_k|f_\text{cut})
\end{align}
We invert the following equation to determine the minimal cutoff frequency $f_{\text{cut},min}$:
\begin{align}
  \sum_{k\in K} w(f_k|f_{\text{cut},min}) = N_\text{eff,min}
  \label{eq:neffmin}
\end{align}
where $N_\text{eff,min}$ is the minimum effective number of grid points used for the regression.
To reduce the risk of numerical issues, the default value for this lower bound is $5P$, where $P$ is the number of model parameters.
In practice, the corresponding minimal cutoff should not be assigned a significant weight by the cutoff criterion discussed below.
The ratio between two neighboring grid points is
\begin{align}
  r = \exp(g_\text{sp} / \beta)
  \label{eq:ratio}
\end{align}
where $g_\text{sp}$ is a dimensionless parameter that controls the spacing of the cutoff grid on a logarithmic scale (default value $1/2$).
The parameter $\beta$ is the exponent in the weight function $w(f|f_\text{cut})$ defined in Eq.~\eqref{eq:wfull}.
A steeper switching function (larger $\beta$) requires a finer grid of cutoff frequencies to ensure adequate sampling, which increases computational cost.

The model parameters are optimized for each cutoff index $j$, yielding parameter estimates $\hat{\mathbf{b}}^{(j)}$ and their covariance matrices $\hat{\mathbf{C}}^{(j)}_\mathbf{b}$.
For each $j$, a cutoff criterion is defined to quantify the discrepancy between the parameters fitted to the first and second halves of the low-frequency region of the spectrum.
These two halves are defined using smooth weight functions to ensure robust cross-validation:
\begin{align}
  \begin{split}
    w_1(f|f_\text{cut}) &=  w(f|g_\text{cv} f_\text{cut}/2)
    \\
    w_2(f|f_\text{cut}) &=  w(f|g_\text{cv} f_\text{cut}) - w(f|g_\text{cv} f_\text{cut}/2)
  \end{split}
  \label{eq:whalf}
\end{align}
where $g_\text{cv} \ge 1$ controls the amount of data used in the cross-validation.
The default value is $g_\text{cv}=1.25$, which means that the criterion is computed with 25\% more data than the parameters were originally fitted to.
The advantage of including more data at this stage is that the criterion will detect sooner when the model is not able to explain the spectrum, which reduces the risk of bias due to underfitting.
This step is illustrated in Fig.~\ref{fig:cutoff-scan}(b).

Cross-validation with more than two splits requires more data and is not considered to make STACIE applicable to situations where data is scarce.
While it may be tempting to compare $\hat{\mathbf{b}}^{(j)}$ to the parameters refitted to the first half only, this approach is suboptimal since these two parameter vectors are strongly correlated by construction.
Fitting to two halves strikes a good balance between robustness and sensitivity.

Instead of performing a full non-linear regression for each halve, we linearize the problem around the cutoff and estimate two first-order corrections to $\hat{\mathbf{b}}^{(j)}$ with linear regression.
For suitable cutoffs, parameters fitted to each half should not deviate much from $\hat{\mathbf{b}}^{(j)}$, which justifies the linear approximation.
Linear regression has the advantage of being computationally efficient and numerically robust.

The expected values for the linear regression are the difference between the rescaled sampling PSD and the optimized model, $\hat{I}_k - I^\text{model}(f_k, \hat{\mathbf{b}}^{(j)})$.
Linear basis functions for the regression are constructed as the first-order derivatives of the spectrum model, computed with the optimized parameters:
\begin{align}
  \left.
    \frac{\partial I^\text{model}(f_k,\mathbf{b})}{\partial b_p}
  \right|_{\mathbf{b}=\hat{\mathbf{b}}^{(j)}}
  \quad\forall p \in \{1, \ldots, P\}
\end{align}
where $P$ is the number of model parameters.
STACIE has a robust and efficient implementation of the difference, $\hat{\mathbf{d}}^{(j)}$ between the linear regression results for the two weight functions in Eq.~\eqref{eq:whalf}.
It also computes the covariance of the difference, $\hat{\mathbf{C}}^{(j)}_{\mathbf{d}}$.
If there is no underfitting, both halves should yield the same linear parameters, and the expected value of the difference between the two solutions is the null vector, as visualized in Fig.~\ref{fig:cutoff-scan}(c).
The cutoff criterion is then defined as the negative log-likelihood of observing the difference between the two halves of the spectrum below the cutoff:
\begin{align}
    \operatorname{criterion}^\text{CV2L}_j
      = -\ln \mathcal{L}^\text{CV2L}\left(
          \hat{\mathbf{d}}^{(j)},
          \hat{\mathbf{C}}^{(j)}_{\mathbf{d}}
        \right)
      = \frac{P}{2}\ln(2\pi)
        + \frac{1}{2}\ln\left|\hat{\mathbf{C}}^{(j)}_{\mathbf{d}}\right|
        + \frac{1}{2}
          \bigl(\hat{\mathbf{d}}^{(j)}\bigr)^\top
          \bigl(\hat{\mathbf{C}}^{(j)}_{\mathbf{d}}\bigr)^{-1}
          \hat{\mathbf{d}}^{(j)}
    \label{eq:cv2l}
\end{align}
The superscript ``CV2L'' stands for cross-validation with two halves and a linearized model.

When starting from the lowest cutoff frequency and moving to higher cutoffs,
the CV2L criterion will first decrease, because more data points result in a smaller variance, lowering the second term in Eq.~\eqref{eq:cv2l}.
When the cutoff increases further, the third term in Eq.~\eqref{eq:cv2l} will increase rapidly when the sampling PSD contains features that the model cannot explain.
Good cutoffs represent a compromise between these two terms, which express the variance and the bias of the fitted model, respectively.
The cutoff grid is truncated when the CV2L criterion reaches a value of $g_\text{incr}$ above the incumbent minimum (to avoid analyzing irrelevant cutoffs) or when $N_\text{eff}$ exceeds $N_\text{eff,max}$ (to limit the computational cost).
The default values for these two parameters in STACIE are
\begin{align}
  g_\text{incr} = 100 \quad \text{and} \quad N_\text{eff,max} = 1000
  \label{eq:g4g5}
\end{align}

After optimizing the parameters for a range of cutoffs $f_{\text{cut},j}\,\forall j\in\{1, \ldots, J\}$, it would be misleading to select only the cutoff with the lowest criterion.
Instead, the uncertainty of the cutoff should be accounted for by computing a weighted average of the model parameters (and their uncertainties) over the cutoff grid:
\begin{align}
  \begin{split}
    \hat{\mathbf{b}} &= \sum_{j=1}^J W_j\, \hat{\mathbf{b}}^{(j)}
    \\
    \hat{C}_{\mathbf{b},\mathbf{b}} &= \sum_{j=1}^J W_j\, \left(
      \hat{C}_{\mathbf{b}^{(j)},\mathbf{b}^{(j)}}
      + (\hat{\mathbf{b}} - \hat{\mathbf{b}}^{(j)}) (\hat{\mathbf{b}} - \hat{\mathbf{b}}^{(j)})^\top
    \right)
  \end{split}
  \label{eq:weightedaverage}
\end{align}
Here, $\hat{\mathbf{b}}^{(j)}$ and $\hat{C}_{\mathbf{b}^{(j)},\mathbf{b}^{(j)}}$ represent the parameters and their covariance, respectively, for cutoff $j$.
The weights $W_j$ are proportional to $\mathcal{L}^\text{CV2L}$ and sum to one.
This last step is illustrated in Fig.~\ref{fig:cutoff-scan}(d).
This weighted averaging is inspired by Bayesian marginalization over a hyperparameter, but it uses a model for the likelihood instead of the true likelihood of the cutoff frequency.
When using the exppoly model, Eq.~\eqref{eq:lognormal} is applied to the weighted averages from Eq.~\eqref{eq:weightedaverage} to compute the autocorrelation integral and its variance.

\subsection{Summary of algorithmic hyperparameters}

In the preceding sections, we have introduced six dimensionless algorithmic hyperparameters.
They have fixed default values in STACIE, but they can be changed by the user.
However, we recommend using the default values, which have been used for nearly all results in this work, in a follow-up publication,~\cite{Toraman2025viscosity} and for all examples in the STACIE documentation (covering a broad range of applications).~\cite{STACIE}
For the sake of transparency, we summarize all hyperparameters, how they affect the algorithm, their default values, and ranges of reasonable values in Table~\ref{tab:hyperparameters}.

\begin{table}
  \centering
  \begin{tabular}
    {l | l l l l}
    &
    \textbf{Affects} &
    \textbf{Default} &
    \textbf{Range} &
    \textbf{Eq.}
    \\
    \hline
    $\beta$ &
    Steepness of the switching function &
    $8$ &
    $[8, 20]$ &
    \eqref{eq:wfull}
    \\
    $N_\text{eff,min}$ &
    Start cutoff grid &
    $5$ &
    $[5, 40]$ &
    \eqref{eq:neffmin}
    \\
    $g_\text{sp}$ &
    Ratio between neighboring cutoff frequencies &
    $0.5$ &
    $[0.1, 0.5]$ &
    \eqref{eq:ratio}
    \\
    $g_\text{cv}$ &
    Amount of data used in cross-validation &
    $1.25$ &
    $[1.0, 2.0]$ &
    \eqref{eq:whalf}
    \\
    $g_\text{incr}$ &
    End of cutoff grid (max.\ increase CV2L criterion) &
    $100$ &
    $[10, 100]$ &
    \eqref{eq:g4g5}
    \\
    $N_\text{eff,max}$ &
    End of cutoff grid &
    $1000$ &
    $[1000, 10000]$ &
    \eqref{eq:g4g5}
    \\
    \hline
  \end{tabular}
  \caption{Overview of dimensionless algorithmic hyperparameters in STACIE.}
  \label{tab:hyperparameters}
\end{table}

The only scenario in which it may be useful to change these hyperparameters is when longer-than-necessary time series are used, resulting in a very high resolution frequency grid.
In this case, the model can be fit to more than the default $N_\text{eff,max}=1000$ points to potentially obtain a lower variance.
One can then also increase $N_\text{eff,min}$ if the sampling PSD has few degrees of freedom $\nu_k$ (and is therefore very noisy).
This reduces the risk that a model will inadvertently overfit for the smallest cutoffs.
While adjusting hyperparameters may improve the results in this case,
it generally means that shorter and more independent time series could have been used instead.
This has two practical advantages: (i) the analysis will be faster, and (ii) the plot of the sampling PSD will be less noisy.

\subsection{Preparation of sufficient inputs}\label{subsec:inputs}

The previous subsections explain how STACIE estimates the autocorrelation integral for given input data consisting of $M$ time series comprising $N$ steps each.
In order to use STACIE effectively, the user must prepare sufficient data to obtain an accurate estimate with minimal bias.
This is difficult because the amount of data required depends on how strongly it is time-correlated,
which is only known after the analysis.
This subsection proposes a practical workflow for data generation and analysis to address this chicken-and-egg problem.

Let's assume that the integral $\mathcal{I}$ must be estimated with a relative error below a certain threshold $\epsilon_\text{rel}$.
This target is first used to estimate the number of independent time series required, $M$.
Then, the required length of the time series is found by generating $M$ preliminary inputs, analyzing them with STACIE, and extending them until the effective number of grid points used to fit the model spectrum, $N_\text{eff}$, becomes sufficiently large.

Because the sampling PSD amplitudes are Gamma-distributed with shape parameter $\alpha=\nu_k/2$ and scale parameter $\theta=2C_k/\nu_k$, the relative uncertainty of the sampling PSD, defined as its standard deviation divided by its expected value, is:
\begin{align}
  \operatorname{RelErr}[\hat{I}_k] = \sqrt{\frac{2}{\nu_k}}
\end{align}
where $\nu_k$ is the number of degrees of freedom for the $k$-th component of the sampling PSD defined in Eq.~\eqref{eq:nu_k}.
An important observation is that the relative error does not depend on the spectrum amplitude and is therefore known in advance.
Except for grid points at the boundaries of the sampling PSD we have $\nu_k=2M$, and for the remainder of the analysis, we will use this value for all points.

For simplicity, consider the case where the model fitted to the spectrum is a constant (a white noise model).
This constant amplitude is then also the autocorrelation integral.
One can estimate the constant amplitude by simply taking the average of the sampling PSD below the cutoff,
and the relative uncertainty of this average for sufficiently large $N_\text{eff}$ is approximately:
\begin{align}
  \operatorname{RelErr}[\mathcal{I}] \approx \sqrt{\frac{1}{N_\text{eff}\,M}}
  \label{eq:relerr}
\end{align}
To obtain a robust fit, we recommend aiming for 20 (or more) frequency grid points per parameter in the regression.
By substituting $N_\text{eff}=20\,P$, where $P$ is the number of parameters, we can estimate the number of inputs $M$ required to achieve a relative error below a certain threshold $\epsilon_\text{rel}$:
\begin{align}
  M \approx \frac{1}{20 \, P \, \epsilon_\text{rel}^2}
  \label{eq:numseries}
\end{align}
This is only a coarse estimate due to the drastic assumptions made, but it provides a good starting point.

\begin{figure}
  \centering
  \includegraphics{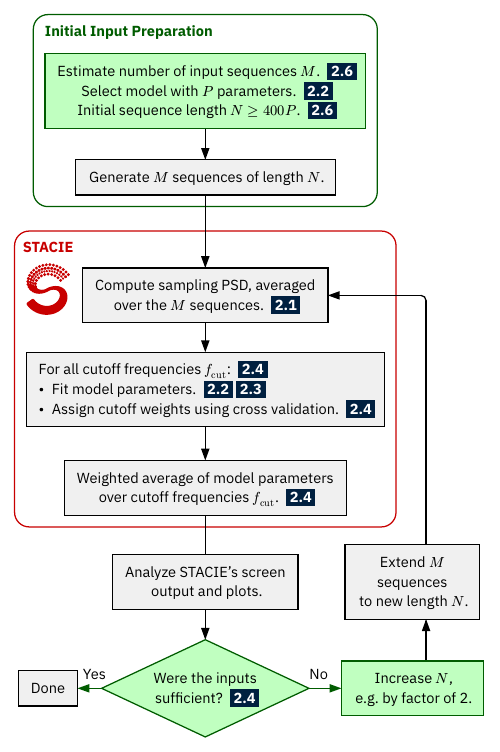}
  \caption{
    STACIE's usage flowchart.
    The green blocks represent steps where the user must decide how the time series are generated and which model to fit to the spectrum.
    Where relevant, sentences are labeled with a section number (in white text on blue background) where the corresponding steps are explained in detail.
    All steps inside the red frame are internal to STACIE and do not require user intervention.
  }
  \label{fig:flowchart}
\end{figure}

Once the number of independent time series, $M$, is fixed, the next step is to find the necessary length of each time series, $N$.
As mentioned earlier, this is unavoidably an iterative process because the required length depends on the slowest timescales in the system, which are unknown \textit{a priori}.
Therefore, we propose the workflow depicted in Fig.~\ref{fig:flowchart} to determine $N$.
We recommend starting with $M$ initial time series of length $N \ge 400 P$.
If the model can be fitted to the first $20 P$ points, then the higher frequencies included in the fit are still about a factor of $10$ below the Nyquist frequency.
This should prevent aliasing artifacts and provide a reasonable initial estimate of the low-frequency part of the spectrum.
Analyzing the initial inputs with STACIE reveals the number of points used in the fit and the relative error of the autocorrelation integral.
If the results are satisfactory, the analysis is complete.
If not, the length of the time series must be increased and the analysis repeated.
Since the frequency grid resolution is $1/(hN)$, longer time series introduce more grid points below the cutoff.
These additional points will facilitate the regression and reduce uncertainties.

STACIE also computes two Z-scores, in addition to the relative error and the effective number of points used in the fit, to help detect insufficient inputs.
The first Z-score indicates how many standard deviations the cost function in Eq.~\eqref{eq:cost} deviates from its mean.
The expected values (mean and standard deviation) are computed for a given parameter vector $\mathbf{b}$ by assuming that the data is generated by the model, representing the ideal scenario where the model fully explains the data.
This first Z-score serves as a goodness-of-fit metric and should ideally be close to zero in well-behaved cases, with deviations on the order of 1 being acceptable.
The second Z-score is computed in a similar way, but for the CV2L criterion in Eq.~\eqref{eq:cv2l}.
Significant deviations from zero suggest that the balance between bias and variance still exhibits a notable bias.
Closed-form expressions for both Z-scores are derived in STACIE's documentation.~\cite{STACIE}
In practice, high Z-score values (above 2 or higher) indicate insufficient input, hindering the identification of suitable cutoff frequencies.
Additionally, elevated Z-scores also arise when the selected model fails to adequately explain the spectrum, thereby assisting in model selection.

In hard cases, e.g.\ when the preliminary time series are too short by orders of magnitude, the extension of the time series may need to be repeated to obtain a sufficient number of grid points below the cutoff.
In addition, the relative error in Eq.~\eqref{eq:relerr} may be too optimistic, in which case either more independent time series or longer ones could be of interest.
It is therefore desirable to use a workflow for MD simulations (or other time series) in which extending the simulation time (e.g.\ with restart files) and extending the number of time series is convenient.

\subsection{Efficient trajectory storage with block averages}\label{subsec:efficient-storage}

When performing MD simulations, it is rarely useful to write out the state of the system and its properties at every time step.
At first sight, the effect of the time step $h$ on aliasing in Eq.~\eqref{eq:wiener_khinchin_disc_time} suggests that the inputs for the DFT cannot be subsampled without perturbing the sampling PSD and thus biasing the results.
This problem can be circumvented by storing block averages of the input sequences instead of subsampling them.
For a block size $B$, every block average stored on disk replaces $B$ individual values it averages over.
Blocks should not overlap in time to avoid redundancies and should be contiguous to not lose relevant information.
The block averages filter out high-frequency oscillations that would otherwise cause aliasing artifacts when subsampling.

The sampling PSD of block averages will be a good substitute for the original sampling PSD under certain conditions.
Consider a sequence length $N=B L$, where $L$ is the number of consecutive blocks.
The original sequence is $\hat{x}_n$ and we will investigate the effect on the sampling PSD when replacing $\hat{x}_n$ by a piecewise constant sequence of block averages $\hat{a}_m$:
\begin{align}
  \begin{split}
    \hat{I}_k
    &= \frac{F h}{2} \sum_{\Delta=0}^{N-1} \hat{c}_\Delta \omega_N^{-k\Delta}
    = \frac{F h}{2 N}\left|\sum_{n=0}^{N-1} \hat{x}_n \omega_N^{-kn}\right|^2
    \\
    &\approx \frac{F h}{2 N} \left|\sum_{n=0}^{N-1} \hat{a}_{\lfloor n/B\rfloor} \omega_N^{-kn}\right|^2
    \approx \frac{F h}{2 N} \left| \sum_{\ell=0}^{L-1} B \hat{a}_\ell \omega_N^{-k\ell B}\right|^2
    \approx
    \frac{F h B}{2 L} \left| \sum_{\ell=0}^{L-1} \hat{a}_\ell \omega_L^{-k\ell }\right|^2
  \end{split}
\end{align}
where
\begin{align}
  \omega_N &= \exp(2 \pi i / N)
  \\
  \omega_L &= \exp(2 \pi i / L) = \omega_N^B
\end{align}
The final result of the derivation is the sampling PSD of block averages, in which $hB$ takes the role of the new time step and $L$ is the new sequence length.
The approximations made will be small when the factor $\omega_N^{kn}$ is nearly independent of $n$ for values of $n$ within one block, e.g., $n \in [0, B]$.
This is the case when $B \ll N/k$.
By substituting $k=N_\text{eff}$ into this expression, a block size can be determined that will not cause aliasing artifacts in the part of the spectrum used to fit the model.
Just like the sufficient simulation length, an appropriate block size is determined by the frequency cutoff.
This underscores the importance of a preliminary analysis, possibly followed by extended data generation and refined analysis.

Note that in some cases, block averages can be computed more easily as finite differences of the antiderivative of $\hat{x}(t)$.
This is primarily useful for particle velocities, for which block averages can be computed as finite differences of the position.

\section{Example Applications}\label{sec:examples}
\subsection{Minimal example}

We first illustrate the usage of STACIE in Python with a minimal example that includes both data generation and analysis.
By not relying on external data sources, this self-contained example is easy to replicate, helping prospective users get started with STACIE.

The input sequences were generated using a simple Markov chain:
\begin{align}
  \hat{x}_{n+1} = \phi \hat{x}_{n} + \xi \hat{z}_n
\end{align}
where $\hat{z}_n$ is drawn from a standard normal distribution.
The parameter $\xi$ controls the magnitude of the noise term and $\phi$ controls the decay rate of the chain, with $0 < \phi < 1$.
The discrete ACF of this chain decays exponentially:
\begin{align}
  c_\Delta = \frac{\phi^{|\Delta|} \xi^2}{1-\phi^2}
\end{align}
and the ground truth of the autocorrelation integral (with $F=1$) is:
\begin{align}
  \mathcal{I} = \frac{1}{2} \sum_{\Delta=-\infty}^{\infty} c_\Delta = \frac{\xi^2}{2(1-\phi)^2}
\end{align}
Finally, the integrated correlation time, which is also estimated by default in STACIE, is:
\begin{align}
  \tau_\text{int} = \frac{\mathcal{I}}{c_0} = \frac{1 + \phi}{2(1-\phi)}
\end{align}

For the minimal example, we used parameters $\phi=\frac{31}{33}$ and $\xi=\sqrt{\frac{8}{1089}}$, which are chosen to yield $\mathcal{I}=1$ and $\tau_\text{int}=16$.
Because the ACF decays exponentially, we expect the PSD to have a Lorentzian shape with a maximum at the origin.
Hence, we used the model in Eq.~\eqref{eq:exppoly} with degrees $S=\{0, 2\}$ to fit the low-frequency part of the spectrum, meaning that $P=2$.
For this example, we aimed to estimate the autocorrelation integral with an error of about 2\%,
for which the suggested number of input time series, see Eq.~\eqref{eq:numseries}, is $M\approx1/(20\,P\,0.02^2)\approx64$.

Starting with 1024 steps, and doubling the number of steps until the desired accuracy is reached, the required number of steps was found to be 32768.
In the final run, we obtained $N_\text{eff} \approx 117 > 20\, P$, which is more than originally planned.
Note that this example has a negligible computational cost: it completed on an Intel i7 CPU in about 1 second (plotting excluded).

\begin{figure}
  \centering
  \includegraphics{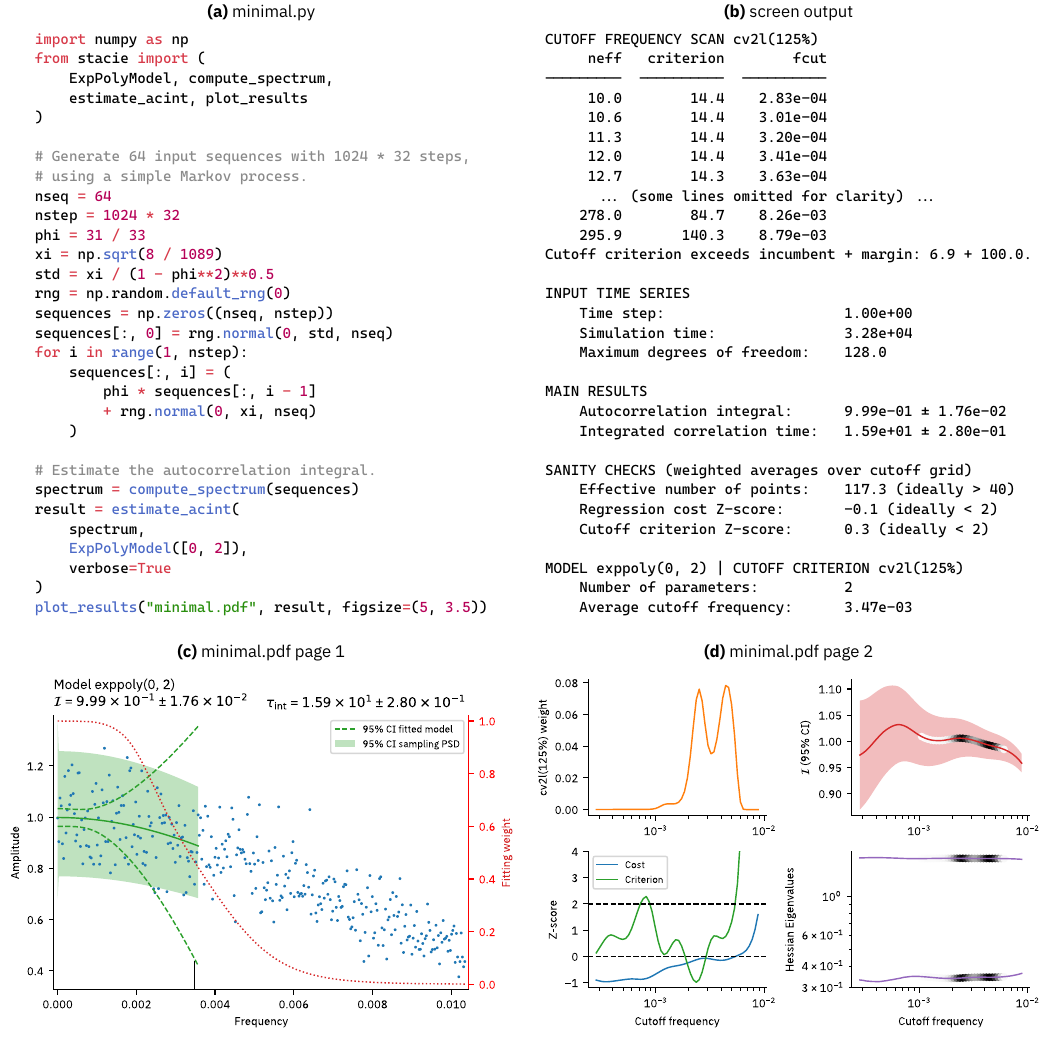}
  \caption{
    Minimal example of how to use STACIE.
    (a) Python source code to sample the Markov chain and compute the autocorrelation integral.
    (b) The screen output of the script.
    (c) Low-frequency part of the sampling PSD (blue dots), fitted model (green curve), its 95\% confidence interval (green dashed lines), 95\% confidence interval of the spectral data according to the model (green band) and the weighted average of the switching function over all cutoff frequencies (dotted red curve).
    (d) Intermediate results of the analysis: Cutoff weight $W_j$ (top-left), estimated autocorrelation integral and its 95\% confidence interval as a function of cutoff frequency (top-right), Z-scores to help detecting regression issues (bottom-left), and eigenvalues of the Hessian of the cost function in pre-conditioned parameter space (bottom-right).
    See text for a detailed explanation.
  }
  \label{fig:minimal}
\end{figure}

Figure~\ref{fig:minimal}(a) shows the Python source code for this example.
The first code block imports the required packages: NumPy and STACIE.
The second block implements the Markov chain using vectorized NumPy operations.
The chain is initialized with samples from the stationary distribution, eliminating the need for an equilibration run.
The third block demonstrates the use of STACIE, which typically involves three steps.
First, the sampling PSD is computed, internally using NumPy's FFT library.
If not specified, the default time step $h=1$ and factor $F=1$ are assumed.
Second, the autocorrelation integral is estimated using the algorithm described in Section~\ref{sec:algorithm}.
Finally, A 2-page PDF document is generated, showing the model fitted to the spectrum and some intermediate results, which will be discussed below.

Figure~\ref{fig:minimal}(b) displays the screen output of the Python script.
First, STACIE prints the progress of the cutoff frequency scan with three columns: $N_\text{eff}$, $-\ln \mathcal{L}^\text{CV2L}$ and $f_\text{cut}$.
After completing the scan, all results are marginalized over the cutoff frequency and a summary of the results is printed on screen.
Figures~\ref{fig:minimal}(c) and \ref{fig:minimal}(d) are the plots saved to ``minimal.pdf''.

Figure~\ref{fig:minimal}(c) shows the main results of the regression after marginalization over the cutoff frequency.
The sampling PSD is shown as blue dots.
The dotted red curve represents the weighted average of the switching function in Eq.~\eqref{eq:wfull} over all cutoffs.
The solid green curve is the fitted model, computed using the weighted average of the parameters over all cutoffs.
The green band represents the 95\% confidence interval of the sampling PSD.
In the frequency domain where the model is fitted, most data points are expected to lie in the green band.
The dashed green curves represent the 95\% confidence interval of the fitted model, derived from the covariance of the parameters.
Finally, the estimated integral and integrated correlation time, displayed in the plot title, agree with the ground truth within the predicted uncertainties.

Figure~\ref{fig:minimal}(d) shows several intermediate results, which can be used to analyze potentially failed fits and gain deeper insight into how STACIE works.
The top-left plot shows the weights $W_j$ from Eq.~\eqref{eq:weightedaverage}, which were used to average over the cutoff frequency.
This plot highlights that there was an entire range of viable cutoffs instead of just one single suitable cutoff.
The top-right plot shows the predicted integral and its 95\% confidence interval as a function of the cutoff frequency.
Cutoffs with higher weights are represented as darker dots.
This plot demonstrates how selecting a single cutoff frequency can lead to biased results and an underestimation of uncertainty.
The bottom-left plot shows the Z-scores which remain sufficiently low for cutoff frequencies where the cutoff weight is high.
Finally, the bottom-right plot shows the eigenvalues of the Hessian of the cost function in Eq.~\eqref{eq:cost}, in a pre-conditioned parameter space.
This plot is occasionally helpful for detecting conditioning problems in the fit, though no such issues were evident in this case.

In summary, this minimal example demonstrates how to use STACIE with just a few lines of Python code.
The default screen output and plots offer a comprehensive overview of the results, including the fitted model, its uncertainties, and intermediate outputs that provide deeper insights.
In more realistic use cases, the preparation and conversion of data into NumPy arrays also requires some effort, while the actual analysis is straightforward.
Because data preparation is very application specific, we made sure STACIE is completely agnostic to the data source, and provide examples in the documentation for several popular MD engines.

\subsection{Ionic electrical conductivity of an electrolyte}\label{sec:electrolyte}

Electrolytes play a crucial role in energy storage, catalysis, and biological processes, enabling technologies such as batteries, fuel cells, and desalination systems~\cite{Chu2016}.
Accurately calculating their transport properties, particularly ionic conductivity, through MD simulations is essential for designing sustainable energy solutions.
In this section, we demonstrate the application of STACIE to estimate the ionic conductivity of a benchmark electrolyte, specifically an aqueous NaCl solution.
Our results are directly compared to those of Gullbrekken \textit{et al.}~\cite{Gullbrekken2023}, who derived transport properties using the OCTP plug-in~\cite{Jamali2019} of LAMMPS.
For this example, their LAMMPS input files~\cite{GullbrekkenZenodo2023} were adapted to prepare inputs for STACIE~\cite{md-electrolyte}.

\begin{figure}
    \centering
    \includegraphics[width=3.333in]{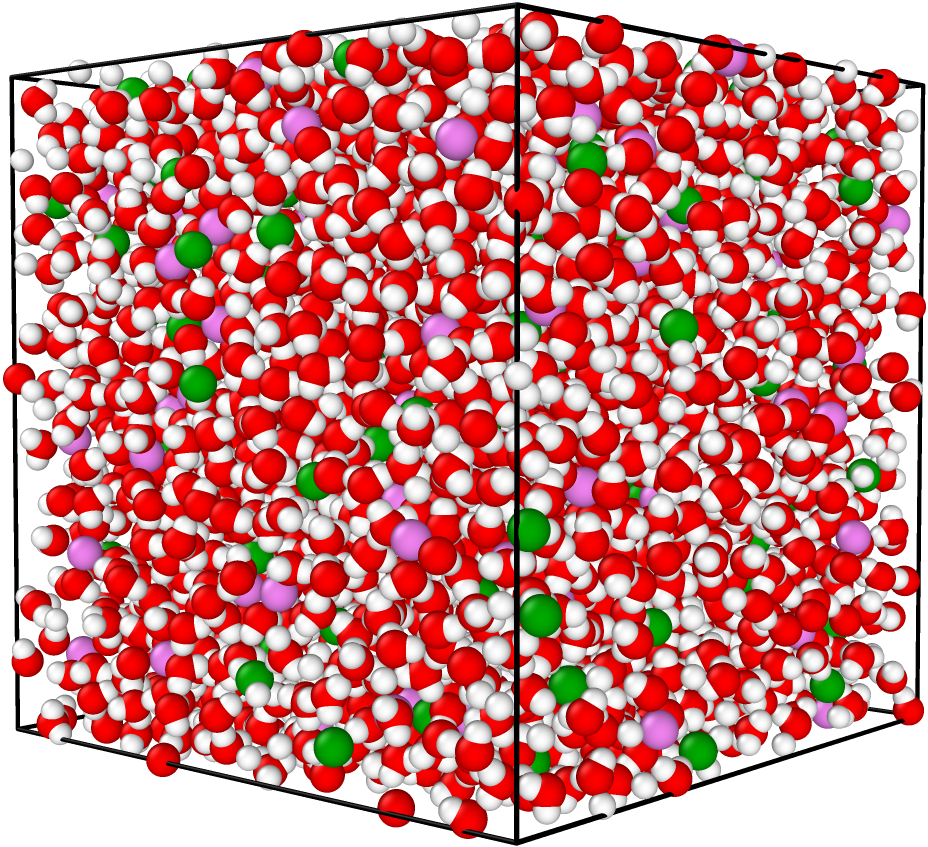}
    \caption{
      Water-NaCl electrolyte system used in our simulations.
      Hydrogen atoms are depicted in white, oxygen atoms in red, sodium ions in violet, and chloride ions in green.
    }
    \label{fig:water-nacl}
\end{figure}

EMD simulations were performed under ambient conditions (\SI{293}{\kelvin}, \SI{1}{\atm}) on a system consisting of 3000 water molecules and 140 $\text{Na}^+$ and $\text{Cl}^-$ ions each,
corresponding to a concentration of \qty{2.5}{\mol\per\liter}, as shown in Figure~\ref{fig:water-nacl}.
The interactions between water molecules were modeled using the SPC/E force field~\cite{Berendsen1987}, with bond lengths and angles constrained via the SHAKE algorithm~\cite{Ryckaert1977}.
Ion parameters for $\text{Na}^+$ and $\text{Cl}^-$ were taken from the work of Weerasinghe and Smith~\cite{Weerasinghe2003}.

As in the work of Gullbrekken \textit{et al.}~\cite{Gullbrekken2023}, NpT equilibration and production runs were used to determine the average volume of the system.
The system is then further equilibrated in the NVT ensemble, followed by a \qty{1}{\nano\second} NVT production run from which the conductivity is derived.
Unlike their work, we repeated this process 5 times starting from different initial ion insertions to obtain independent trajectories, instead of heating and cooling the system before performing additional NVT runs.
Our approach is easier to run in parallel, but should not affect the results significantly.

In the NVT production runs, the charge current is computed as a function of time:
\begin{align}
  \hat{\mathbf{J}}^\text{c}(t) = \sum_{n=1}^{N_q} q_n \hat{\mathbf{v}}_n(t),
  \label{eq:charge-current}
\end{align}
where $\hat{\mathbf{v}}_n(t)$ are the ion velocities and $q_n$ are the corresponding charges.
During the LAMMPS simulations, charge current components were block-averaged every \qty{50}{\femto\second} using the ``fix ave/atom'' command to provide input for STACIE.
The electrical conductivity of the system is defined as~\cite{Hansen2013,Fong2020}:
\begin{align}
    \sigma = F\,\frac{1}{2}
        \int_{-\infty}^{+\infty}
        \left\langle
        \hat{J}^\text{c}_i(t_0) \,,\, \hat{J}^\text{c}_i(t_0 + \Delta_t)
        \right\rangle
        \,\mathrm{d}\Delta_t,
    \label{eq:conductivity}
\end{align}
where $\hat{J}^\text{c}_i(t)$ represents a Cartesian component $i$ of the charge current
and the angle brackets $\langle \cdot \rangle$ denote the ensemble average.
Practically, we average over the three Cartesian components, the time origins $t_0$, and the replicas of MD simulations with consistent settings.
The factor $F$ is given by:
\begin{align}
  F = \frac{1}{V k_\text{B} T},
\end{align}
where $V$ is the simulation box volume, $k_\text{B}$ is the Boltzmann constant, and $T$ is the temperature.
Note that the usual factor $1/3$ is not needed in $F$ because in STACIE it is implied in the average over Cartesian components in $\langle \cdot \rangle$.

Since the five replicas were equilibrated independently using NpT simulations, the box sizes differed slightly during the subsequent NVT production runs.
The replica volumes differ by about 0.01\%, and the average of these volumes was used to compute $F$.
The number of time series used as input in Eq.~\eqref{eq:wiener_khinchin_fully_disc} is $M = 3 \times 5$, as there are three components of the charge current and five MD trajectories.

STACIE was used to construct an appropriate sampling PSD from the charge current data,
employing the factor $F$ discussed earlier.
Due to the low value of $M$, the spectrum was fitted using a model with degrees $S=\{0,1\}$.
According to Eq.~\eqref{eq:numseries}, $M=15$ time series is approximately sufficient for a relative error of \qty{4}{\percent} when fitting a model to $20\, P = 40$ points, with $P=2$ for the number of model parameters.

\begin{figure}
  \centering
  \includegraphics{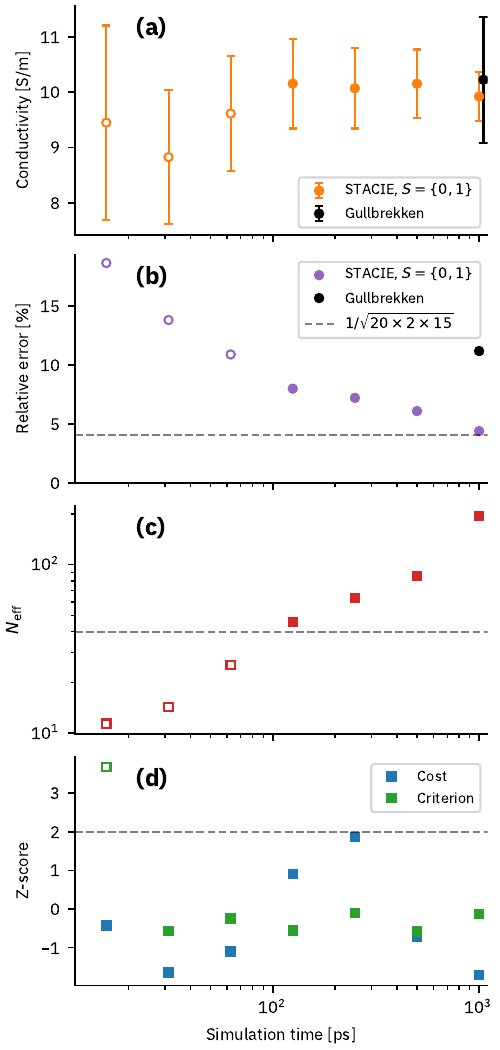}
  \caption{
    STACIE results for the ionic conductivity as a function of simulation time.
    (a) Ionic conductivity with its standard error estimated from five EMD trajectories.
    The value from Gullbrekken \textit{et al.}~\cite{Gullbrekken2023} is shown as a black dot (slightly shifted to the right for clarity).
    STACIE results that pass all sanity checks (i.e. $N_\text{eff} > 20 P$ and Z-scores $> 2$) are shown as filled dots.
    (b) Relative error of the estimated conductivity.
    (c) Effective number of grid points $N_\text{eff}$ used in the fit.
    The dashed line represents the recommended $N_\text{eff} = 20 \,P = 40$.
    (d) The Z-scores derived from the regression cost function (green) and the cutoff criterion (yellow).
  }
  \label{fig:simtime}
\end{figure}

To facilitate a direct comparison, we maintained the same total simulation time as Gullbrekken \textit{et al.}.
However, we advise starting with shorter trajectories, and progressively extending the simulations (or incorporating additional trajectories) until the statistical errors are acceptably small.
To demonstrate this approach, we conducted the STACIE analysis using truncated trajectories with simulation times $\in \{15.6, 31.25, 62.5, 125, 250, 500, 1000\}$ \unit{\pico\second}.
Figure~\ref{fig:simtime}(a) shows the evolution of the estimated conductivity and its standard error for each simulation time, and Figure~\ref{fig:simtime}(b) presents the corresponding relative errors.
Filled dots indicate results that satisfied all sanity checks depicted in the subsequent two figures.
Figure~\ref{fig:simtime}(c) displays the effective number of grid points $N_\text{eff}$ employed in the fit, reaching the recommended value of 40 for simulation times exceeding \SI{100}{\pico\second}.
Lastly, Figure~\ref{fig:simtime}(d) shows the Z-scores derived from both the regression cost function and the cutoff criterion.
Z-scores above 2 suggest potentially questionable fits, typically resulting from insufficient input data, though this is infrequent here due to the model's simplicity.
Supplementary plots for these analyses are available in Section S2 of the Supporting Information.

For shorter simulation times, the estimated conductivity exhibited larger uncertainties.
As the simulation time increased, the uncertainty progressively decreased, and the number of effective grid points $N_\text{eff}$ increased, indicating improving statistical sampling.
The relative error from the \qty{1}{\nano\second} trajectories approached \qty{4.4}{\percent} for the full simulation, and STACIE used about 195 effective points instead of 40, demonstrating that Eq.~\eqref{eq:numseries} should be treated as a guideline rather than a strict rule.

There are no compelling physical reasons to specifically choose polynomial degrees $S=\{0, 1\}$ and our selection was primarily motivated by the model's simplicity.
To demonstrate STACIE's ability to handle more complex models, identify poor fits, and produce reliable results in other cases, we also tested models with various set of polynomial degrees $\{0\}$, $\{0, 1, 2\}$, $\{0, 1, 2, 3\}$, $\{0, 2\}$, and $\{0, 2, 4\}$.
Plots analogous to Figure~\ref{fig:simtime} for these models are available in Section S3.1 of the Supporting Information.
In all cases where the analysis passes the sanity checks, which consistently occurs when using the full simulation time, the results align with those obtained using the model with $S=\{0, 1\}$.
The QQ-plot in Section S3.2 of the Supporting Information confirms the validity of the error estimates.

Table~\ref{tab:electrolyte-results} summarizes the ionic conductivities obtained in this work and compares them with the results of Gullbrekken \textit{et al.}~\cite{Gullbrekken2023} and experimental values reported by Bešter-Rogač \textit{et al.}~\cite{BeterRoga2000}.
STACIE's estimate for all models considered align well with the findings of Gullbrekken \textit{et al.}, who used equivalent EMD trajectories as input.
Notably, STACIE achieved smaller error bars, in particular for $S=\{0, 2, 4\}$, demonstrating its ability to extract relevant information with improved statistical efficiency.
All simulations results underestimated the experimental value, which was expected due to the limitations of the SPC/E water model and the force field used for ions.~\cite{Gullbrekken2023}

\begin{table}
    \centering
    \caption{
      Comparison of ionic electrical conductivities for the water-NaCl electrolyte system at $T=\qty{20}{\celsius}$ and $p=\qty{1}{\atm}$.
      The STACIE results were obtained with different models and five replica EMD trajectories as input.
      The value from Gullbrekken \textit{et al.}~\cite{Gullbrekken2023} is derived using comparable MD inputs, employing the OCTP method~\cite{Jamali2019}.
      The experimental value is taken from the work of Bešter-Rogač~\cite{BeterRoga2000}.
    }
    
\begin{tabular}{
  lc
  S[
    table-format = 2.4,
    table-number-alignment = center,
    separate-uncertainty = true,
    table-figures-uncertainty = 2
  ]
}
  &
  \textbf{Concentration} [\unit{\mol/\liter}] &
  \multicolumn{1}{c}{\textbf{Conductivity} [\unit{\siemens/\meter}]}
  \\ \hline
  $S=\{0\}$ & 2.5 & 10.05 \pm 0.65 \\
  $S=\{0, 1\}$ & 2.5 & 9.92 \pm 0.44 \\
  $S=\{0, 1, 2\}$ & 2.5 & 9.94 \pm 0.50 \\
  $S=\{0, 1, 2, 3\}$ & 2.5 & 10.22 \pm 0.46 \\
  $S=\{0, 2\}$ & 2.5 & 10.62 \pm 0.49 \\
  $S=\{0, 2, 4\}$ & 2.5 & 10.82 \pm 0.27 \\
  Gullbrekken &
  2.5 &
  10.23 \pm 1.14
  \\
  Bešter-Rogač &
  2.0 &
  13.196 \pm 0.026
  \\
  Bešter-Rogač &
  3.0 &
  16.723 \pm 0.033
  \\ \hline
\end{tabular}

    \label{tab:electrolyte-results}
\end{table}

In summary, this application of STACIE to the ionic conductivity of an aqueous NaCl solution demonstrates its effectiveness in analyzing EMD data.
We recommend starting with relatively short trajectories and gradually increasing the simulation time until the desired accuracy is achieved.
STACIE automatically identifies the portion of the sampling PSD to be included in the fit, and implements simple sanity checks to ensure the reliability of the results.
This is a significant advantage over the OCTP plugin in LAMMPS, where users must still make subjective decisions to determine a suitable part of the MSD to fit.
Since such choices are not unique and can influence the final outcome, they represent a source of uncertainty that is difficult to quantify and control.
By eliminating such subjective decisions, STACIE enhances the reproducibility and robustness of the results.

\section{Validation}\label{sec:validation}
This section first introduces a massive data set, called the AutoCorrelation Integral Drill (ACID), for validating algorithms that estimate the autocorrelation integral.
It was designed to accurately quantify three types of limitations of such algorithms: (i) systematic errors in the prediction of the integral, (ii) systematic errors in the predicted uncertainty, and (iii) suboptimal scaling of these errors with the amount of input data.
With the ACID test set, we validated STACIE's performance, which is discussed below.

The ACID test set consists of time-correlated sequences, sampled from GPs with 12 different covariance kernels.
The kernels consist of linear combinations of three kernel models, defined in continuous time and frequency domain as follows:
\begin{enumerate}
    \item
    The White Noise model assumes uncorrelated data, for which the ACF is a Dirac delta function and the PSD is constant:
    \begin{align}
        c(\Delta_t) = C_0 \delta(\Delta_t) \quad \text{and} \quad C(f) = C_0
    \end{align}
    This model is denoted as $\operatorname{W}(C_0)$.

    \item
    The Exponential model features an exponentially decaying ACF with a characteristic time $\tau$:
    \begin{align}
      c(\Delta_t) = \frac{C_0}{2 \tau}\exp\left(-\frac{|\Delta_t|}{\tau}\right)
    \end{align}
    and the corresponding PSD is
    \begin{align}
      C(f) = \frac{C_0}{1 + (2 \pi f \tau)^2}
    \end{align}
    This model is denoted as $\operatorname{E}(C_0,\tau)$.

    \item
    The Stochastic Harmonic Oscillator is adapted from the work of Foreman--Mackey \textit{et al.}~\cite{ForemanMackey2017} and has the following ACF:
    \begin{align}
       c(\Delta_t) = C_0 \pi f_0 Q \exp\left(-\frac{\pi f_0 \Delta_t}{Q}\right) \begin{cases}
          \cosh(\eta 2\pi f_0 \Delta_t) + \frac{1}{2\eta Q} \sinh(\eta 2\pi f_0 \Delta_t)
          &
          \qquad \text{if}\quad 0 < Q < \frac{1}{2}
          \\[1em]
          1 + 2 \pi f_0 \tau
          &
          \qquad \text{if}\quad Q = \frac{1}{2}
          \\[1em]
          \cos(\eta 2\pi f_0 \Delta_t) + \frac{1}{2\eta Q} \sin(\eta 2\pi f_0 \Delta_t)
          &
          \qquad \text{if}\quad Q > \frac{1}{2}
       \end{cases}
    \end{align}
    where $\eta$ is defined as:
    \begin{align}
        \eta = \left|\frac{1}{4Q^2} - 1\right|^{\frac{1}{2}}
    \end{align}
    The corresponding PSD is:
    \begin{align}
        C(f) = \frac{C_0 f_0^4}{(f^2 - f_0^2)^2 + (f f_0/Q)^2}
    \end{align}
    $Q$ represents the quality of the oscillator, $f_0$ is the resonant frequency, and $C_0$ is the zero-frequency limit of the spectrum.
    This model is denoted as $\operatorname{S}(C_0,f_0,q)$.
    The notation and conventions deviate from those of Foreman--Mackey \textit{et al.}~\cite{ForemanMackey2017} to ensure consistency within this article.
\end{enumerate}

Using these three models, covariance kernels were either a single kernel model with specific parameters or a linear combination of two, as shown in Table~\ref{tab:kernels}.
While the kernels were designed to produce diverse time series, they do share some common traits.
In all cases, the theoretical PSD at zero frequency is exactly one,
which is also the expected value of the autocorrelation integral (from $-\infty$ to $+\infty$) with $F=2$.
Furthermore, for each test case, it was numerically verified that the first 20 amplitudes of the discrete PSD of the analytical kernel deviate by no more than 2.5\% RMS from an even quadratic function, and the first 40 amplitudes deviate by no more than 10\%.
This ensures that the sampled sequences are all sufficiently long to resolve the spectrum of the slowest variations.

\begin{table}
    \centering
    \begin{tabular}{llc}
    \textbf{Kernel} &
    \textbf{Definition} &
    $\tau_\text{int}$
    \\
    \hline
    exp1p & $\operatorname{E}(1.0, 5.0)$ & 5.000 \\
exp1w & $\operatorname{E}(0.9, 5.0) + \operatorname{W}(0.1)$ & 2.632 \\
exp2 & $\operatorname{E}(0.5, 2.0) + \operatorname{E}(0.5, 5.0)$ & 2.857 \\
sho1pcrit & $\operatorname{S}(1.0, 0.04, 0.5)$ & 7.958 \\
sho1pover & $\operatorname{S}(1.0, 0.15, 0.2)$ & 5.305 \\
sho1punder & $\operatorname{S}(1.0, 0.03, 1.4)$ & 3.789 \\
sho1wcrit & $\operatorname{S}(0.9, 0.04, 0.5) + \operatorname{W}(0.1)$ & 3.194 \\
sho1wover & $\operatorname{S}(0.9, 0.15, 0.2) + \operatorname{W}(0.1)$ & 2.705 \\
sho1wunder & $\operatorname{S}(0.9, 0.03, 1.4) + \operatorname{W}(0.1)$ & 2.286 \\
sho2crit & $\operatorname{S}(0.8, 0.04, 0.5) + \operatorname{S}(0.2, 0.35, 0.1)$ & 6.920 \\
sho2over & $\operatorname{S}(0.8, 0.15, 0.3) + \operatorname{S}(0.2, 0.35, 0.1)$ & 3.701 \\
sho2under & $\operatorname{S}(0.8, 0.03, 1.4) + \operatorname{S}(0.2, 0.35, 0.1)$ & 3.920 \\
\hline

    \end{tabular}
    \caption{
        Covariance kernels and their parameters used to generate the ACID test set.
        The integrated correlation time is derived from the analytical models with Eq.~\ref{eq:tauint_cont}.
    }
    \label{tab:kernels}
\end{table}

For each kernel listed in Table~\ref{tab:kernels}, sequences with lengths $N =$ 1024,  4096, 16384, or 65536 were generated, with a dimensionless time step of 1.
For each kernel and sequence length, different test sets were included, comprising $M =$ 1, 4, 16, 64, or 256 independent one-dimensional sequences each.
The systematic increase of $M$ and $N$ by powers of 4 allows for a straightforward validation of the decrease in uncertainty: ideally, a four-fold increase in data, decreases the error by a factor of two.
Finally, for every combination of kernel, $N$ and $M$, 64 independent cases were constructed by applying a Fourier filter to white noise.~\cite{Yamazaki1988}
Each test case was generated with a unique random seed to eliminate any correlation between different test cases.
By applying an algorithm to a set of 64 test cases that only differ by their random seed, one can accurately compare the estimated uncertainty with the distribution of predicted autocorrelation integrals.
The total number of test cases is 12 (covariance kernels) $\times$ 4 (sequence lengths) $\times$ 5 (different numbers of independent 1D sequences) $\times$ 64 (unique random seeds) = 15360.
All time series, along with metadata such as the analytical PSD, can be regenerated (and reanalyzed) with Python scripts and StepUp workflows~\cite{StepUp} provided in the ACID test set~\cite{ACID}.
The workflow stores the timeseries in ZARR files~\cite{ZARR2025}, which occupy approximately 80 GB of disk space in total.
The validation of an algorithm with the ACID test set is obviously only feasible if it can be applied to all test cases without manual user intervention.

\begin{figure}
  \includegraphics{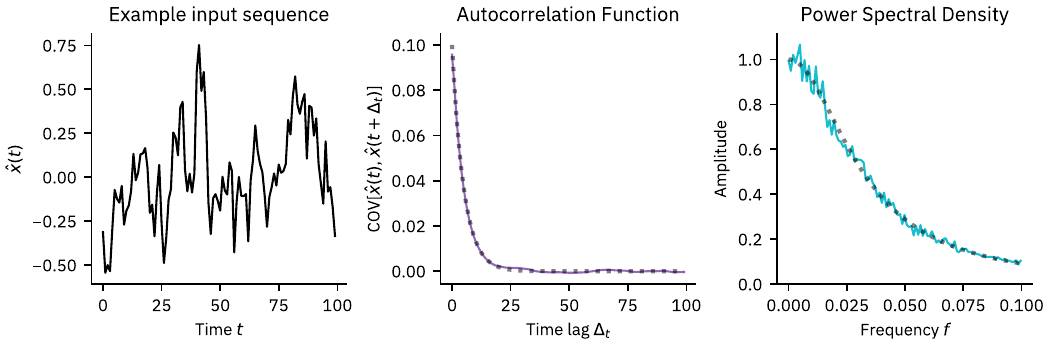}
  \caption{
      Example test case in the ACID test set for kernel exp1p with $N=1024$ and $M=256$.
      (a) The first 100 steps of a single time series.
      (b) Ground truth of the ACF (dotted) and sampled ACF (solid purple) up to $\Delta_t = 100$.
      (c) Ground truth of the PSD (dotted) and sampled PSD (solid turquoise) up to $f=0.1$.
  }
  \label{fig:example-exp1p}
\end{figure}

To validate STACIE, it was applied to all test cases, using a similar approach to the analysis part of the minimal example, see the third code block of Figure~\ref{fig:minimal}(a).
The only differences from the minimal example were: (i) $F=2$ instead of $1$, and (ii) the maximum number of points included in the fit was set to $N/8$.
STACIE completed all test cases without error messages.
In the main text, we only present results for the kernel exp1p.
Results for all other kernels, which exhibit comparable trends, are provided in Section S4 of the Supporting Information.
An example test case for the kernel exp1p is visualized in Figure~\ref{fig:example-exp1p}.

\begin{figure}
    \centering
    \includegraphics{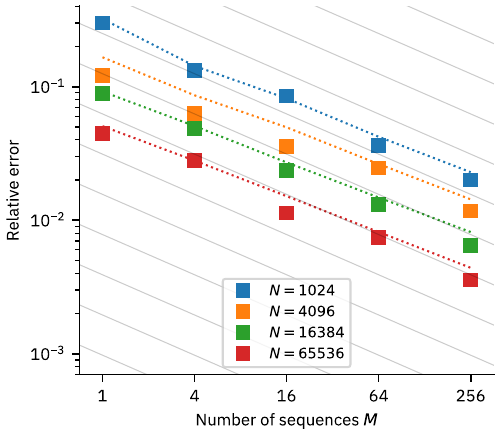}
    \caption{
        Scaling of the errors of STACIE's predictions for kernel exp1p in Table~\ref{tab:kernels}.
        Each data point represents a statistic over 64 independent tests for a given sequence length ($N$) and number of sequences averaged over ($M$, horizontal axis).
        Filled squares: the standard deviation on the predicted autocorrelation integral.
        Dotted lines: the predicted uncertainty.
        The slope of the light gray grid lines corresponds to the ideal decrease in error with increasing amount of data ($\propto 1/\sqrt{M}$) .
        Their spacing corresponds to the ideal decrease in error when quadrupling the sequence length.
    }
    \label{fig:error-scaling-exp1p}
\end{figure}

Figure~\ref{fig:error-scaling-exp1p} illustrates the scaling of errors as a function of $N$ and $M$ for the kernel exp1p.
The filled squares show the standard deviation of STACIE's estimates of the autocorrelation integral over 64 independent test cases.
These deviations decreased as more data was included, closely approximating the ideal scaling ($\propto 1/\sqrt{NM}$) indicated by the slanted grid lines.
For the most data-rich cases, relative errors below 1\% were achieved.
The dotted lines represent the root-mean-square value of the errors on the integral predicted by STACIE over all 64 tests.
Notably, the predicted uncertainty aligned closely with the actual uncertainty (represented by the filled squares).

\begin{figure}
    \centering
    \includegraphics{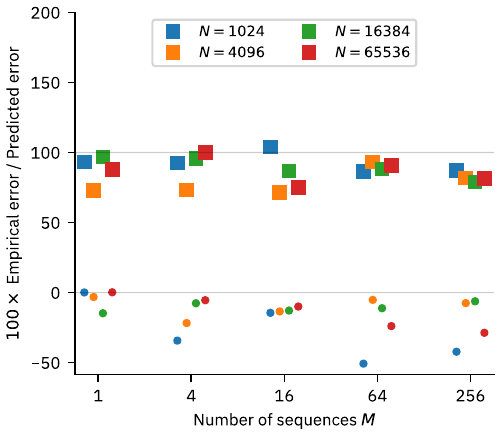}
    \caption{
        Assessment of the uncertainty quantification of STACIE for the kernel exp1p.
        Each data point represents a statistics over 64 independent tests for a given sequence length ($N$) and number of sequences averaged over ($M$, horizontal axis).
        Filled squares: the standard deviation of the predicted autocorrelation integral divided by the predicted uncertainty.
        Dots: the mean error divided by the predicted uncertainty.
    }
    \label{fig:error-ratios-exp1p}
\end{figure}

To further clarify the quality of the uncertainty quantification, Figure~\ref{fig:error-ratios-exp1p} shows the standard deviation of the estimates and the mean error, both normalized by the predicted uncertainty.
These should not be confused with relative errors, which were generally much smaller.
This plot shows that the predicted uncertainty closely matched the standard deviation of STACIE's output over 64 independent tests for different values of $M$ and $N$.
The mean error (plotted as dots) revealed a small bias, in particular for the shortest sequences ($N=1024$), but remained smaller than the predicted uncertainty.
Ideally, these points should be closer to 0, but, given their small magnitude, STACIE's uncertainty estimates are still practically valuable.
As the tables in Section S4 of the Supporting Information show, the effective number of points used in the fits for $N=1024$ was nearly always below $20P$, which explains the poorer performance for these inputs.

\begin{figure}
    \centering
    \includegraphics{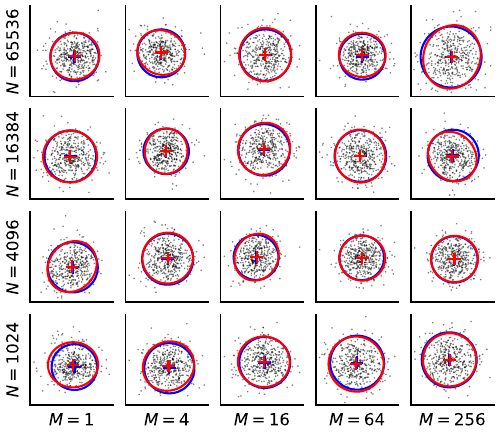}
    \caption{
        Monte Carlo sampling of the model parameters for kernel exp1p,
        for different combinations of $M$ and $N$.
        The MAP results are plotted in blue and the MC results in red.
        Crosses represent the mode (in case of MAP) and mean (in case of MC) of the distribution.
        The ellipses represent the 2$\sigma$ confidence region.
        Scatter points represent the ensemble at the end of the MC chain used for analysis.
        Tick marks are omitted for clarity and because the data are plotted in a reduced parameter space, in which the MAP covariance becomes the identity matrix.
    }
    \label{fig:monte-carlo-exp1p}
\end{figure}

One possible explanation for the small observed bias is that STACIE makes a maximum a posteriori (MAP) estimate of the model parameters, which may not necessarily provide the best estimate of the mean.
To test this hypothesis, we performed a Monte Carlo (MC) sampling of the model parameters for all kernels, and all combinations of $N$ and $M$.
To limit the computational cost, we only considered one out of 64 independent tests for each combination of kernel, $N$ and $M$. Additionally, we only performed the sampling for the frequency cutoff with the lowest value of the CV2L criterion in the frequency scan.
All MC simulations were performed with the emcee Python package, \cite{ForemanMackey2013} using an ensemble of 400 walkers.
The number of steps was chosen to ensure that the integrated correlation time of the last 90\% of the Markov chain was less than 2\% of the number of MC iterations.
The final state of the ensemble of 400 samples was used for further analysis.
Figure \ref{fig:monte-carlo-exp1p} shows the results for the kernel exp1p.
To clarify the visual comparison, all parameters are transformed to a new basis in which the covariance matrix of the MAP estimate is the identity matrix.
In general, the MAP (blue) and MC (red) results agreed well with each other for all cases considered.
The minute deviations between the two were too small to explain the mean error observed in Figure~\ref{fig:error-ratios-exp1p}.

\begin{figure}
    \centering
    \includegraphics{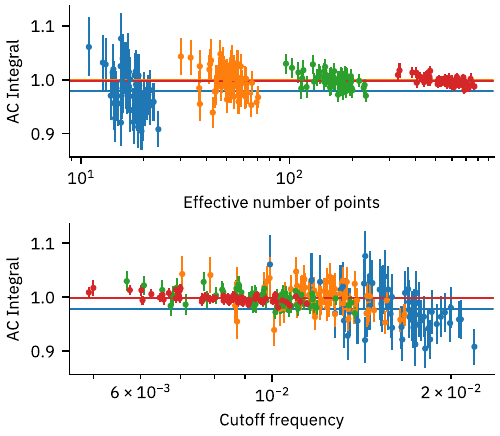}
    \caption{
        The estimated autocorrelation integral as a function of the effective number of fitting points (top panel) and the cutoff frequency $f_\text{cut}$ (bottom panel), for all 64 test cases with kernel exp1p, for $M=64$, and for sequence lengths $N\in\{1024, 4096, 16384, 65536\}$.
        The color of the points indicates the sequence length $N$, using the same color scheme as in Figures~\ref{fig:error-scaling-exp1p} and \ref{fig:error-ratios-exp1p}.
    }
    \label{fig:cutoff-exp1p}
\end{figure}

To gain more insight into the bias of the predicted integral, Figure~\ref{fig:cutoff-exp1p} shows how the estimate of $\mathcal{I}$ depends on the cutoff frequency for kernel exp1p.
Only tests with $M=64$ are included for clarity.
This plot reveals that larger cutoffs systematically lead to biased results, simply because the model was not able to fit the data well when higher-frequency part of the spectrum was included.
In this case, the prediction of $\mathcal{I}$ exhibited a negative correlation with the cutoff frequency, but for some other kernels a positive correlation was found.
(See Section S4 of the Supporting Information.)
This was most notable for the shortest sequences because the corresponding sampling PSD had a lower resolution of the frequency grid, making it harder for the algorithm to identify a suitable cutoff frequency.
This is a known minor limitation of the current implementation of STACIE, and is a topic of ongoing research.

In summary, the ACID test set effectively identified STACIE's strengths and weaknesses.
The predicted autocorrelation integral and its uncertainty were compared to the ground truth for a wide range of sequence lengths ($N$) and number of sequences ($M$), revealing minor limitations that would not be apparent if the algorithm were only tested with illustrative applications.
This validation showed that STACIE's predictions efficiently converged to the ground truth as the amount of input data increased, with reliable error bars, but with a bias smaller than the predicted uncertainty, especially for the shortest sequences.
While the ACID test revealed potential areas of improvements for STACIE,
we would like to stress again that these do not undermine the utility of STACIE.
At this stage, no other algorithms have been tested with the same level of scrutiny to offer a fair comparison.
We hope this work will encourage more systematic testing in the development of future algorithms, so their performance can be compared to STACIE's.

\section{Conclusions}\label{sec:conclusions}
We unveiled a novel algorithm and its open-source Python implementation for estimating transport properties and their uncertainties from Equilibrium Molecular Dynamics simulations, called the STable AutoCorrelation Integral Estimator (STACIE).
While STACIE was primarily designed for transport properties, the implementation is completely general and can be applied to any time-correlated data that can be loaded into NumPy arrays.
We demonstrated its use with a minimal abstract example and a more complex application to estimate the electrical conductivity of an electrolyte.
These examples illustrated how one can easily plan data generation to target a desired relative error and obtain state-of-the-art results with minimal effort.
Moreover, in the electrolyte example, STACIE's estimate had a significantly smaller uncertainty than the state-of-the-art estimate obtained with the OCTP plugin for LAMMPS.
Additional examples for other transport properties are available in STACIE's documentation.
Finally, we validated STACIE against a massive dataset comprising 15360 synthetic test cases for which the ground truth is known.
This analysis confirmed the reliability of STACIE's (error) estimate of the autocorrelation integral, but it also revealed a slight bias smaller than the error estimate.

STACIE features two distinct advantages over the state of the art.
First, users do not need to adjust tunable hyperparameters to obtain a final estimate.
Given time-correlated data and a model to fit to the spectrum, STACIE proceeds without manual intervention.
Second, STACIE comes with a simple protocol for planning the number of time series to be generated to achieve a desired relative error.
The sufficiency of the length of the simulations can be checked by the algorithm itself, which may indicate that longer sequences need to be generated to obtain reliable results.

While we believe that STACIE in its current form can already greatly benefit the scientific community, we envision several avenues for future improvement.
In an upcoming publication, we will present a more sophisticated model for the power spectrum to deduce the exponential correlation time.
Moreover, STACIE currently computes so-called ``diagonal'' (scalar) transport properties.
An extension to ``off-diagonal'' properties, e.g., the Seebeck coefficient or partial ionic conductivities, would allow for the calculation of complete tensorial transport properties.

\section*{Data and Software Availability}\label{sec:availability}
The algorithm introduced in this work is implemented in an open-source Python package called STACIE, which stands for ``STable AutoCorrelation Integral Estimator''.~\cite{STACIE}
The package is available on PyPI (pip install stacie) and on GitHub (\href{http://github.com/molmod/stacie}{http://github.com/molmod/stacie}).
The documentation can be found at \href{http://molmod.github.io/stacie}{http://molmod.github.io/stacie} and includes theoretical background, skeletons of Python scripts and worked examples for various transport properties (shear and bulk viscosity, ionic electrical and thermal conductivity, diffusivity) and other use cases (uncertainty quantification and identification of correlation times).

The following datasets have been made available on Zenodo:
\begin{itemize}
  \item
  The molecular dynamics input, output and workflow files used in the electrolyte conductivity example.
  \cite{md-electrolyte}
  \item
  The AutoCorrelation Integral Drill (ACID) Test Set, with which STACIE was formally validated.
  \cite{ACID}
  \item
  Example trajectory data and Jupyter Notebooks showing how to compute various properties with STACIE.
  \cite{worked-examples}
\end{itemize}

\section*{Supporting Information}\label{sec:supporting}
The Supporting Information is available free of charge at \url{https://pubs.acs.org/doi/10.1021/acs.jsim.TODO}.
PDF document with additional display items:
\begin{itemize}
  \item
  Section S1: Table of software packages implementing algorithms for estimating transport properties from EMD simulations.
  \item
  Section S2: Additional plots for the electrolyte conductivity example, using different models and simulation times.
  \item
  Section S3: Summary plots of the electrolyte conductivity example computed using different spectrum models.
  \item
  Section S4: Plots and tables showing the validation results for the ACID test for all kernels.
\end{itemize}

\section*{Author Information}\label{sec:authorinfo}
\subsection*{Author Contributions}
G.T. performed and analyzed simulations, designed and validated algorithms and contributed to software development.
D.F. and T.V. conceptualized and supervised the project.
T.V. performed and analyzed simulations, designed and validated algorithms, and led software development.
All authors wrote the original draft of the manuscript.

\subsection*{Notes}
The authors declare no competing financial interest.

\section*{Acknowledgments}\label{sec:acknowledgments}
This research was funded by the Research Board of Ghent University with grant number BOF/24J/2021/118.
Computational resources were provided by Ghent University’s Stevin High-Performance Computing (HPC) infrastructure and the Flemish Supercomputer Center (VSC), funded by the Research Foundation Flanders (FWO).
We extend our gratitude to Øystein Gullbrekken and Sondre Kvalvåg Schnell for their insightful responses to our questions regarding their electrolyte simulations.

\bibliography{references}



\section*{Supplementary material}

# STable AutoCorrelation Integral Estimator (STACIE): Robust and accurate transport properties from molecular dynamics simulations

Gözdenur Toraman,[†] Dieter Fauconnier,[†‡] and Toon Verstraelen*[¶]

† Soete Laboratory, Ghent University, Technologiepark-Zwijnaarde 46, 9052 Ghent, Belgium

‡ FlandersMake@UGent, Core Lab EEDT-MP, 3001 Leuven, Belgium

¶ Center for Molecular Modeling (CMM), Ghent University, Technologiepark-Zwijnaarde 46, B-9052, Ghent, Belgium

∗E-mail: toon.verstraelen@ugent.be

## Contents

# S1. Algorithms and Implementations for the Computation of Transport Properties From Equilibrium Molecular Dynamics (EMD) simulations

The following table summarizes the algorithms and implementations for the computation of transport properties from equilibrium molecular dynamics (EMD) simulations. All contained information and hyperlinks were last updated on May 5, 2025. The algorithm categories are described in more details in the main text:

- **Green-Kubo (GK)**: Numerical quadrature of the autocorrelation function (ACF)
- **Einstein-Helfand (EH)**: Determination of the slope of the mean square displacement (MSD)
- **Cepstral (C)**: Numerical quadrature of the cepstrum
- **Spectral (S)**: Zero-frequency limit of the power spectrum

| Name | Availability | Category | Usage | Properties |
|------|-------------|----------|-------|------------|
| Large-scale Atomic/Molecular Massively Parallel Simulator (LAMMPS) [1] | Open source (Apr 2, 2025) | GK, EH | Built-in with `fix ave/correlate` | Tutorials for thermal conductivity, shear viscosity, diffusivity, extendable to ionic electrical conductivity and bulk viscosity |
| On-the-fly calculation of Transport Properties (OTCP) plugin for LAMMPS [2] | Open source, unversioned (last Git commit June 19, 2023) | EH | Built-in with `fix ordern` | diffusivity, shear and bulk viscosity, thermal conductivity, ionic electrical conductivity |
| Python LAMMPS Analysis Tool (PyLAT) | Open source, 0.1.1 (Oct 13, 2022) | GK, EH (depends on property) | Post process | shear viscosity, diffusivity, ionic electrical conductivity |
| GROMACS [3] | Open source, 2025.1 (March 11, 2025) | GK, EH | Post process with `gmx energy` or `gmx velacc`, optionally with regression and extrapolation [4] | Tutorials for shear and bulk viscosity, diffusivity, extendable to ionic electrical conductivity |
| MDAnalysis [5], [6] | Open source, 2.9.0 (March 11, 2025) | EH | Post process | diffusivity |
| tidynamics [7] | Open source, 1.1.2 (July 11, 2023) | GK, EH | Post process | Generally applicable, users must load data into NumPy arrays |
| Amsterdam Modeling Suite (AMS) [8] | Commercial, 2025.102 (April 28, 2025) | GK, EH | Post process | Tutorials for diffusivity, ionic conductivity, shear viscosity and thermal conductivity |
| QuantumATK [9] | Commercial, W-2024.09 (September, 2024) | GK, EH | Post process | Tutorials for diffusivity and viscosity |

| Name | Availability | Category | Usage | Properties |
|------|-------------|----------|-------|-----------|
| nMOLDYN [10] | Open Source, 3.0.12 (August 13, 2022) | GK, EH | Post process | diffusivity |
| Trajectory Analyzer and Visualizer (TRAVIS) [11] | Open Source, July 29, 2022 | EH | Post process | diffusivity |
| Time Decomposition Method (TDM) [12] | Not released as a software package | GK | Post process | shear viscosity |
| Liu *et al.* [13] | Not released as a software package | GK | Post process | shear viscosity |
| SporTran | Open source, 1.0.0rc3 (December 7, 2022) | C | Post process | Generally applicable, tutorial for thermal conductivity |
| STACIE | Open source, 1.0.0 (???, 2025) | S | Post process | thermal conductivity, shear and bulk viscosity, diffusivity, electrical conductivity, error on the average of a time-correlated property, |

# S2. Additional Plots for the Electrolyte Conductivity Example, Using Different Models and Simulation Times

This section contains plots with models fitted to spectra and the intermediate results as a function of the cutoff frequency, for the ionic electrical conductivity of a NaCl-water solution, computed with different truncations of the MD trajectory ($N$) and different sets of polynomial degrees ($S$). A description of the elements shown in the plots is given in the minimal example in the main text. Since data were collected with block averages over intervals of 50 fs, we have $t_{\text{sim}} = N(50 \text{ fs})$, where $N$ is the truncation of the MD trajectory.

## S2.1. $S = \{0\}$

### S2.1.1. $N = 312$, $t_{\text{sim}} = 15.6$ ps

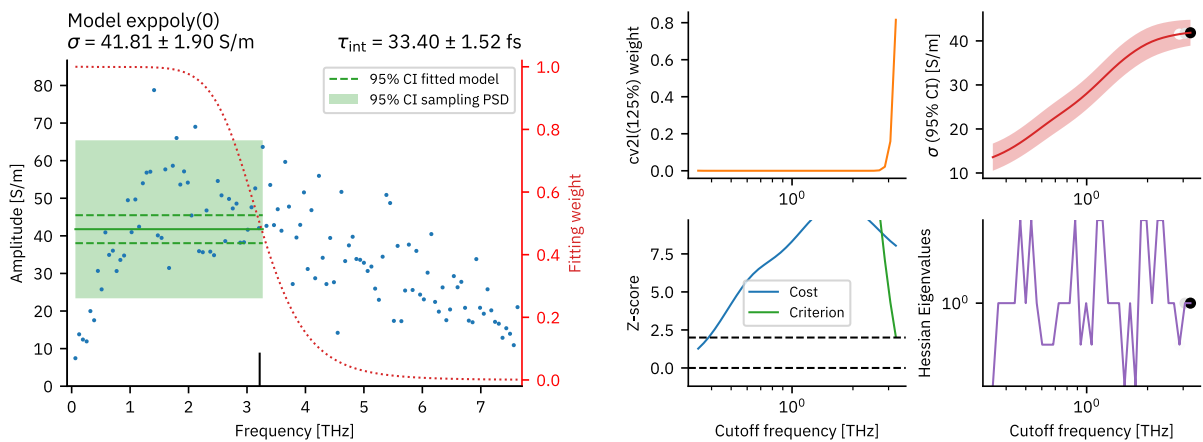

### S2.1.2. $N = 625$, $t_{\text{sim}} = 31.3$ ps

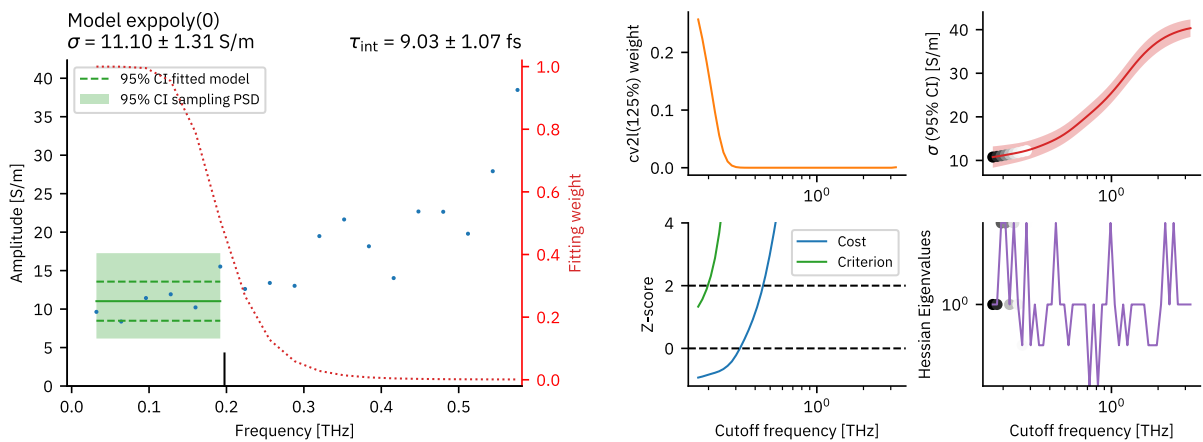

**S2.1.3.** $N = 1250$, $t_{sim} = 62.5$ ps

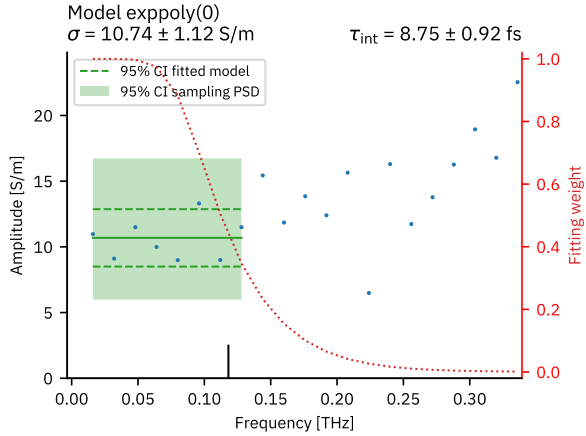
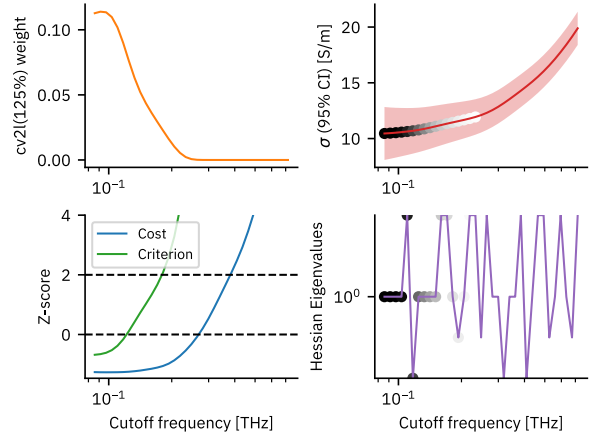

**S2.1.4.** $N = 2500$, $t_{sim} = 125.0$ ps

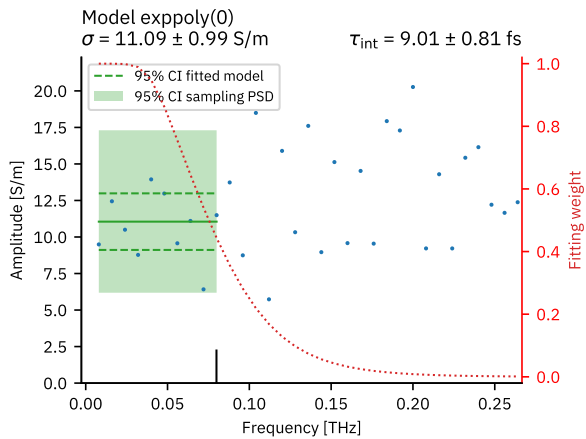
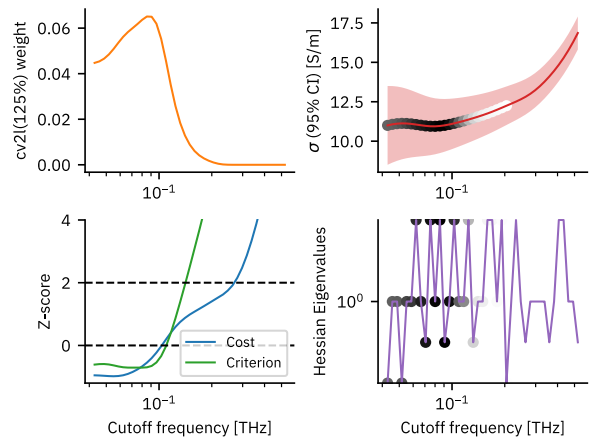

**S2.1.5.** $N = 5000$, $t_{sim} = 250.0$ ps

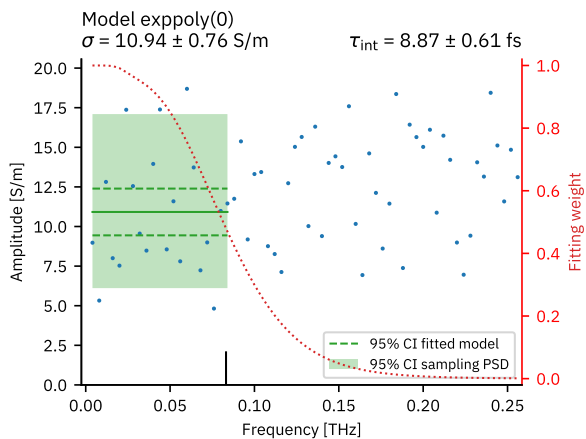
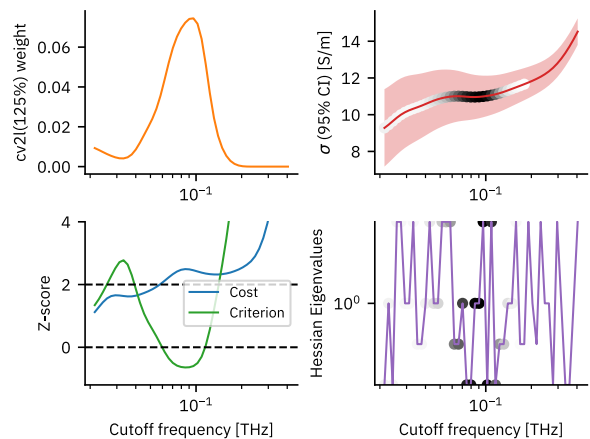

**S2.1.6.** $N = 10000$, $t_{sim} = 500.0$ ps

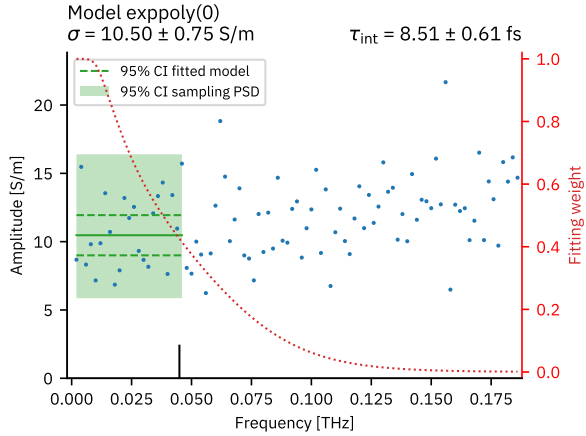
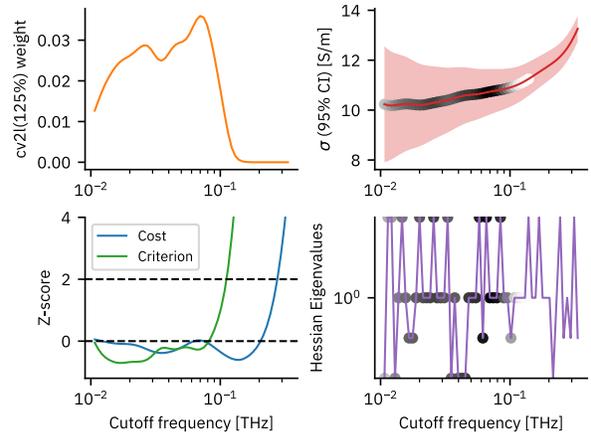

**S2.1.7.** $N = 20000$, $t_{sim} = 1000.0$ ps

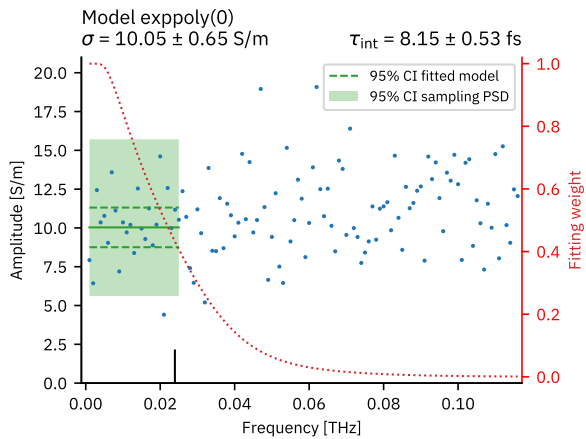
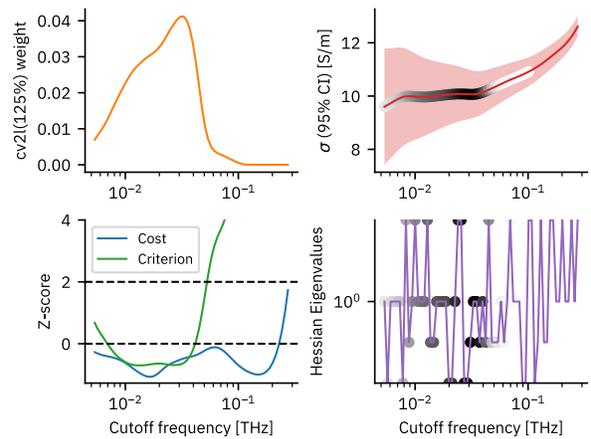

## S2.2. $S = \{0, 1\}$

**S2.2.1.** $N = 312$, $t_{sim} = 15.6$ ps

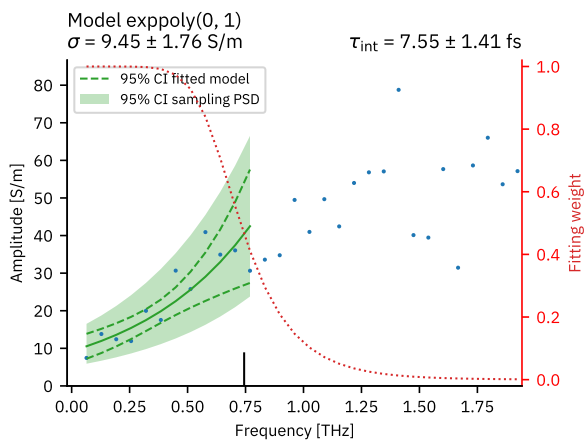
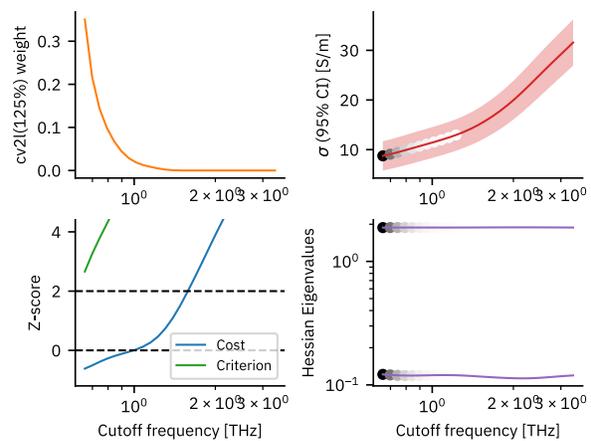

**S2.2.2.** $N = 625$, $t_{\text{sim}} = 31.3$ ps

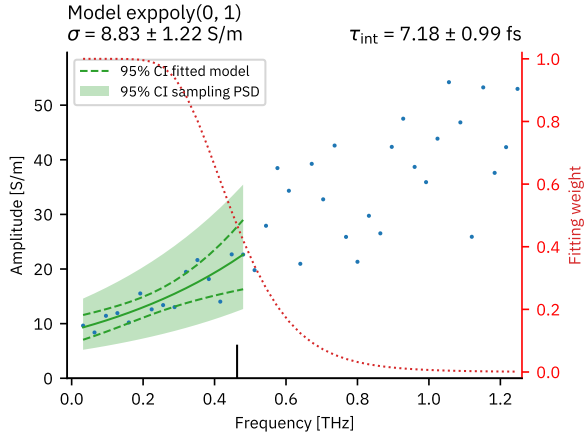
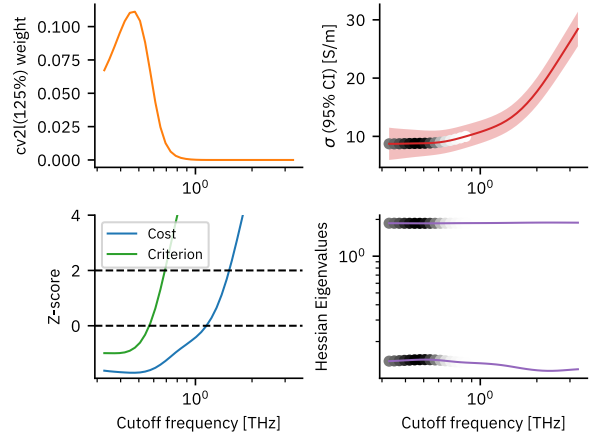

**S2.2.3.** $N = 1250$, $t_{\text{sim}} = 62.5$ ps

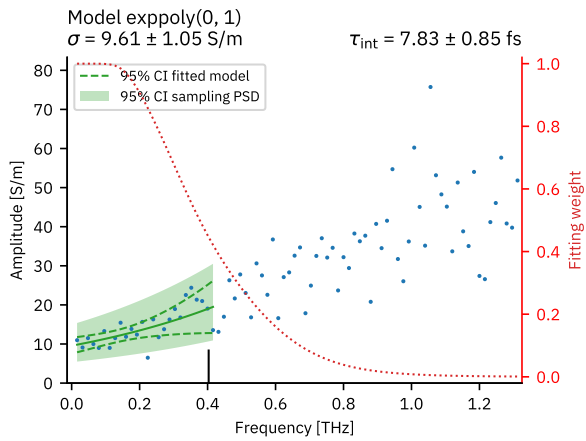
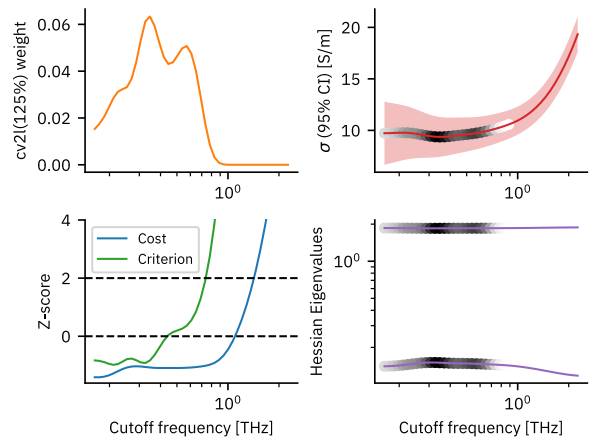

**S2.2.4.** $N = 2500$, $t_{\text{sim}} = 125.0$ ps

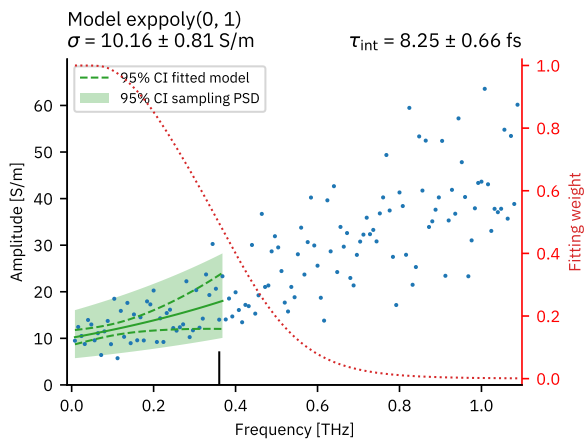
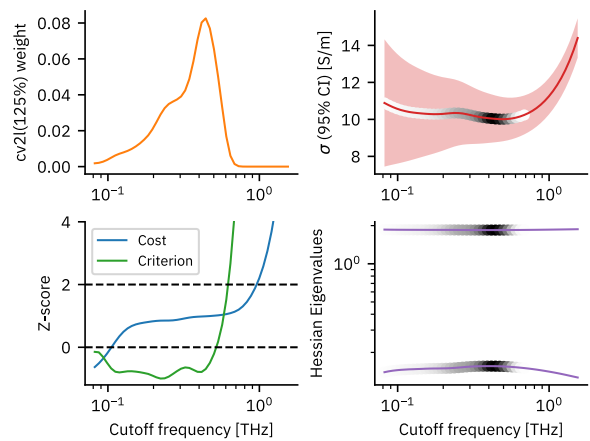

**S2.2.5.** $N = 5000$, $t_{\text{sim}} = 250.0$ ps

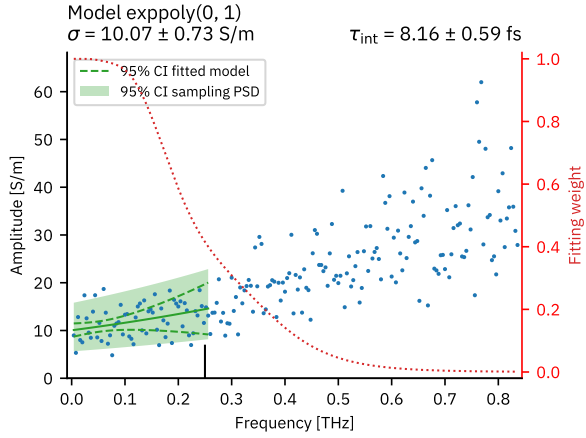
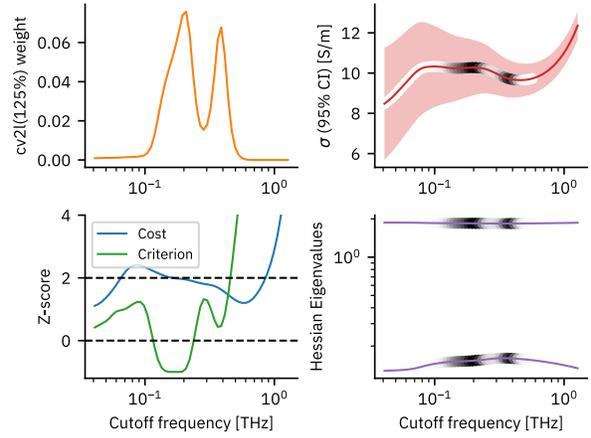

**S2.2.6.** $N = 10000$, $t_{\text{sim}} = 500.0$ ps

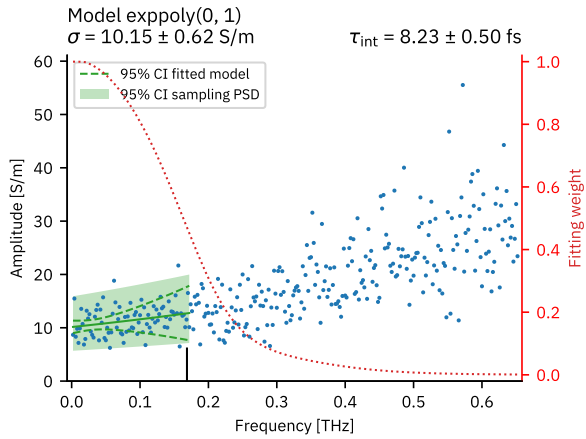
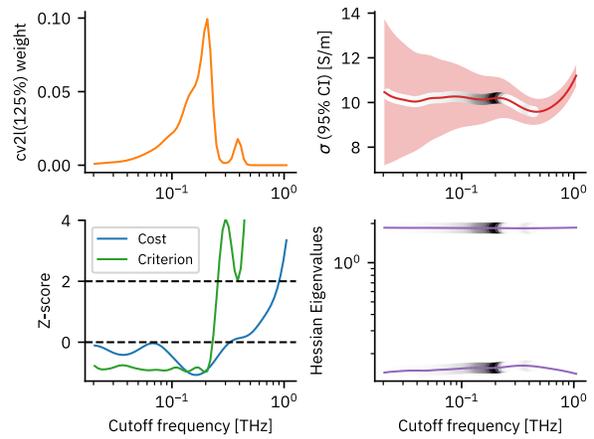

**S2.2.7.** $N = 20000$, $t_{\text{sim}} = 1000.0$ ps

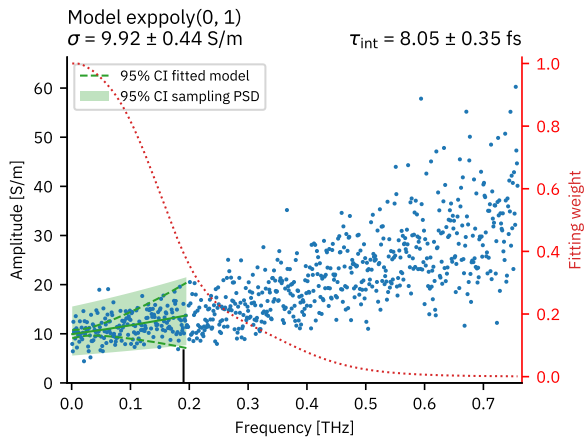
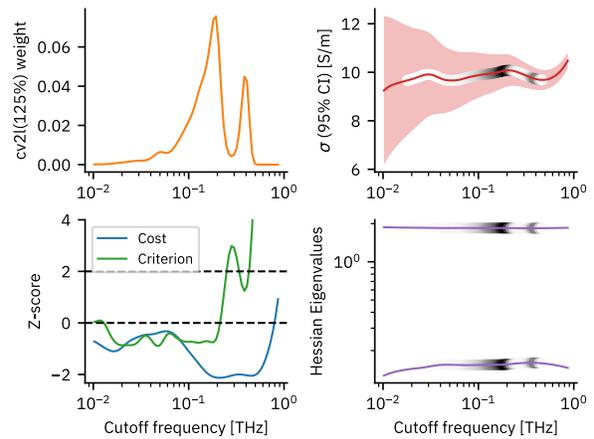

## S2.3. $S = \{0, 1, 2\}$

### S2.3.1. $N = 312$, $t_{sim} = 15.6$ ps

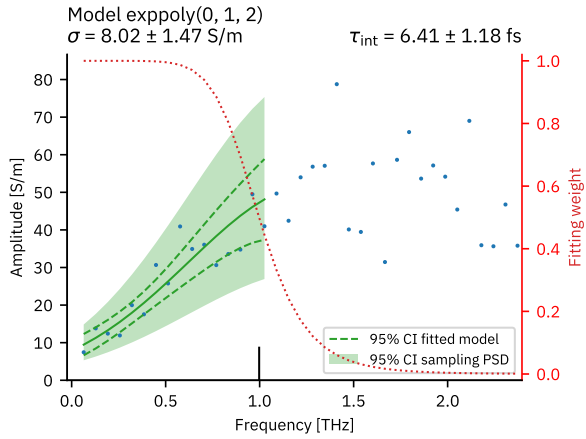
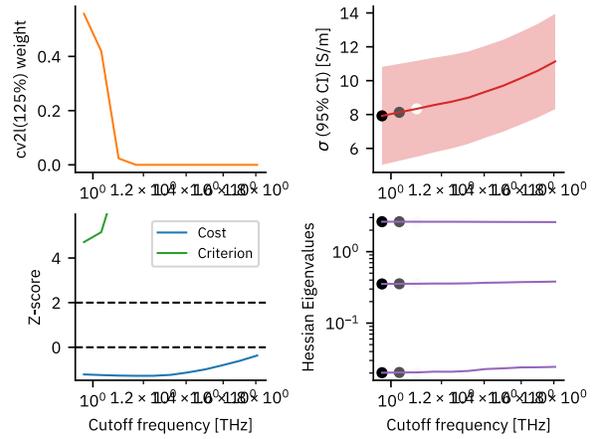

### S2.3.2. $N = 625$, $t_{sim} = 31.3$ ps

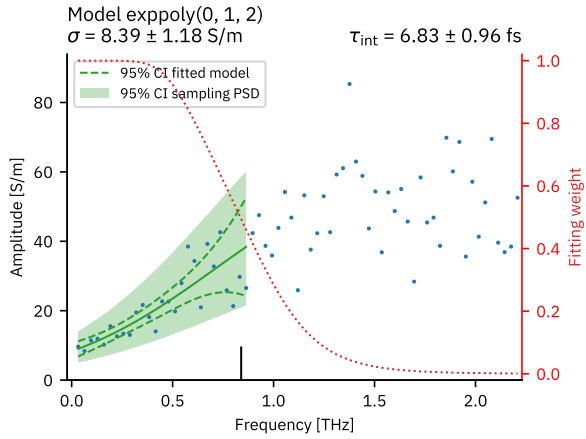
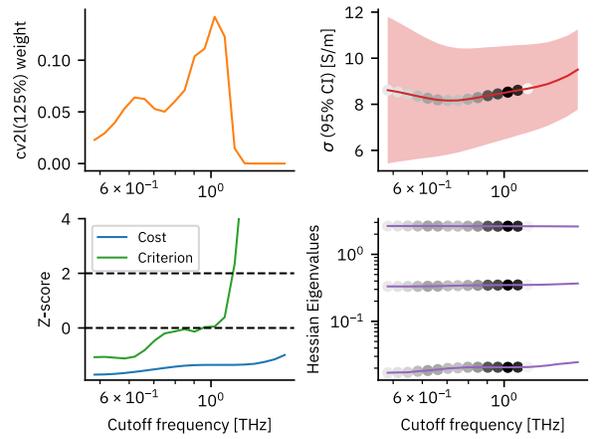

### S2.3.3. $N = 1250$, $t_{sim} = 62.5$ ps

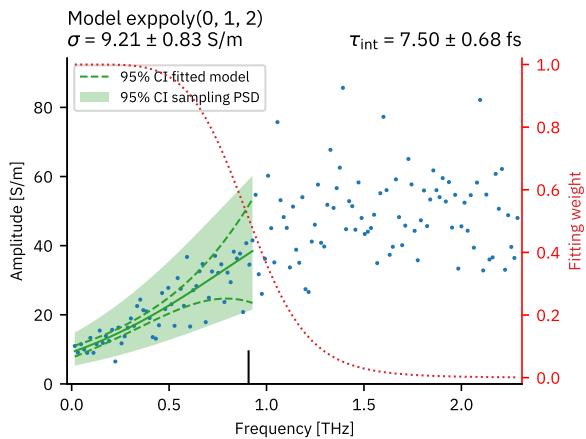
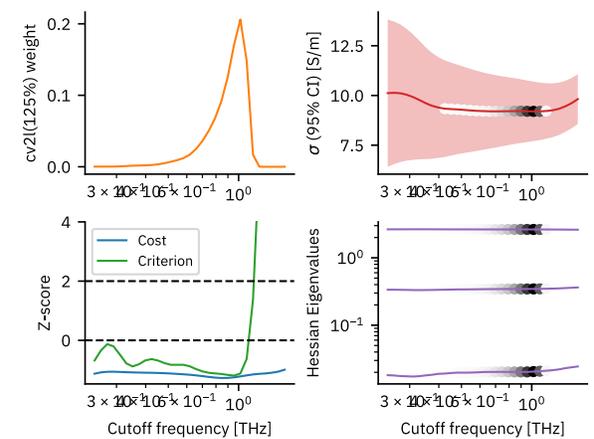

**S2.3.4.** $N = 2500$, $t_{\mathrm{sim}} = 125.0$ ps

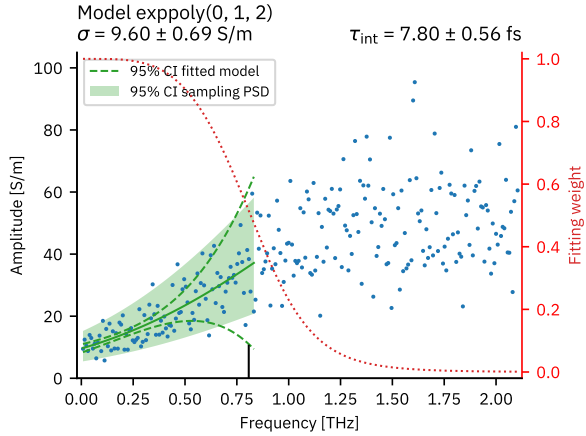
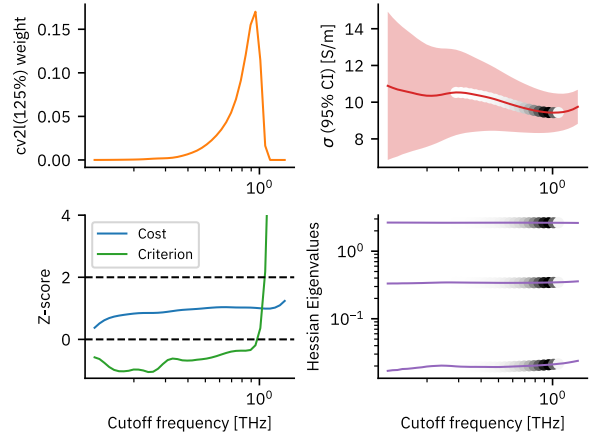

**S2.3.5.** $N = 5000$, $t_{\mathrm{sim}} = 250.0$ ps

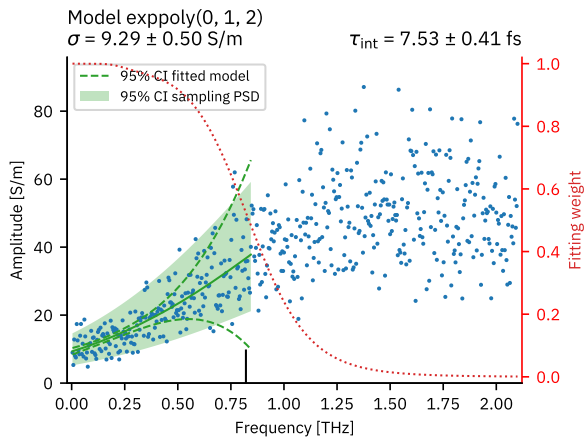
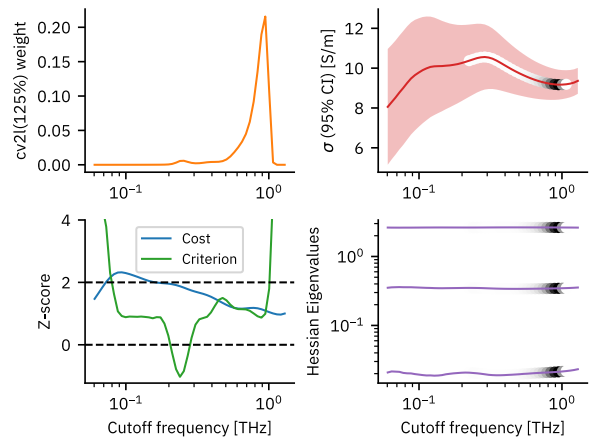

**S2.3.6.** $N = 10000$, $t_{\mathrm{sim}} = 500.0$ ps

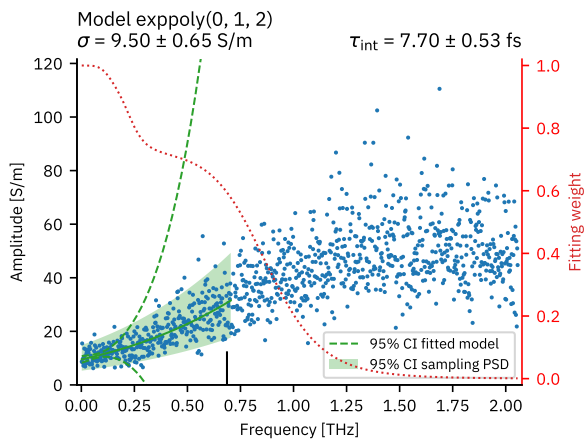
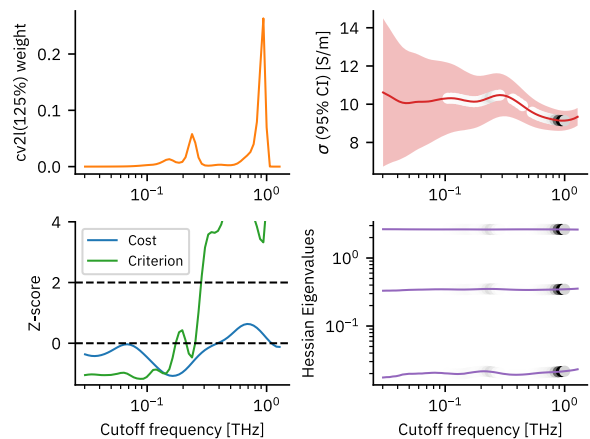

**S2.3.7.** $N = 20000$, $t_{\text{sim}} = 1000.0$ ps

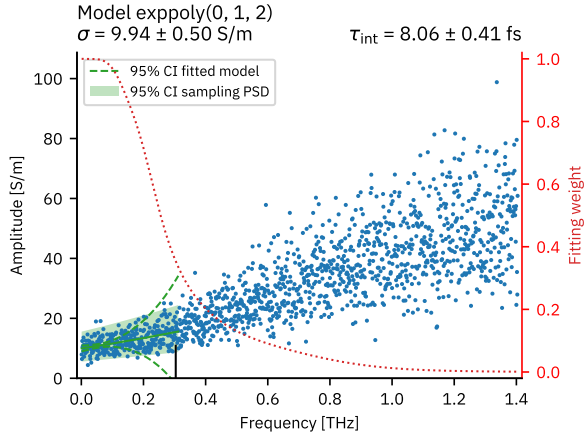
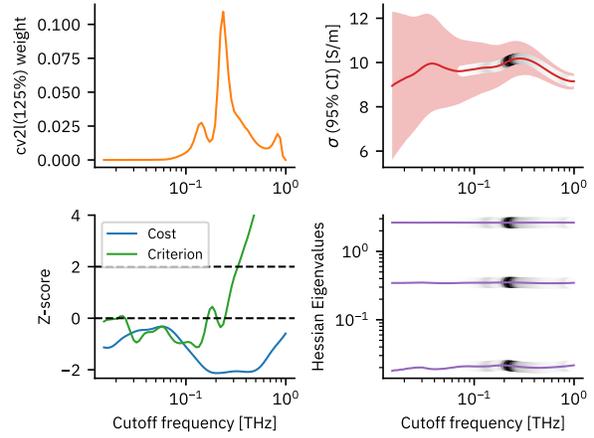

## S2.4. $S = \{0, 1, 2, 3\}$

**S2.4.1.** $N = 312$, $t_{\text{sim}} = 15.6$ ps

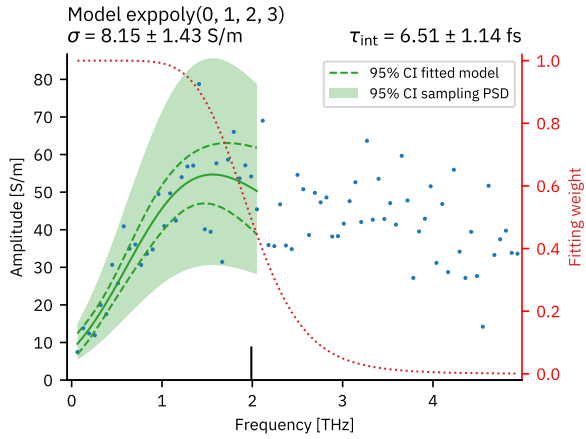
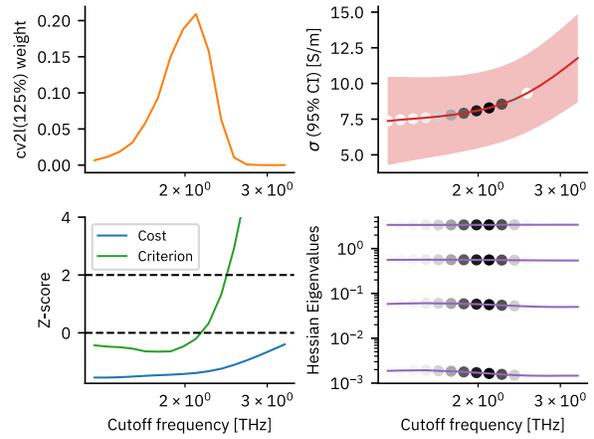

**S2.4.2.** $N = 625$, $t_{\text{sim}} = 31.3$ ps

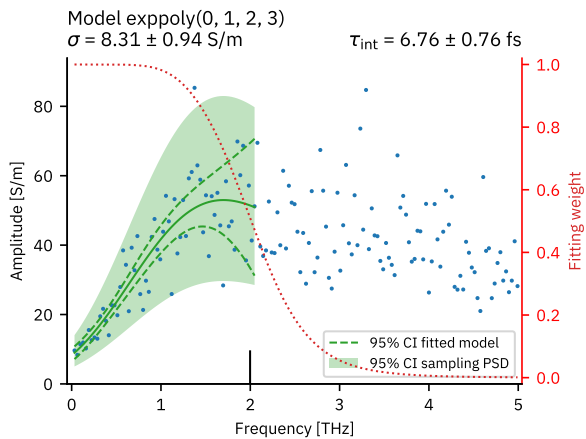
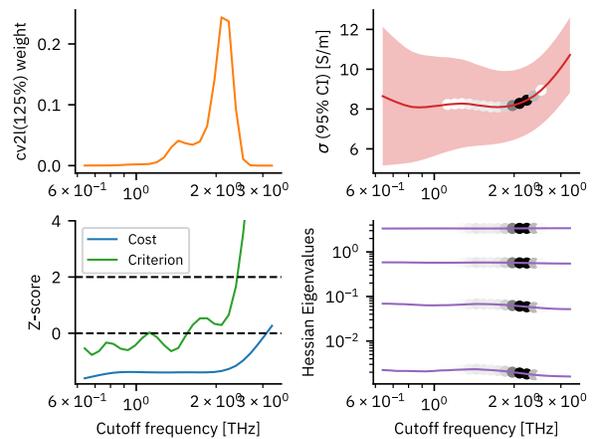

**S2.4.3.** $N = 1250$, $t_{\text{sim}} = 62.5$ ps

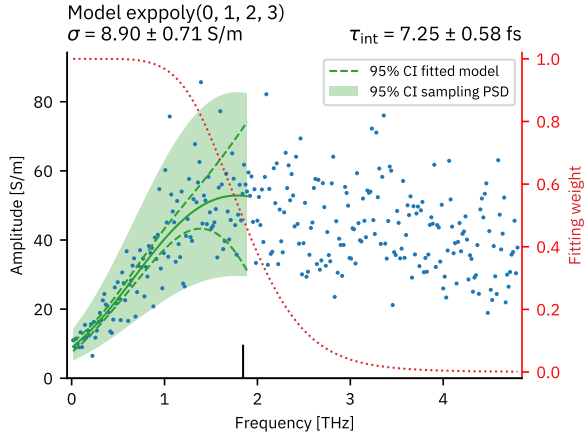
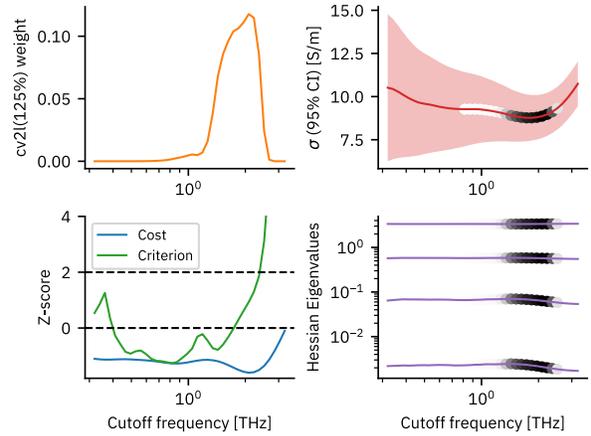

**S2.4.4.** $N = 2500$, $t_{\text{sim}} = 125.0$ ps

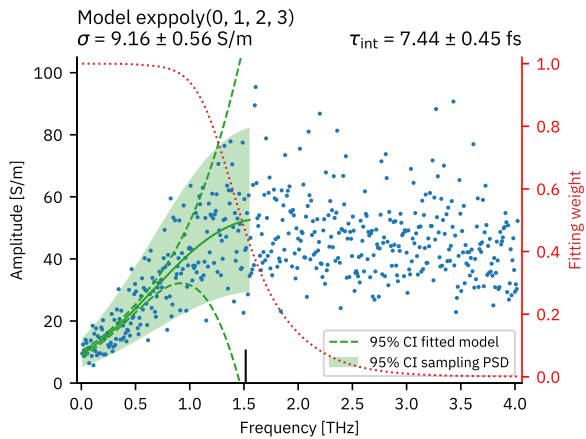
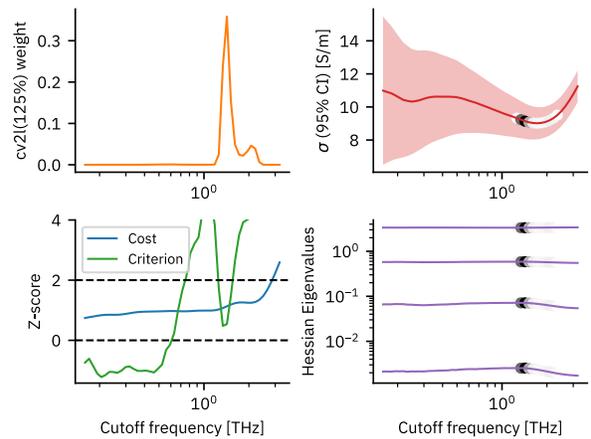

**S2.4.5.** $N = 5000$, $t_{\text{sim}} = 250.0$ ps

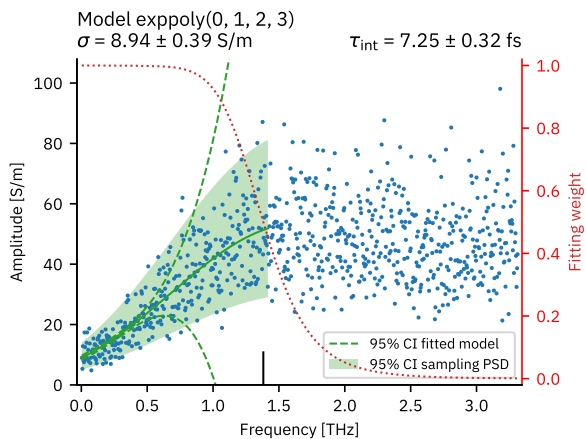
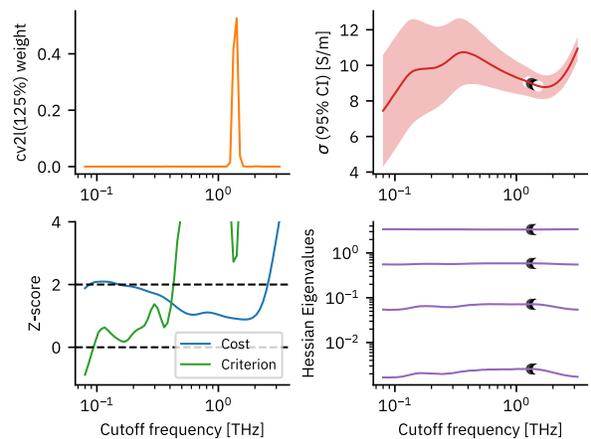

**S2.4.6.** $N = 10000$, $t_{\text{sim}} = 500.0$ ps

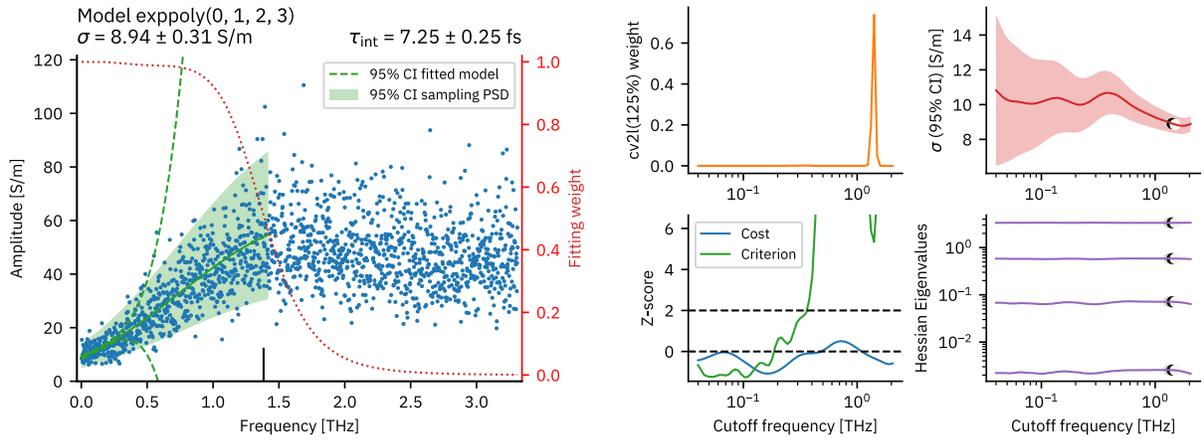

**S2.4.7.** $N = 20000$, $t_{\text{sim}} = 1000.0$ ps

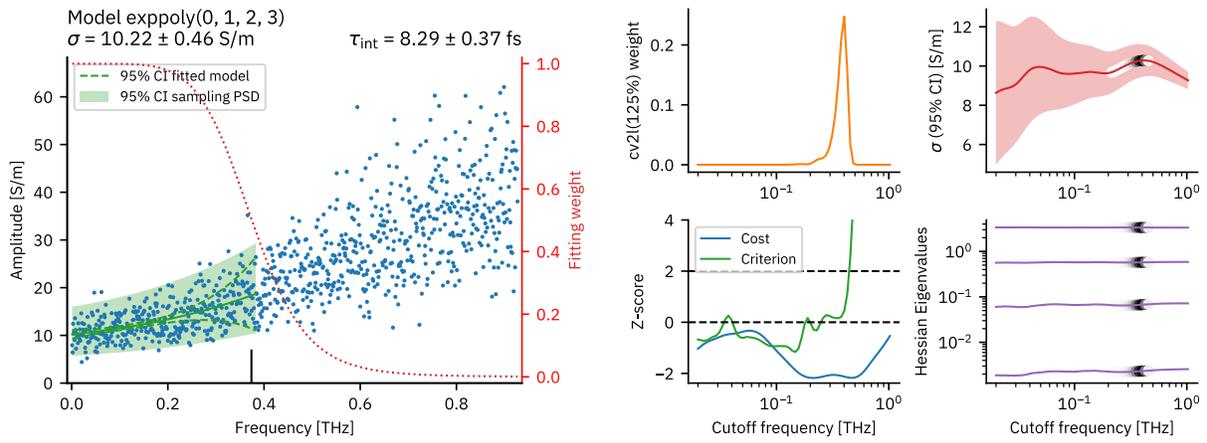

**S2.5.** $S = \{0, 2\}$

**S2.5.1.** $N = 312$, $t_{\text{sim}} = 15.6$ ps

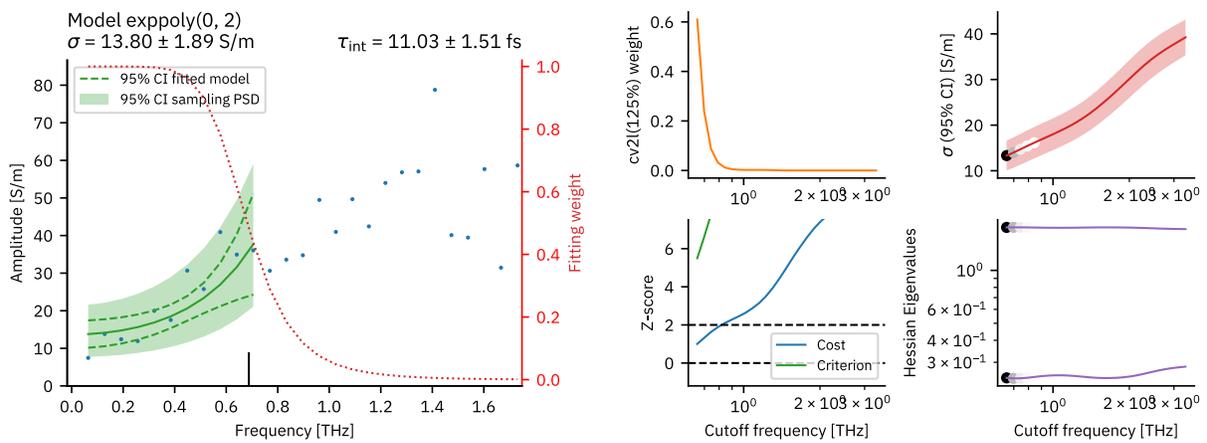

**S2.5.2.** $N = 625$, $t_{\text{sim}} = 31.3$ ps

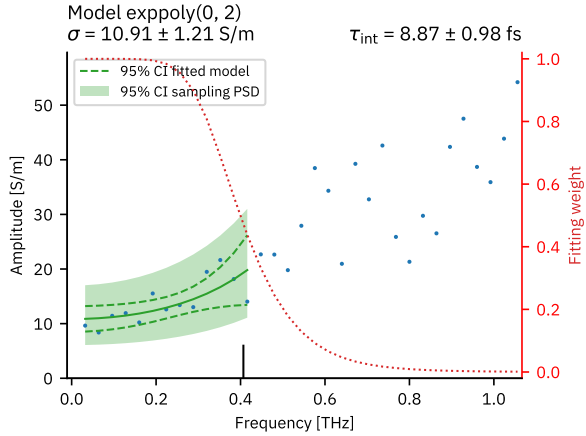
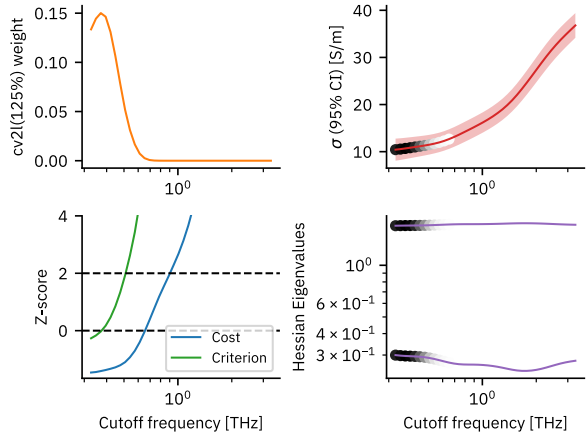

**S2.5.3.** $N = 1250$, $t_{\text{sim}} = 62.5$ ps

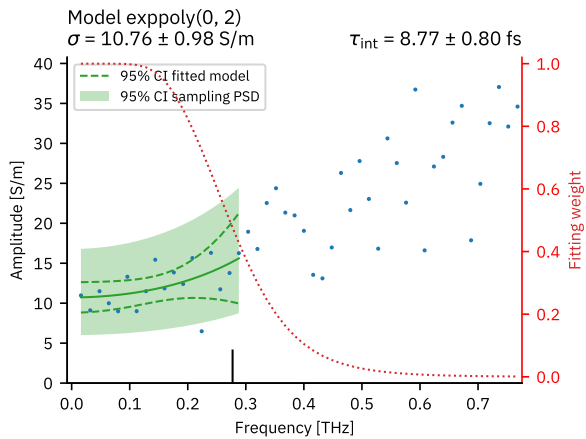
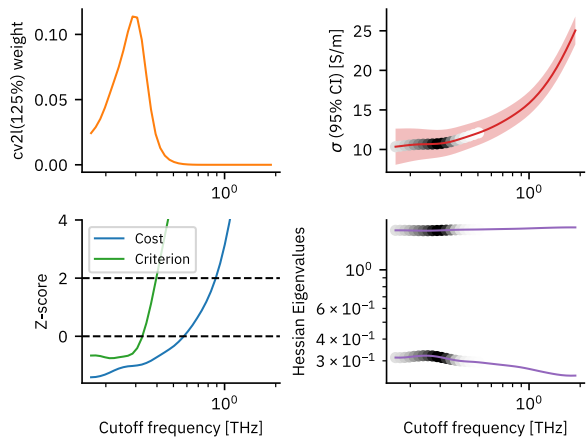

**S2.5.4.** $N = 2500$, $t_{\text{sim}} = 125.0$ ps

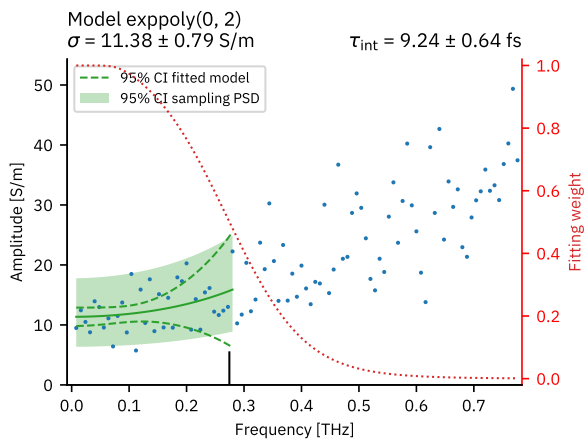
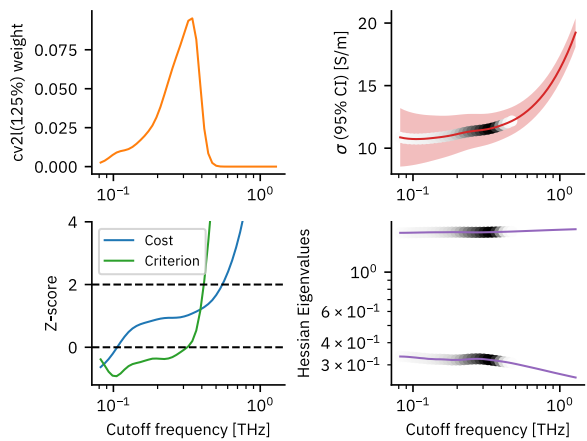

**S2.5.5.** $N = 5000$, $t_{\text{sim}} = 250.0$ ps

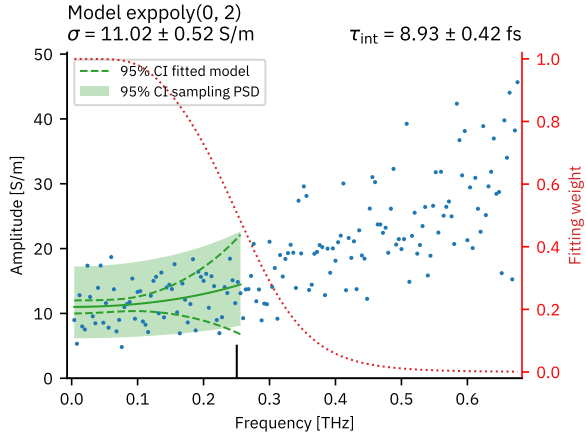

Model exppoly(0, 2)
$\sigma = 11.02 \pm 0.52$ S/m

$\tau_{\text{int}} = 8.93 \pm 0.42$ fs

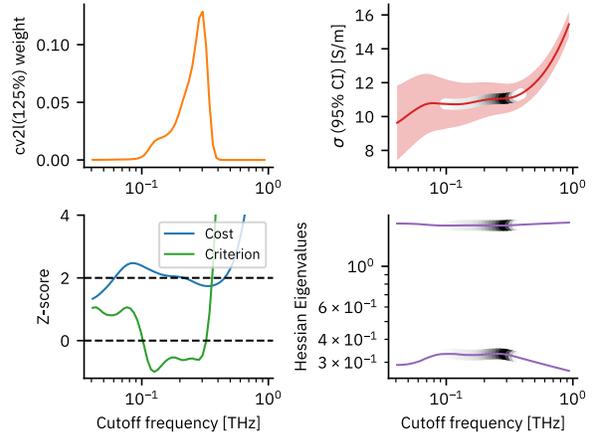

**S2.5.6.** $N = 10000$, $t_{\text{sim}} = 500.0$ ps

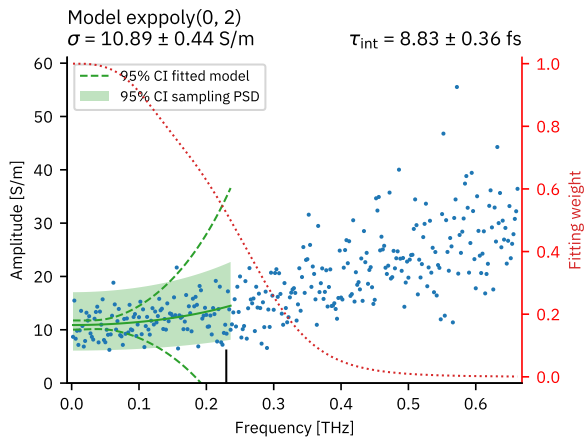

Model exppoly(0, 2)
$\sigma = 10.89 \pm 0.44$ S/m

$\tau_{\text{int}} = 8.83 \pm 0.36$ fs

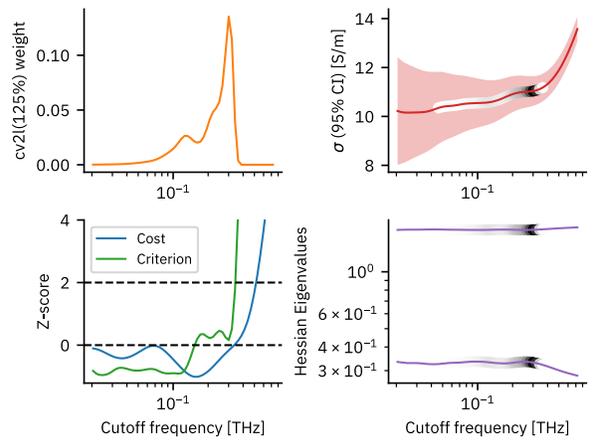

**S2.5.7.** $N = 20000$, $t_{\text{sim}} = 1000.0$ ps

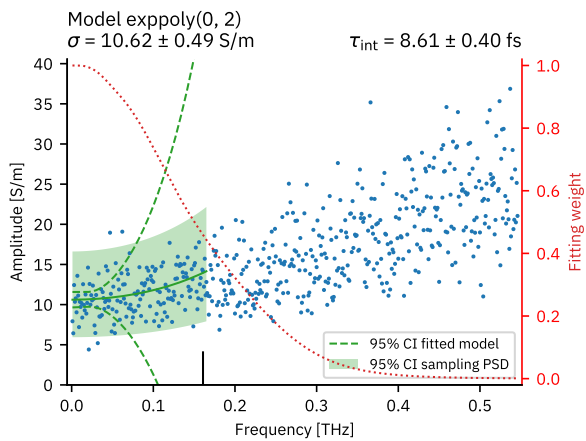

Model exppoly(0, 2)
$\sigma = 10.62 \pm 0.49$ S/m

$\tau_{\text{int}} = 8.61 \pm 0.40$ fs

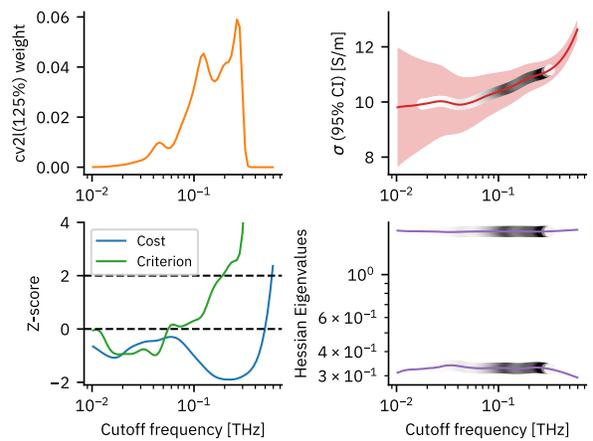

## S2.6. $S = \{0, 2, 4\}$

### S2.6.1. $N = 312$, $t_{sim} = 15.6$ ps

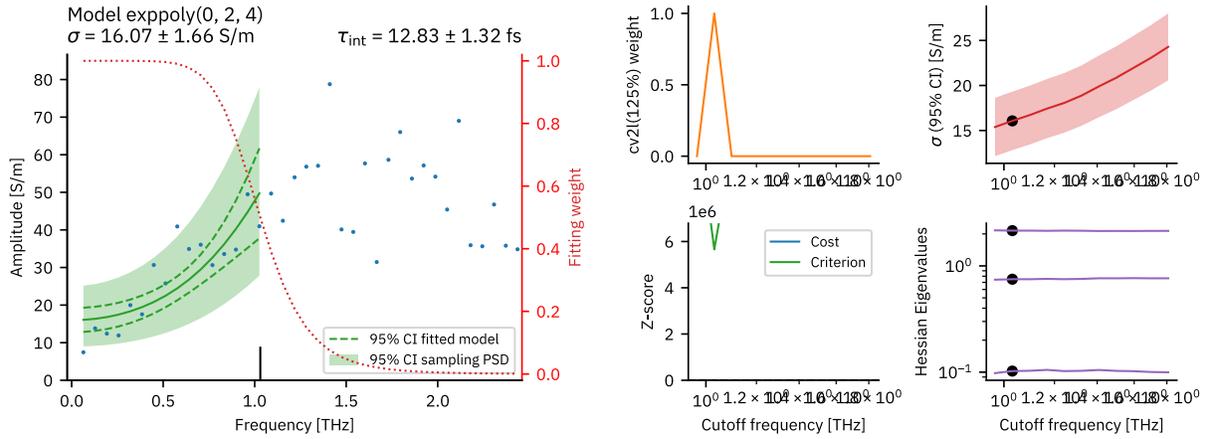

### S2.6.2. $N = 625$, $t_{sim} = 31.3$ ps

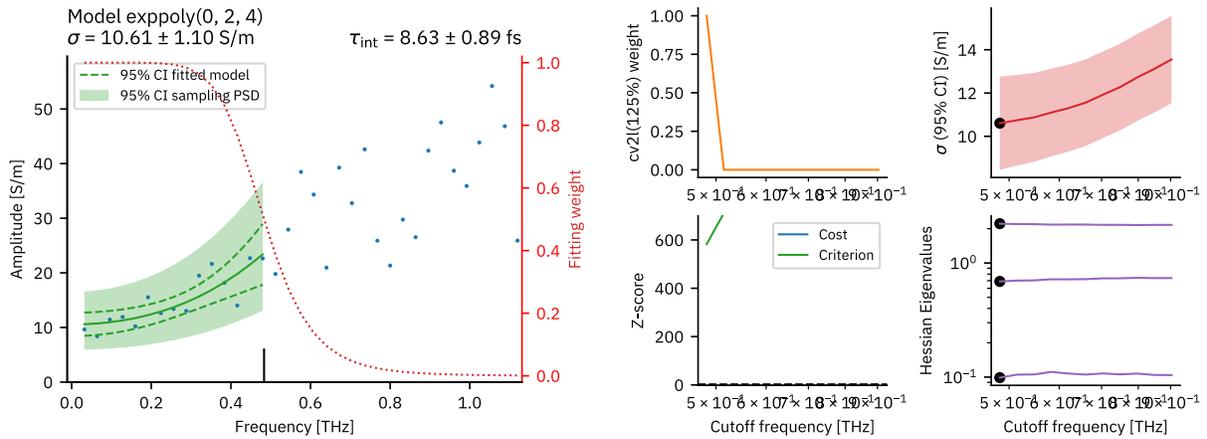

### S2.6.3. $N = 1250$, $t_{sim} = 62.5$ ps

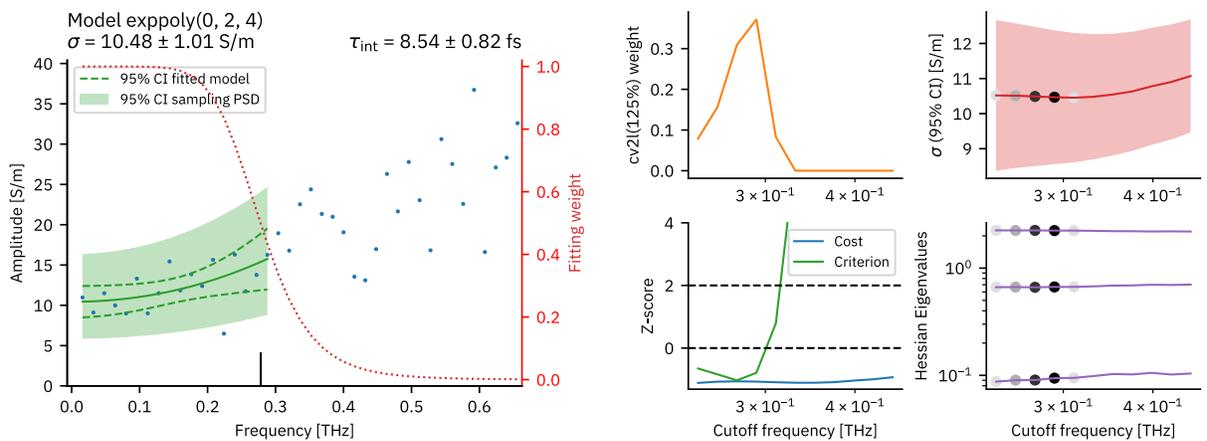

**S2.6.4.** $N = 2500$, $t_{sim} = 125.0$ ps

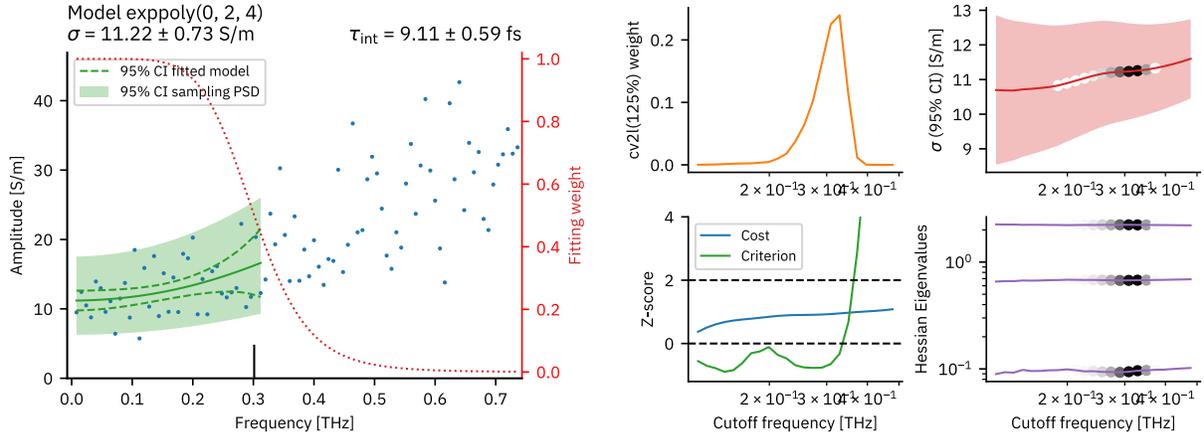

**S2.6.5.** $N = 5000$, $t_{sim} = 250.0$ ps

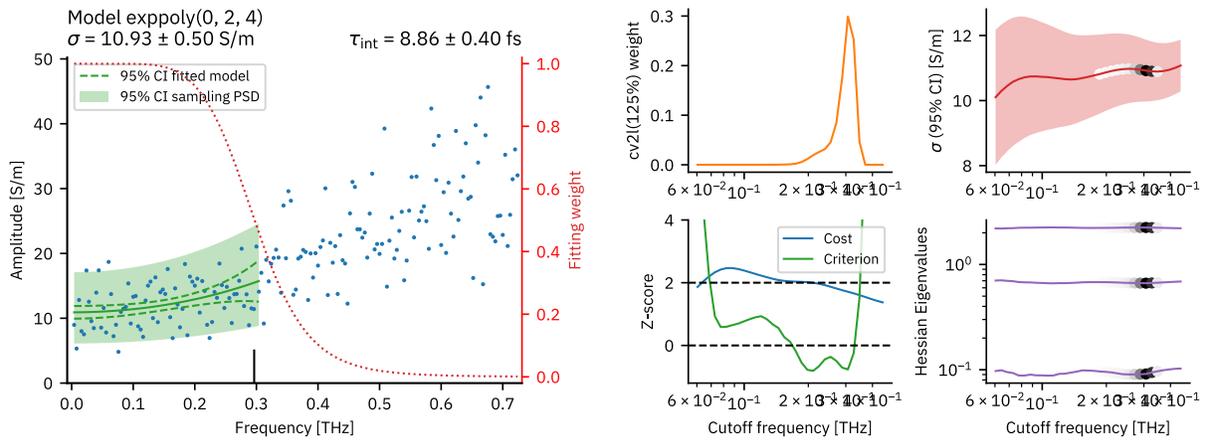

**S2.6.6.** $N = 10000$, $t_{sim} = 500.0$ ps

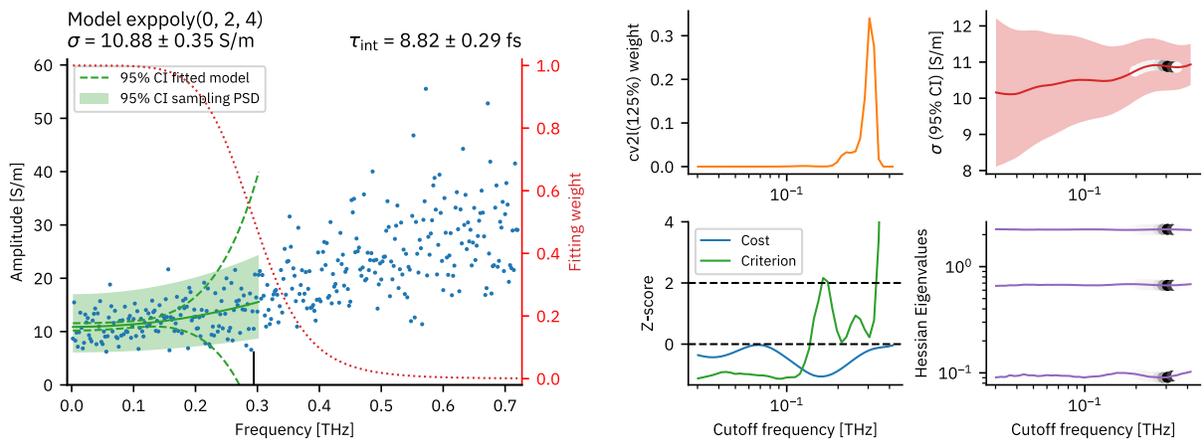

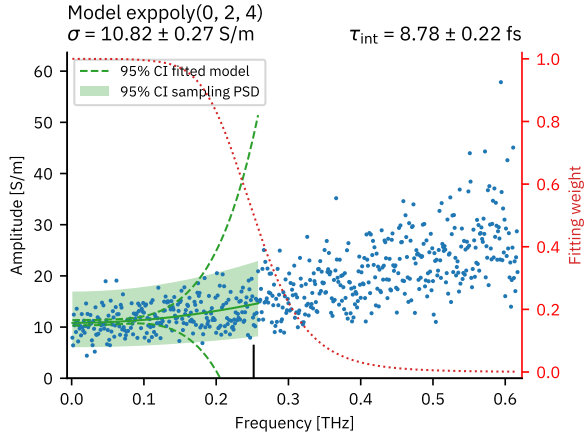

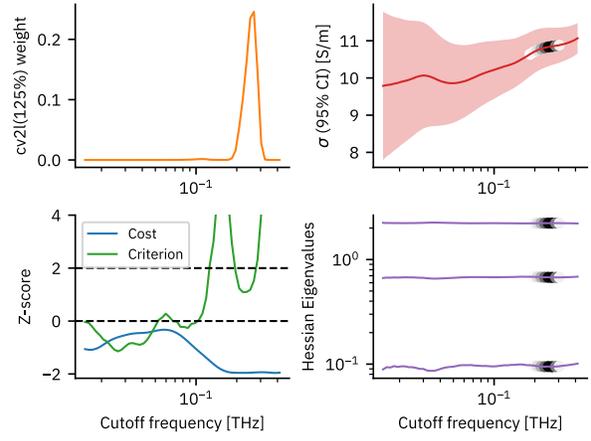

# S3. Summary plots of the electrolyte conductivity example computed using different spectrum models

## S3.1. Overview of Sanity Checks and Conductivity Estimates

The left column shows plots of the sanity checks, with filled squares indicating fits that passed the sanity checks. From top to bottom: the effective number of points used in the fit, the regression cost Z-score, and the cutoff criterion Z-score. The right column displays the conductivity estimates with error bars (top) and their corresponding relative errors (bottom). For clarity, only fits that passed all sanity checks are shown in the conductivity plots.

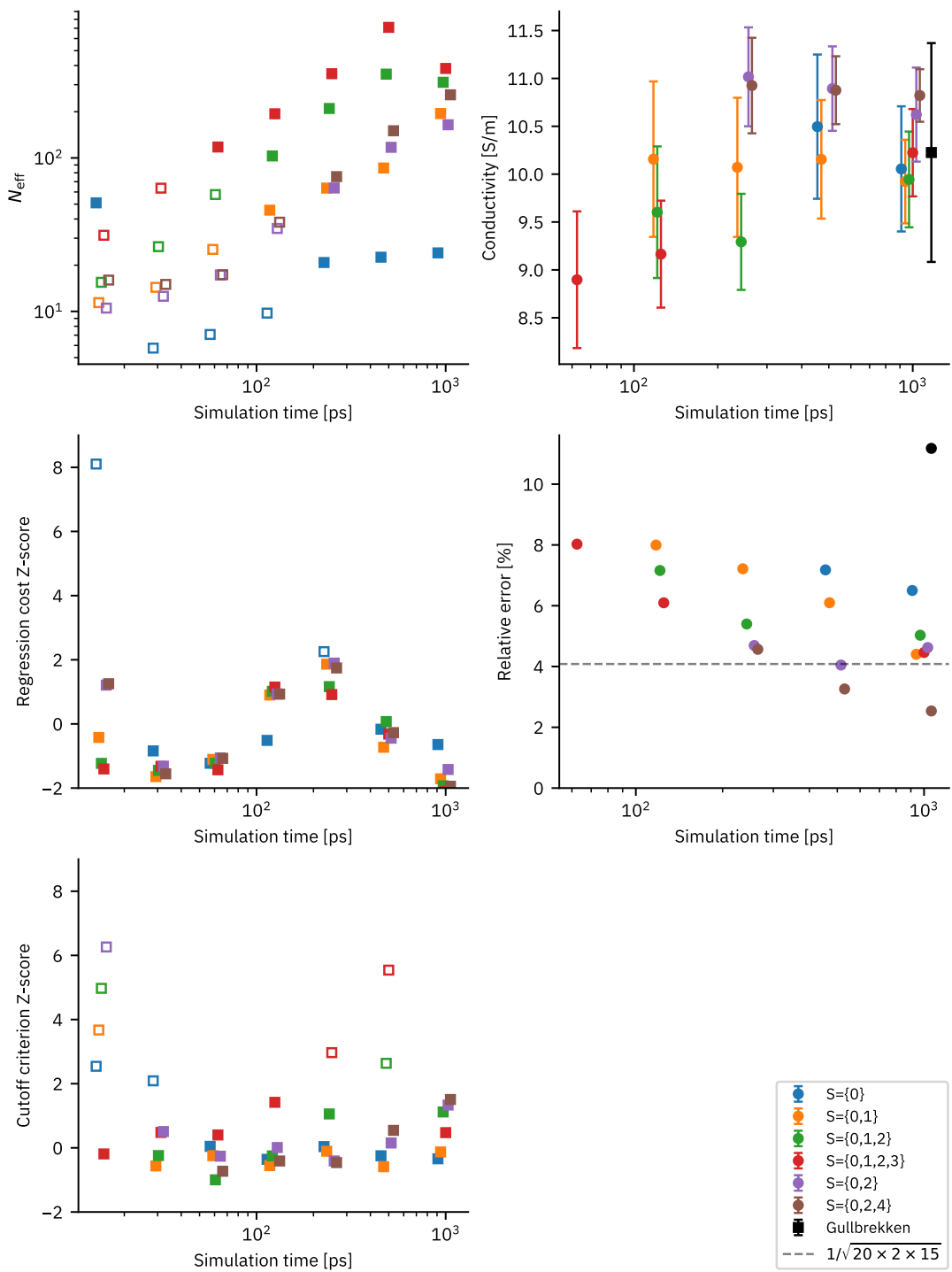

## S3.2. QQ-plot of the Deviations Between Conductivity Estimates From the Full MD Trajectory

The QQ-plot is constructed by computing the differences between all combinations of conductivity estimates obtained with different sets of polynomial degrees $S$, using the full MD trajectory ($N = 20000$). These differences are normalized by dividing them by their estimated standard deviation. If STACIE's error estimates are accurate, the normalized differences should follow a standard normal distribution. This is confirmed by the QQ-plot, which plots the sorted normalized differences (Y-axis) against the corresponding quantiles of a standard normal distribution (X-axis).

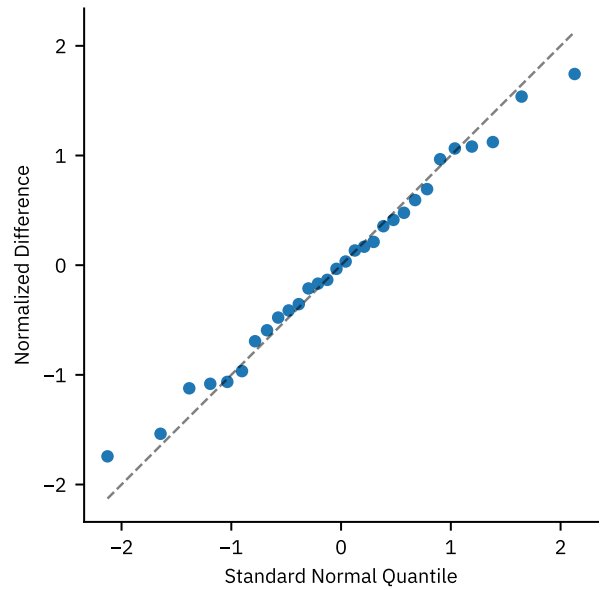

# S4. ACID Test Results for 12 Different Kernels

See main text for a detailed discription of the ACID test and the figures in this section.

## S4.1. Kernel exp1p

**(a) Illustration of input sequences. See Fig. 5 in the main text for details.**

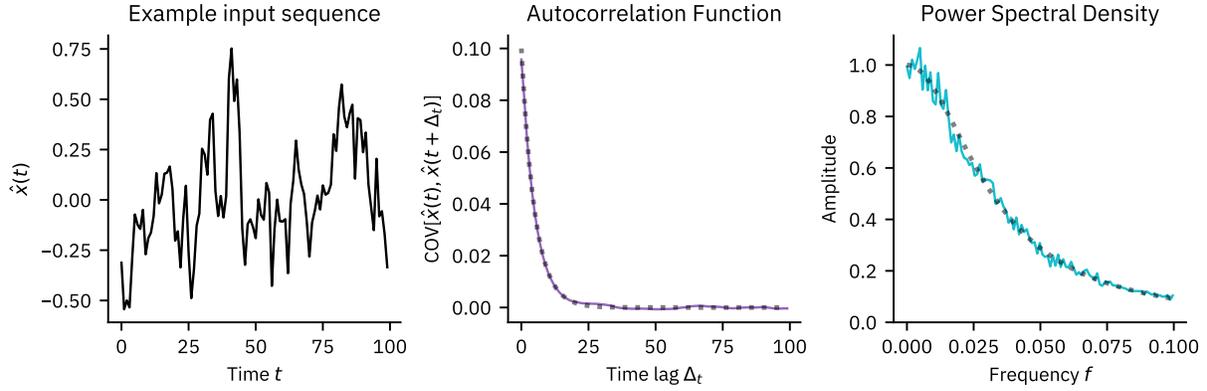

**(b) Scaling of errors with input data. See Fig. 6 in the main text for details.**

**(c) Assessment of the error estimate. See Fig. 7 in the main text for details.**

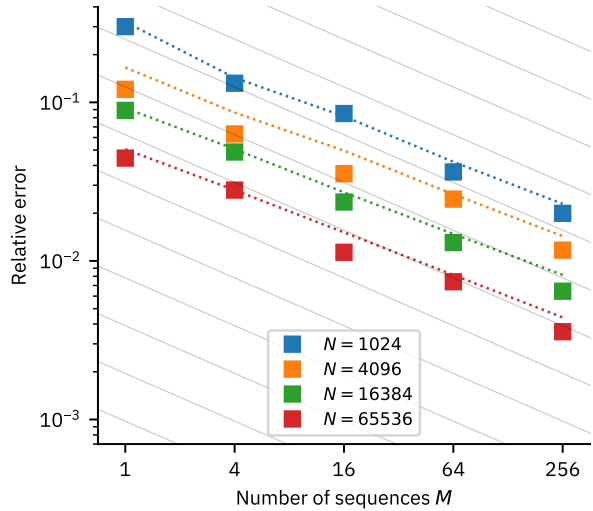
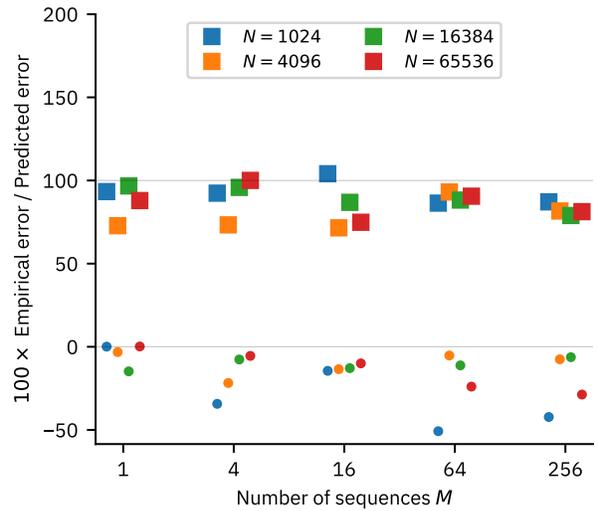

**(d) Validation of the Maximum a Posteriori. See Fig. 8 in the main text for details.**

**(e) Sensitivity of the autocorrelation integral to the cutoff frequency. See Fig. 9 in the main text for details.**

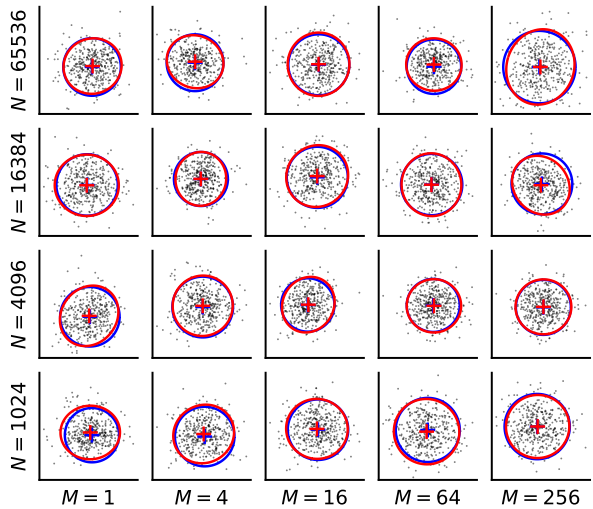
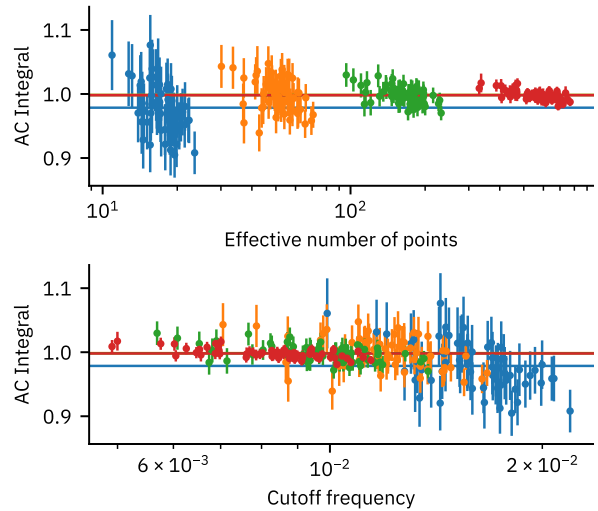

**(f) Sanity check counts for the effective number of points**

|           | $M = 1$ | $M = 4$ | $M = 16$ | $M = 64$ | $M = 256$ |
|-----------|---------|---------|----------|----------|-----------|
| $N = 1024$ | 55 | 64 | 64 | 64 | 64 |
| $N = 4096$ | 0 | 0 | 2 | 5 | 19 |
| $N = 16384$ | 0 | 0 | 0 | 0 | 0 |
| $N = 65536$ | 0 | 0 | 0 | 0 | 0 |

**(g) Sanity check counts for the regression cost z-score**

|           | $M = 1$ | $M = 4$ | $M = 16$ | $M = 64$ | $M = 256$ |
|-----------|---------|---------|----------|----------|-----------|
| $N = 1024$ | 0 | 0 | 0 | 0 | 0 |
| $N = 4096$ | 0 | 1 | 0 | 0 | 1 |
| $N = 16384$ | 0 | 0 | 0 | 0 | 2 |
| $N = 65536$ | 0 | 0 | 2 | 0 | 0 |

**(h) Sanity check counts for the cutoff criterion z-score**

|           | $M = 1$ | $M = 4$ | $M = 16$ | $M = 64$ | $M = 256$ |
|-----------|---------|---------|----------|----------|-----------|
| $N = 1024$ | 1 | 1 | 3 | 2 | 5 |
| $N = 4096$ | 0 | 0 | 1 | 1 | 1 |
| $N = 16384$ | 0 | 0 | 0 | 0 | 2 |
| $N = 65536$ | 0 | 0 | 1 | 0 | 0 |

## S4.2. Kernel exp1w

**(a) Illustration of input sequences. See Fig. 5 in the main text for details.**

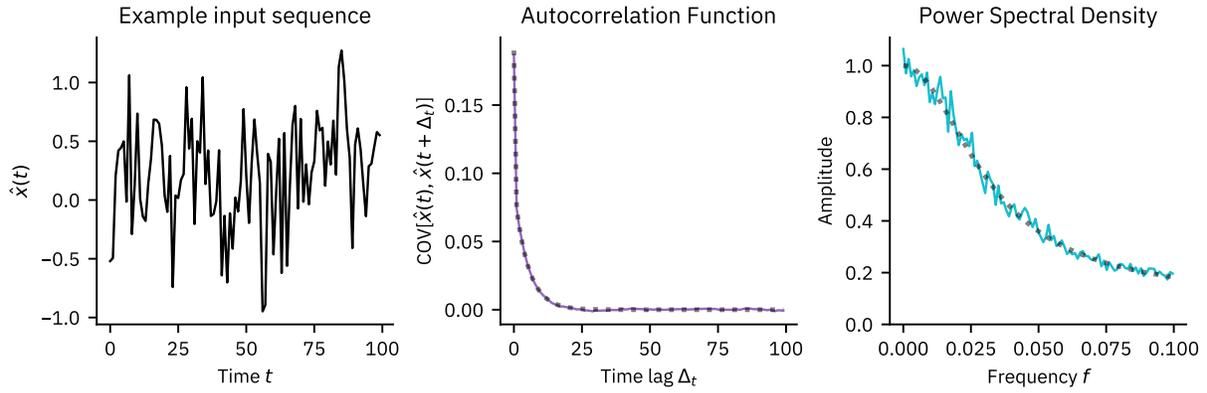

**(b) Scaling of errors with input data.**
**See Fig. 6 in the main text for details.**

**(c) Assessment of the error estimate.**
**See Fig. 7 in the main text for details.**

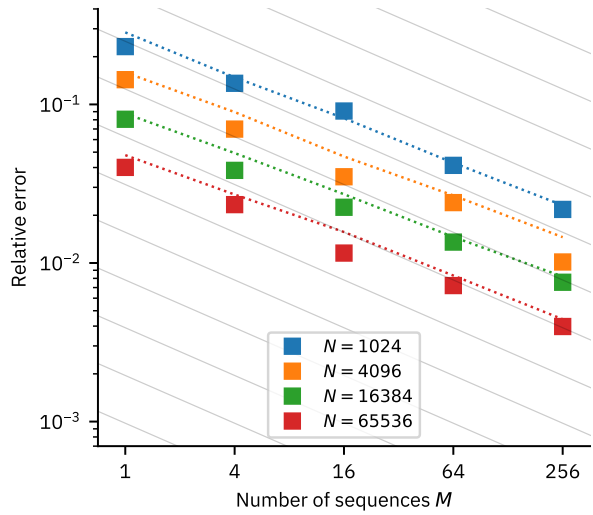

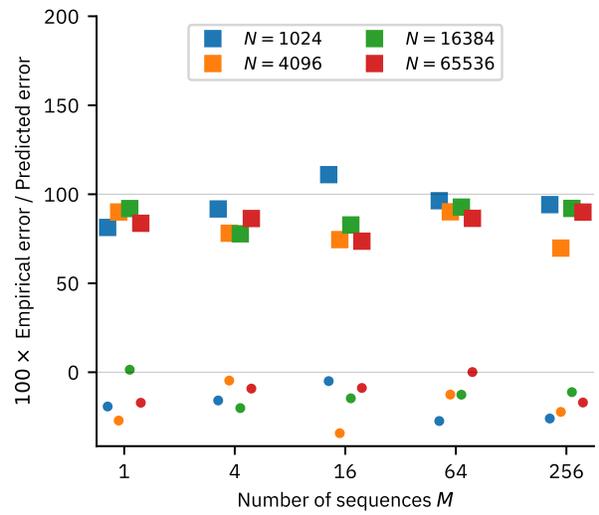

**(d) Validation of the Maximum a Posteriori.**
**See Fig. 8 in the main text for details.**

**(e) Sensitivity of the autocorrelation integral to the cutoff frequency. See Fig. 9 in the main text for details.**

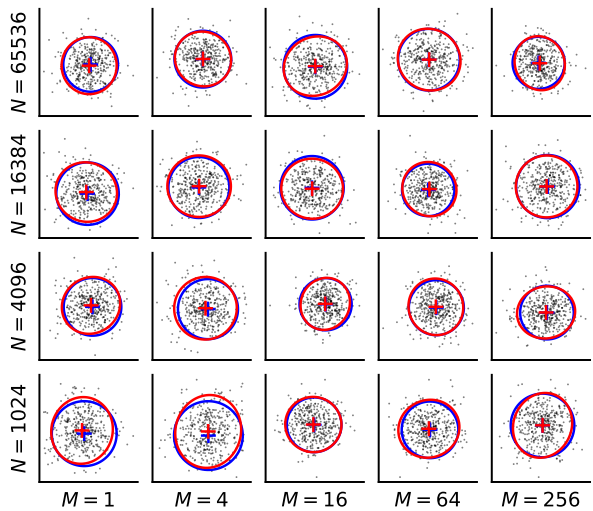

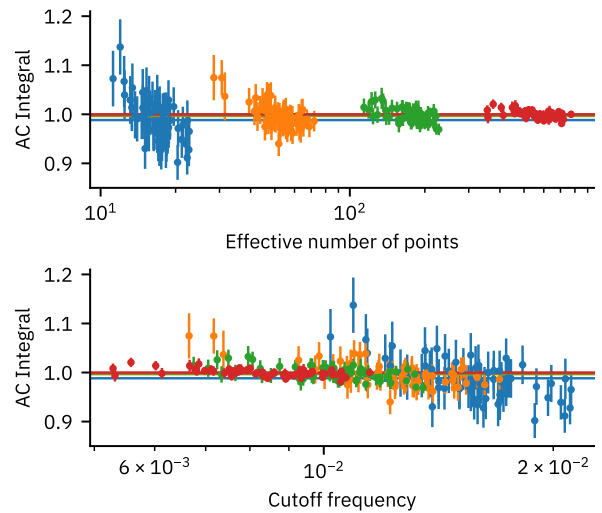

**(f) Sanity check counts for the effective number of points**

|  | $M = 1$ | $M = 4$ | $M = 16$ | $M = 64$ | $M = 256$ |
|---|---|---|---|---|---|
| $N = 1024$ | 51 | 63 | 64 | 64 | 64 |
| $N = 4096$ | 0 | 1 | 1 | 4 | 20 |
| $N = 16384$ | 0 | 0 | 0 | 0 | 0 |
| $N = 65536$ | 0 | 0 | 0 | 0 | 0 |

**(g) Sanity check counts for the regression cost z-score**

|  | $M = 1$ | $M = 4$ | $M = 16$ | $M = 64$ | $M = 256$ |
|---|---|---|---|---|---|
| $N = 1024$ | 0 | 2 | 0 | 0 | 0 |
| $N = 4096$ | 0 | 0 | 2 | 0 | 1 |
| $N = 16384$ | 0 | 1 | 1 | 1 | 2 |
| $N = 65536$ | 0 | 0 | 1 | 1 | 2 |

**(h) Sanity check counts for the cutoff criterion z-score**

|  | $M = 1$ | $M = 4$ | $M = 16$ | $M = 64$ | $M = 256$ |
|---|---|---|---|---|---|
| $N = 1024$ | 1 | 2 | 5 | 4 | 4 |
| $N = 4096$ | 0 | 1 | 0 | 0 | 1 |
| $N = 16384$ | 0 | 1 | 0 | 1 | 1 |
| $N = 65536$ | 0 | 0 | 2 | 1 | 0 |

## S4.3. Kernel exp2

**(a) Illustration of input sequences. See Fig. 5 in the main text for details.**

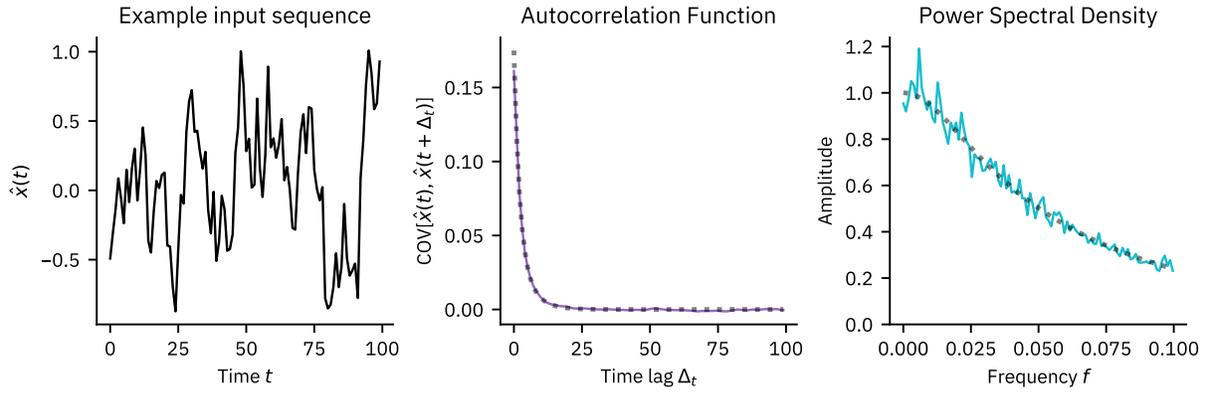

**(b) Scaling of errors with input data.**
**See Fig. 6 in the main text for details.**

**(c) Assessment of the error estimate.**
**See Fig. 7 in the main text for details.**

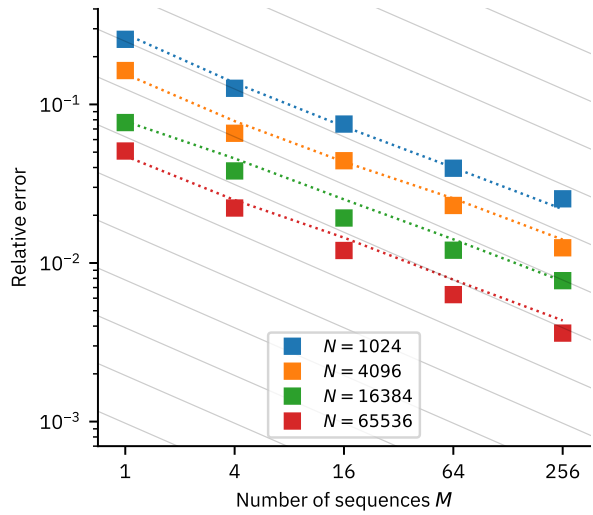
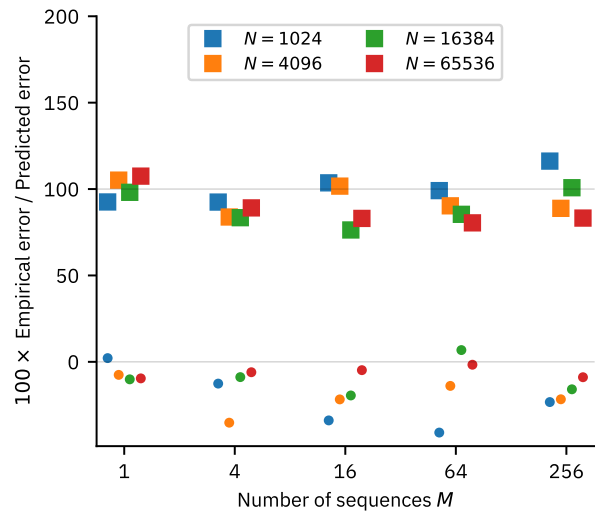

**(d) Validation of the Maximum a Posteriori.**
**See Fig. 8 in the main text for details.**

**(e) Sensitivity of the autocorrelation integral to the cutoff frequency. See Fig. 9 in the main text for details.**

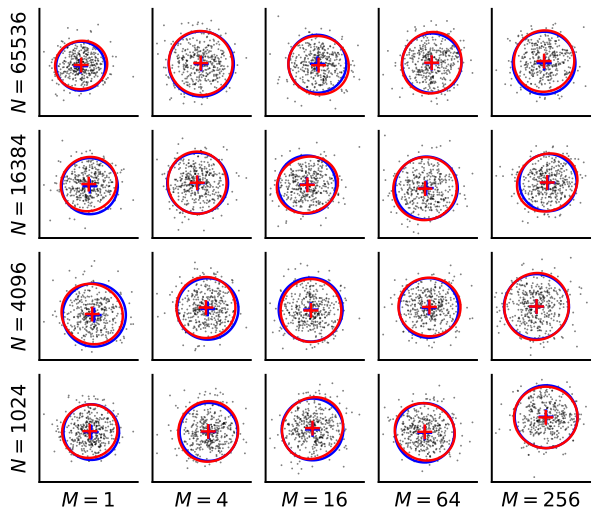
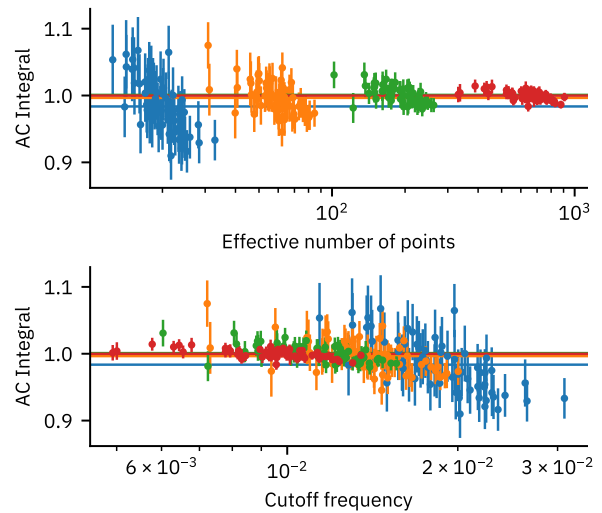

**(f) Sanity check counts for the effective number of points**

|              | $M = 1$ | $M = 4$ | $M = 16$ | $M = 64$ | $M = 256$ |
|--------------|---------|---------|----------|----------|-----------|
| $N = 1024$   | 31      | 48      | 64       | 64       | 64        |
| $N = 4096$   | 0       | 0       | 3        | 3        | 13        |
| $N = 16384$  | 0       | 0       | 0        | 0        | 0         |
| $N = 65536$  | 0       | 0       | 0        | 0        | 0         |

**(g) Sanity check counts for the regression cost z-score**

|              | $M = 1$ | $M = 4$ | $M = 16$ | $M = 64$ | $M = 256$ |
|--------------|---------|---------|----------|----------|-----------|
| $N = 1024$   | 0       | 0       | 0        | 3        | 2         |
| $N = 4096$   | 0       | 0       | 1        | 1        | 1         |
| $N = 16384$  | 0       | 0       | 0        | 1        | 0         |
| $N = 65536$  | 0       | 0       | 1        | 1        | 2         |

**(h) Sanity check counts for the cutoff criterion z-score**

|              | $M = 1$ | $M = 4$ | $M = 16$ | $M = 64$ | $M = 256$ |
|--------------|---------|---------|----------|----------|-----------|
| $N = 1024$   | 3       | 2       | 0        | 0        | 1         |
| $N = 4096$   | 0       | 1       | 0        | 1        | 2         |
| $N = 16384$  | 0       | 0       | 0        | 1        | 1         |
| $N = 65536$  | 0       | 0       | 0        | 1        | 0         |

## S4.4. Kernel sho1pcrit

**(a) Illustration of input sequences. See Fig. 5 in the main text for details.**

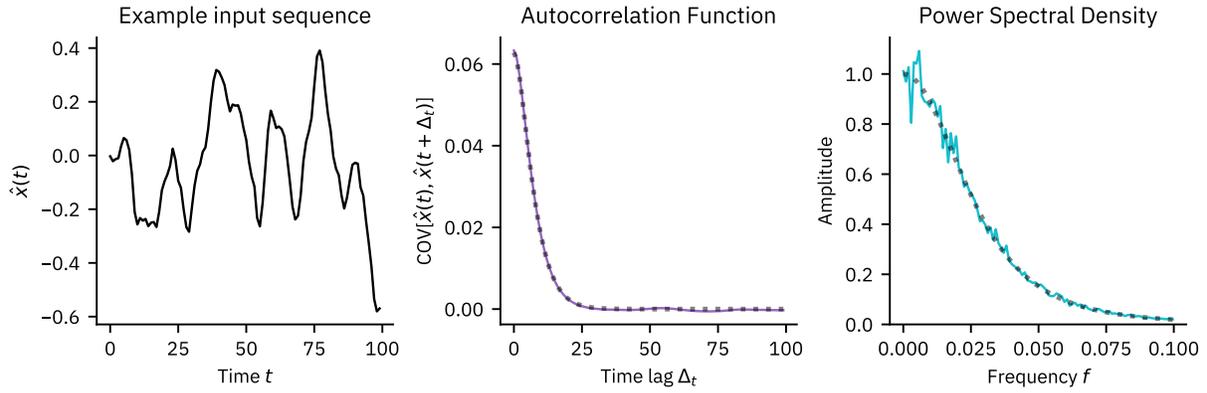

**(b) Scaling of errors with input data.**
**See Fig. 6 in the main text for details.**

**(c) Assessment of the error estimate.**
**See Fig. 7 in the main text for details.**

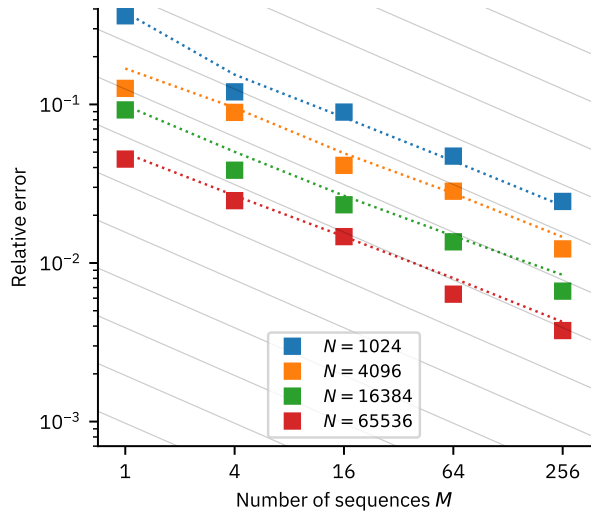

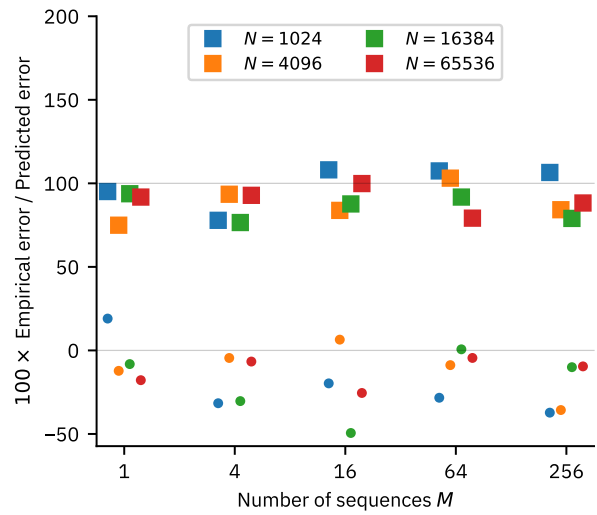

**(d) Validation of the Maximum a Posteriori.**
**See Fig. 8 in the main text for details.**

**(e) Sensitivity of the autocorrelation integral to the cutoff frequency. See Fig. 9 in the main text for details.**

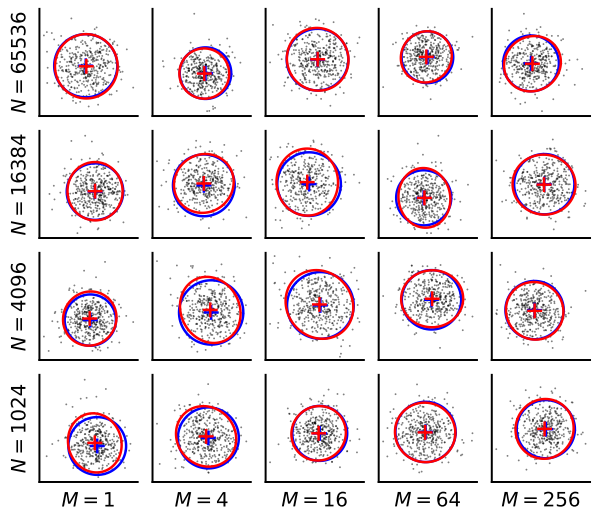

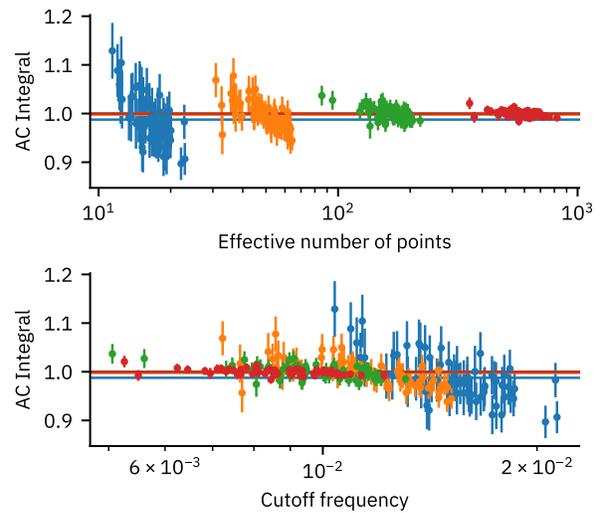

**(f) Sanity check counts for the effective number of points**

|  | $M = 1$ | $M = 4$ | $M = 16$ | $M = 64$ | $M = 256$ |
|---|---|---|---|---|---|
| $N = 1024$ | 64 | 64 | 64 | 64 | 64 |
| $N = 4096$ | 0 | 1 | 1 | 13 | 14 |
| $N = 16384$ | 0 | 0 | 0 | 0 | 0 |
| $N = 65536$ | 0 | 0 | 0 | 0 | 0 |

**(g) Sanity check counts for the regression cost z-score**

|  | $M = 1$ | $M = 4$ | $M = 16$ | $M = 64$ | $M = 256$ |
|---|---|---|---|---|---|
| $N = 1024$ | 0 | 0 | 0 | 2 | 1 |
| $N = 4096$ | 0 | 0 | 1 | 3 | 1 |
| $N = 16384$ | 0 | 0 | 0 | 1 | 2 |
| $N = 65536$ | 0 | 0 | 1 | 0 | 2 |

**(h) Sanity check counts for the cutoff criterion z-score**

|  | $M = 1$ | $M = 4$ | $M = 16$ | $M = 64$ | $M = 256$ |
|---|---|---|---|---|---|
| $N = 1024$ | 1 | 3 | 0 | 2 | 3 |
| $N = 4096$ | 0 | 0 | 0 | 1 | 1 |
| $N = 16384$ | 1 | 0 | 0 | 1 | 0 |
| $N = 65536$ | 0 | 0 | 0 | 1 | 0 |

## S4.5. Kernel sho1pover

**(a) Illustration of input sequences. See Fig. 5 in the main text for details.**

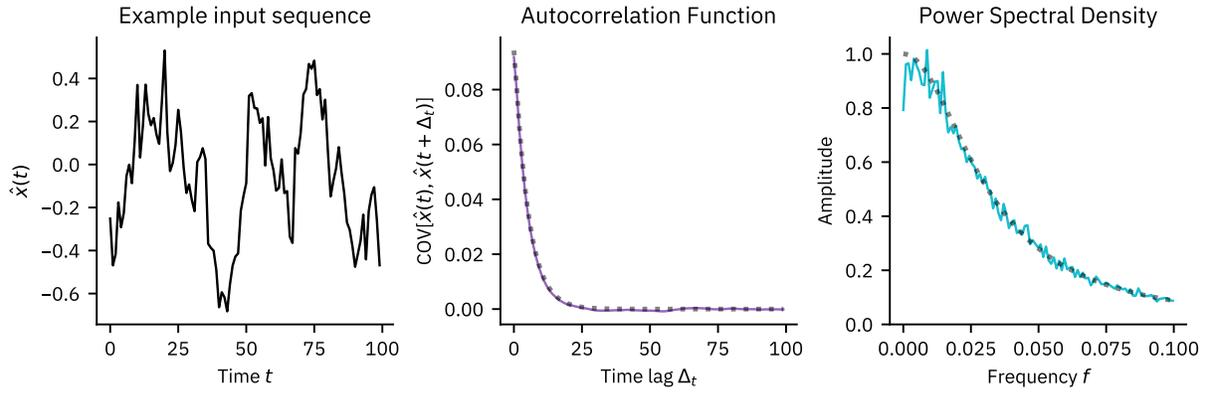

**(b) Scaling of errors with input data. See Fig. 6 in the main text for details.**

**(c) Assessment of the error estimate. See Fig. 7 in the main text for details.**

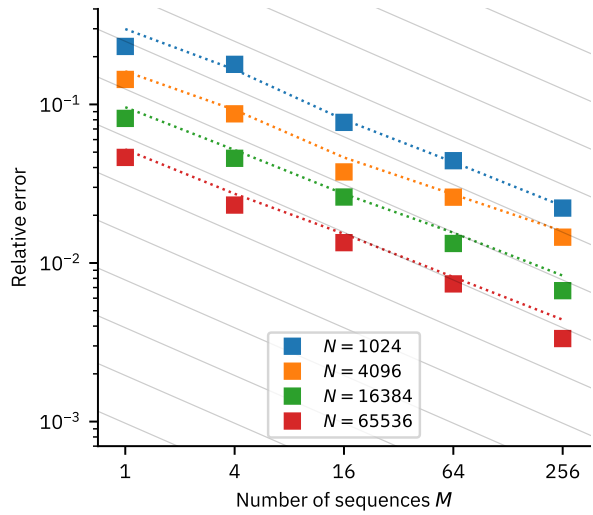
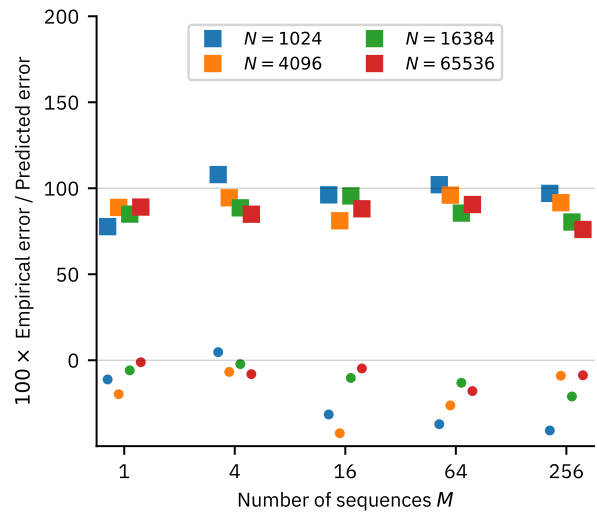

**(d) Validation of the Maximum a Posteriori. See Fig. 8 in the main text for details.**

**(e) Sensitivity of the autocorrelation integral to the cutoff frequency. See Fig. 9 in the main text for details.**

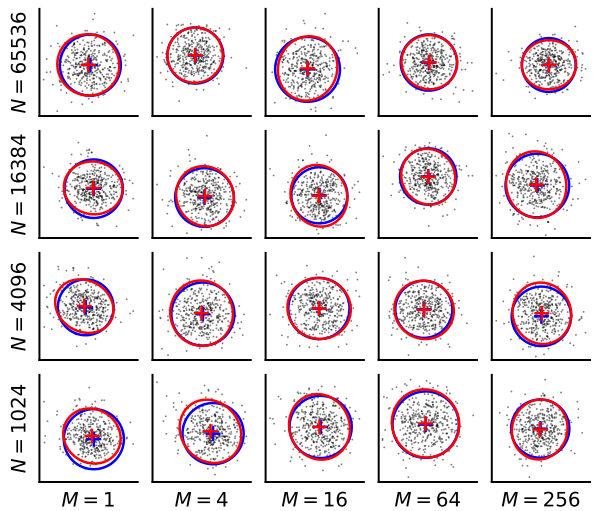
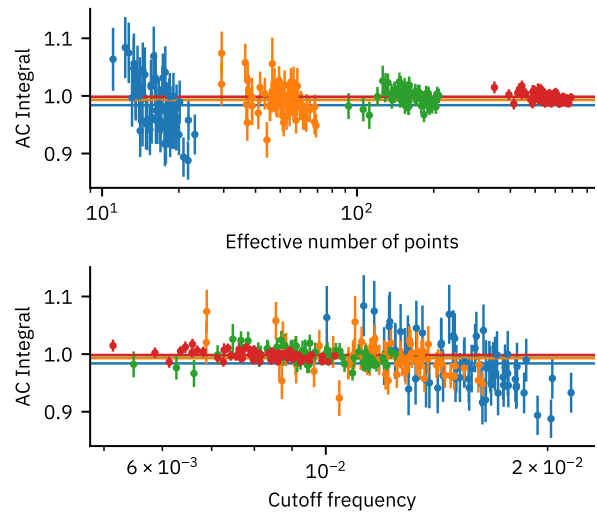

**(f) Sanity check counts for the effective number of points**

|  | $M = 1$ | $M = 4$ | $M = 16$ | $M = 64$ | $M = 256$ |
|---|---|---|---|---|---|
| $N = 1024$ | 55 | 64 | 64 | 64 | 64 |
| $N = 4096$ | 0 | 0 | 0 | 9 | 25 |
| $N = 16384$ | 0 | 0 | 0 | 0 | 0 |
| $N = 65536$ | 0 | 0 | 0 | 0 | 0 |

**(g) Sanity check counts for the regression cost z-score**

|  | $M = 1$ | $M = 4$ | $M = 16$ | $M = 64$ | $M = 256$ |
|---|---|---|---|---|---|
| $N = 1024$ | 0 | 0 | 0 | 1 | 1 |
| $N = 4096$ | 0 | 0 | 2 | 0 | 3 |
| $N = 16384$ | 0 | 0 | 1 | 2 | 1 |
| $N = 65536$ | 0 | 0 | 2 | 3 | 2 |

**(h) Sanity check counts for the cutoff criterion z-score**

|  | $M = 1$ | $M = 4$ | $M = 16$ | $M = 64$ | $M = 256$ |
|---|---|---|---|---|---|
| $N = 1024$ | 3 | 2 | 2 | 1 | 5 |
| $N = 4096$ | 0 | 0 | 0 | 0 | 2 |
| $N = 16384$ | 2 | 0 | 0 | 2 | 0 |
| $N = 65536$ | 0 | 1 | 0 | 0 | 0 |

# S4.6. Kernel sho1punder

**(a) Illustration of input sequences. See Fig. 5 in the main text for details.**

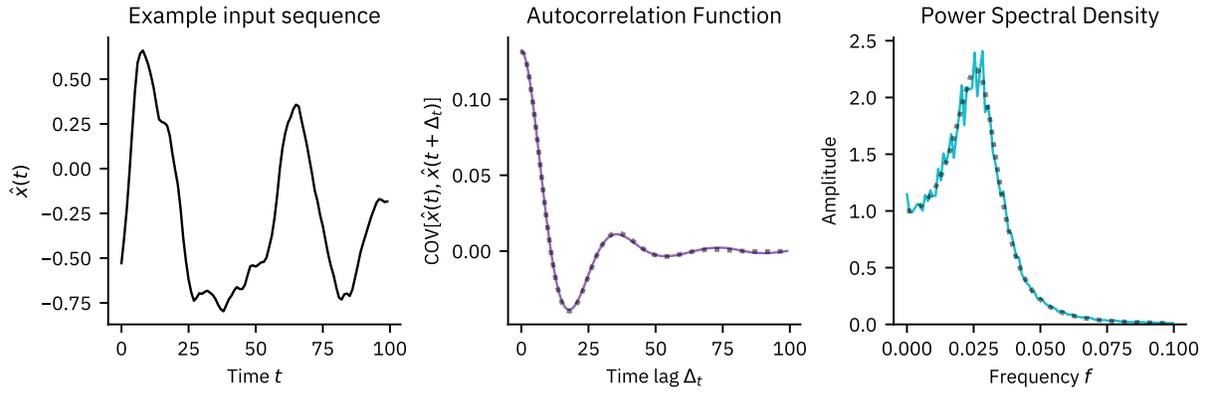

**(b) Scaling of errors with input data.**
**See Fig. 6 in the main text for details.**

**(c) Assessment of the error estimate.**
**See Fig. 7 in the main text for details.**

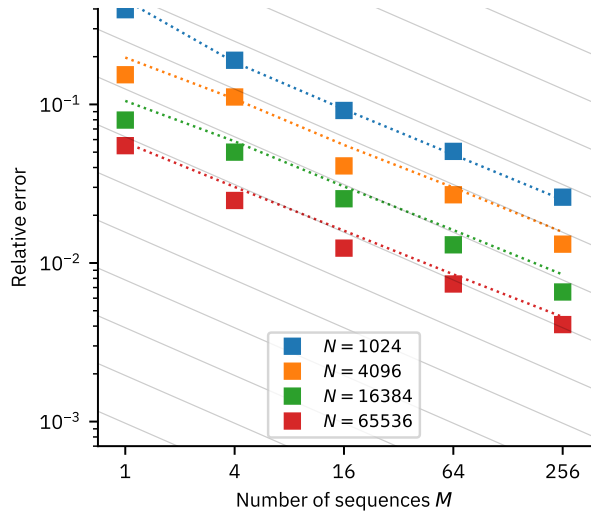

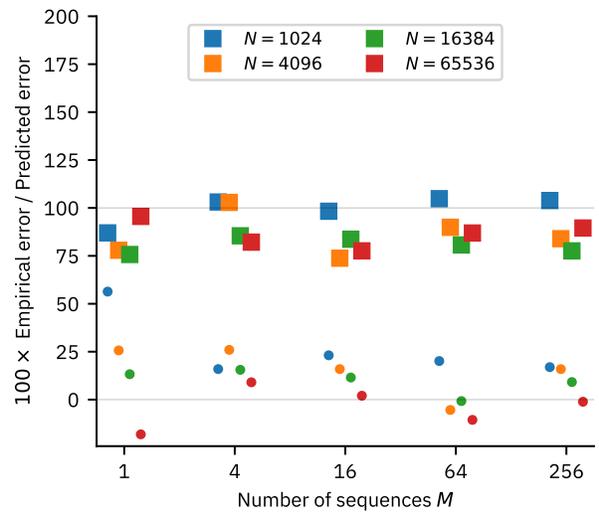

**(d) Validation of the Maximum a Posteriori.**
**See Fig. 8 in the main text for details.**

**(e) Sensitivity of the autocorrelation integral to the cutoff**
**frequency. See Fig. 9 in the main text for details.**

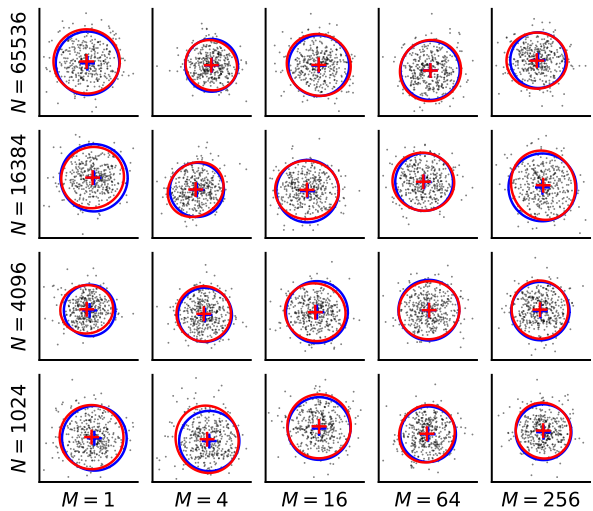

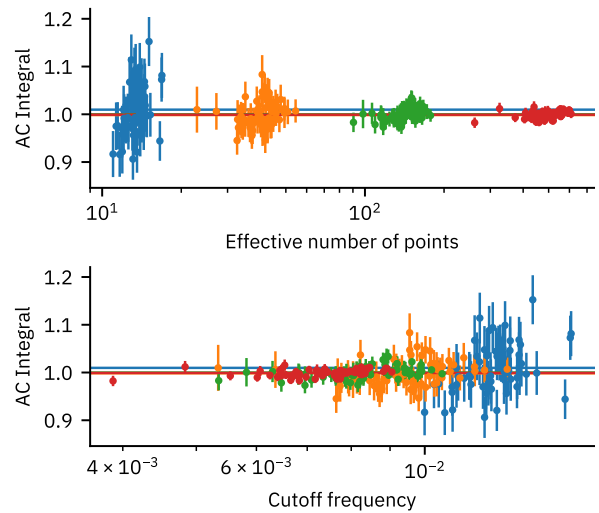

**(f) Sanity check counts for the effective number of points**

|  | $M = 1$ | $M = 4$ | $M = 16$ | $M = 64$ | $M = 256$ |
|---|---|---|---|---|---|
| $N = 1024$ | 63 | 64 | 64 | 64 | 64 |
| $N = 4096$ | 0 | 0 | 7 | 29 | 53 |
| $N = 16384$ | 0 | 0 | 0 | 0 | 0 |
| $N = 65536$ | 0 | 0 | 0 | 0 | 0 |

**(g) Sanity check counts for the regression cost z-score**

|  | $M = 1$ | $M = 4$ | $M = 16$ | $M = 64$ | $M = 256$ |
|---|---|---|---|---|---|
| $N = 1024$ | 0 | 1 | 2 | 0 | 1 |
| $N = 4096$ | 0 | 0 | 0 | 2 | 2 |
| $N = 16384$ | 0 | 0 | 0 | 2 | 2 |
| $N = 65536$ | 0 | 0 | 1 | 3 | 3 |

**(h) Sanity check counts for the cutoff criterion z-score**

|  | $M = 1$ | $M = 4$ | $M = 16$ | $M = 64$ | $M = 256$ |
|---|---|---|---|---|---|
| $N = 1024$ | 1 | 2 | 2 | 3 | 3 |
| $N = 4096$ | 0 | 2 | 0 | 0 | 2 |
| $N = 16384$ | 1 | 0 | 1 | 2 | 0 |
| $N = 65536$ | 1 | 0 | 0 | 0 | 1 |

## S4.7. Kernel sho1wcrit

**(a) Illustration of input sequences. See Fig. 5 in the main text for details.**

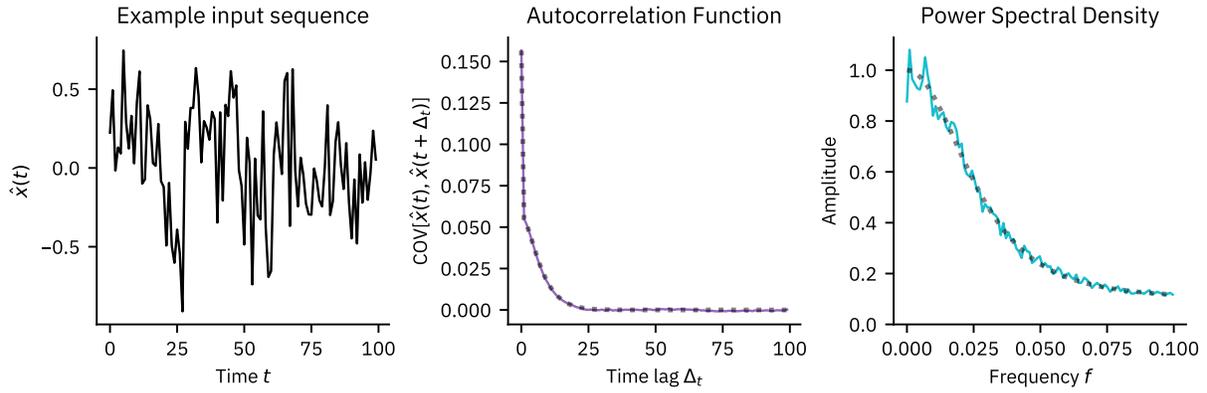

**(b) Scaling of errors with input data. See Fig. 6 in the main text for details.**

**(c) Assessment of the error estimate. See Fig. 7 in the main text for details.**

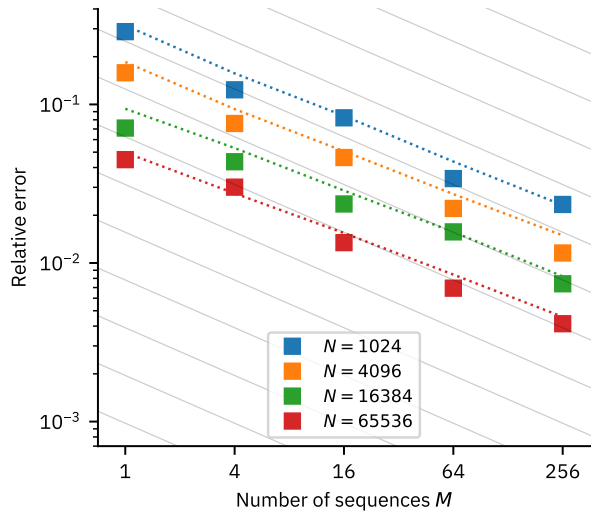

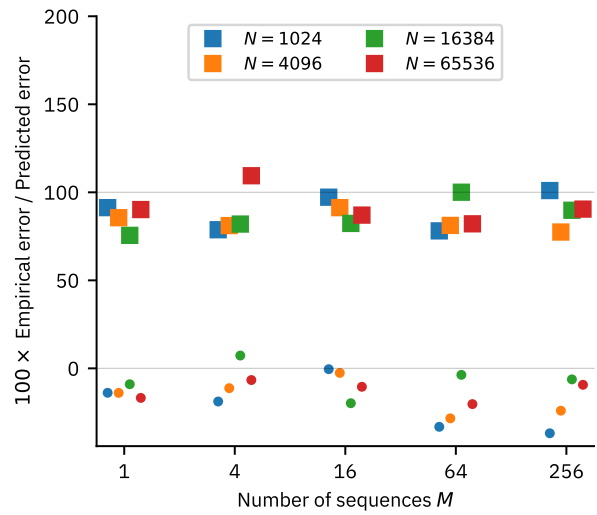

**(d) Validation of the Maximum a Posteriori. See Fig. 8 in the main text for details.**

**(e) Sensitivity of the autocorrelation integral to the cutoff frequency. See Fig. 9 in the main text for details.**

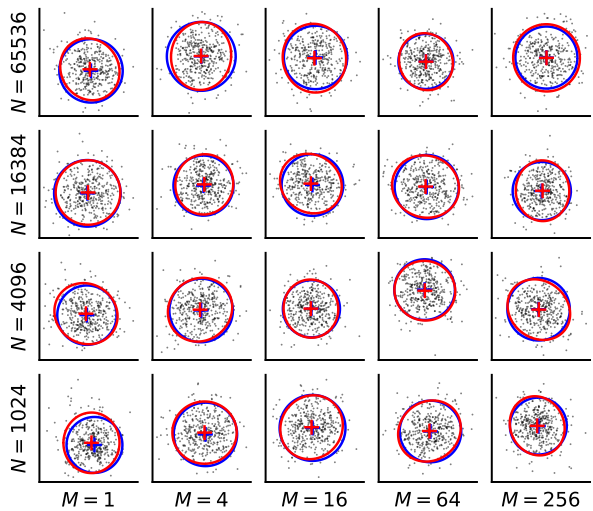

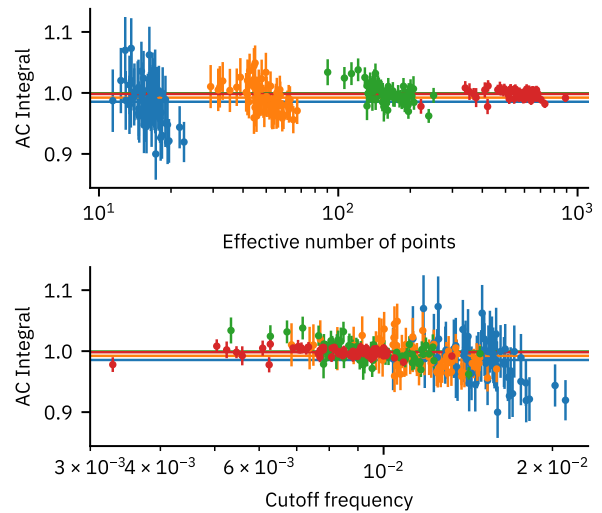

**(f) Sanity check counts for the effective number of points**

|  | $M = 1$ | $M = 4$ | $M = 16$ | $M = 64$ | $M = 256$ |
|---|---|---|---|---|---|
| $N = 1024$ | 62 | 64 | 64 | 64 | 64 |
| $N = 4096$ | 1 | 0 | 3 | 7 | 24 |
| $N = 16384$ | 0 | 0 | 0 | 0 | 0 |
| $N = 65536$ | 0 | 0 | 0 | 0 | 0 |

**(g) Sanity check counts for the regression cost z-score**

|  | $M = 1$ | $M = 4$ | $M = 16$ | $M = 64$ | $M = 256$ |
|---|---|---|---|---|---|
| $N = 1024$ | 0 | 0 | 2 | 0 | 0 |
| $N = 4096$ | 0 | 0 | 0 | 0 | 3 |
| $N = 16384$ | 0 | 1 | 1 | 1 | 1 |
| $N = 65536$ | 0 | 0 | 4 | 0 | 1 |

**(h) Sanity check counts for the cutoff criterion z-score**

|  | $M = 1$ | $M = 4$ | $M = 16$ | $M = 64$ | $M = 256$ |
|---|---|---|---|---|---|
| $N = 1024$ | 4 | 2 | 1 | 3 | 4 |
| $N = 4096$ | 1 | 0 | 0 | 0 | 0 |
| $N = 16384$ | 0 | 0 | 0 | 0 | 0 |
| $N = 65536$ | 0 | 0 | 1 | 1 | 1 |

## S4.8. Kernel sho1wover

**(a) Illustration of input sequences. See Fig. 5 in the main text for details.**

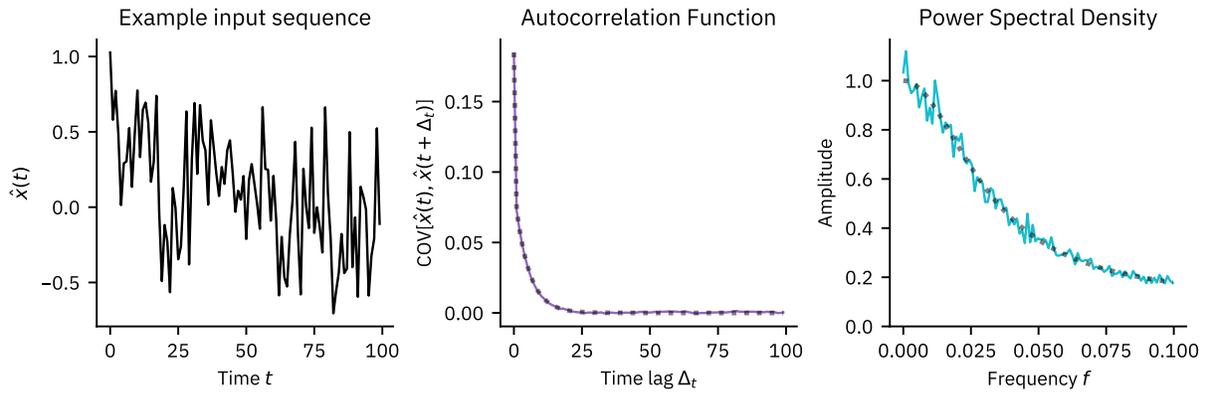

**(b) Scaling of errors with input data.**
**See Fig. 6 in the main text for details.**

**(c) Assessment of the error estimate.**
**See Fig. 7 in the main text for details.**

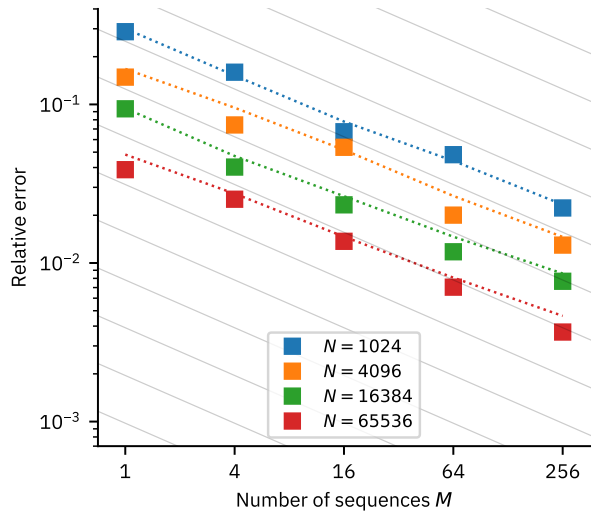

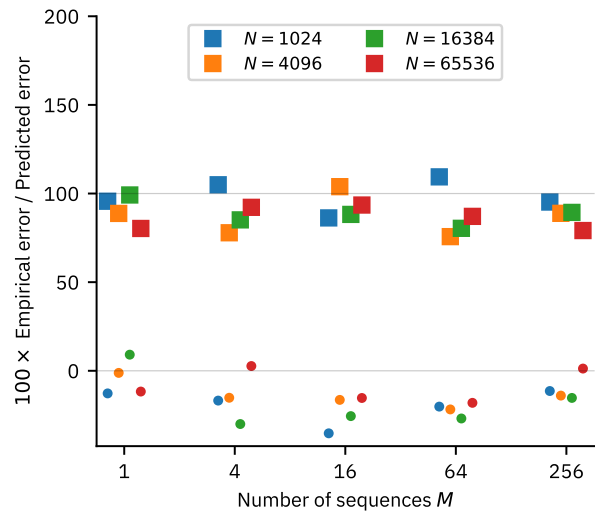

**(d) Validation of the Maximum a Posteriori.**
**See Fig. 8 in the main text for details.**

**(e) Sensitivity of the autocorrelation integral to the cutoff frequency. See Fig. 9 in the main text for details.**

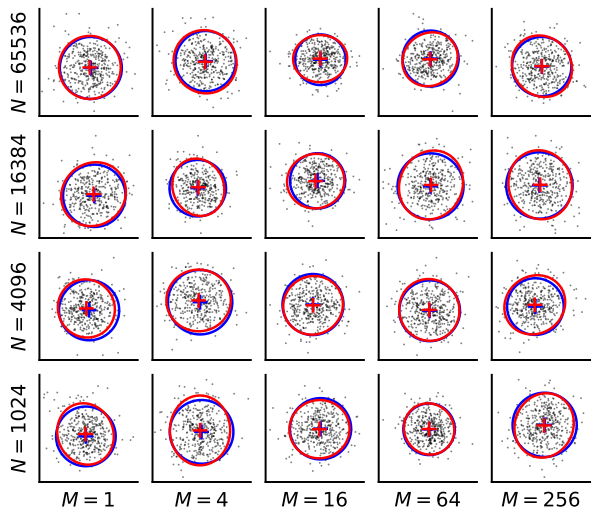

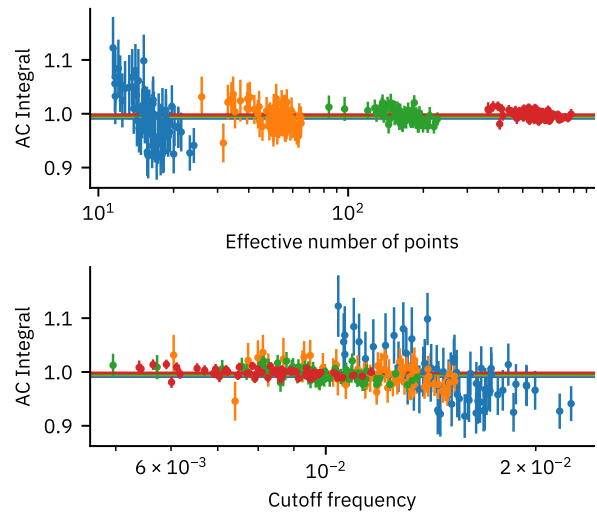

**(f) Sanity check counts for the effective number of points**

|              | $M = 1$ | $M = 4$ | $M = 16$ | $M = 64$ | $M = 256$ |
| ------------ | ------- | ------- | -------- | -------- | --------- |
| $N = 1024$   | 50      | 64      | 64       | 64       | 64        |
| $N = 4096$   | 0       | 0       | 2        | 9        | 24        |
| $N = 16384$  | 0       | 0       | 0        | 0        | 0         |
| $N = 65536$  | 0       | 0       | 0        | 0        | 0         |

**(g) Sanity check counts for the regression cost z-score**

|              | $M = 1$ | $M = 4$ | $M = 16$ | $M = 64$ | $M = 256$ |
| ------------ | ------- | ------- | -------- | -------- | --------- |
| $N = 1024$   | 0       | 1       | 2        | 1        | 1         |
| $N = 4096$   | 0       | 0       | 0        | 2        | 0         |
| $N = 16384$  | 0       | 1       | 1        | 0        | 3         |
| $N = 65536$  | 0       | 1       | 1        | 1        | 1         |

**(h) Sanity check counts for the cutoff criterion z-score**

|              | $M = 1$ | $M = 4$ | $M = 16$ | $M = 64$ | $M = 256$ |
| ------------ | ------- | ------- | -------- | -------- | --------- |
| $N = 1024$   | 4       | 1       | 4        | 2        | 5         |
| $N = 4096$   | 0       | 1       | 1        | 0        | 1         |
| $N = 16384$  | 0       | 0       | 0        | 0        | 2         |
| $N = 65536$  | 1       | 0       | 1        | 0        | 0         |

## S4.9. Kernel sho1wunder

**(a) Illustration of input sequences. See Fig. 5 in the main text for details.**

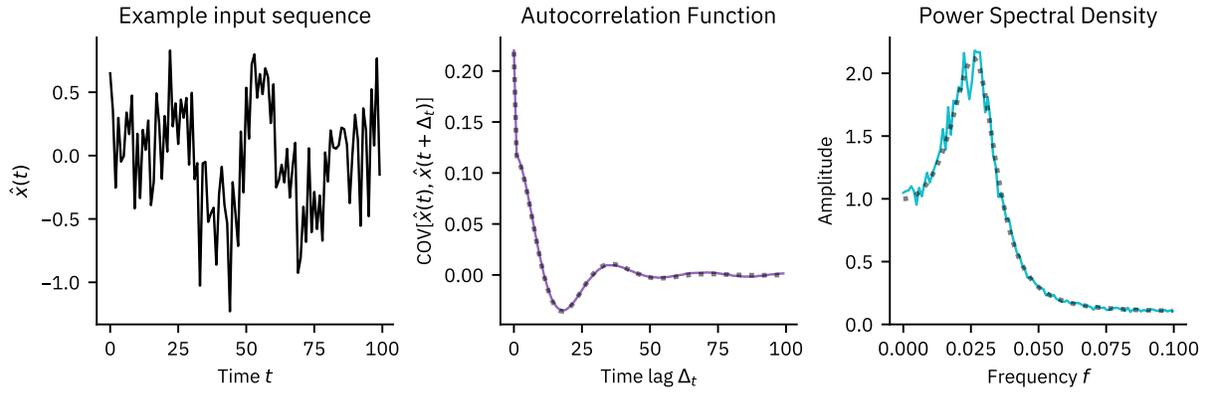

**(b) Scaling of errors with input data.**
**See Fig. 6 in the main text for details.**

**(c) Assessment of the error estimate.**
**See Fig. 7 in the main text for details.**

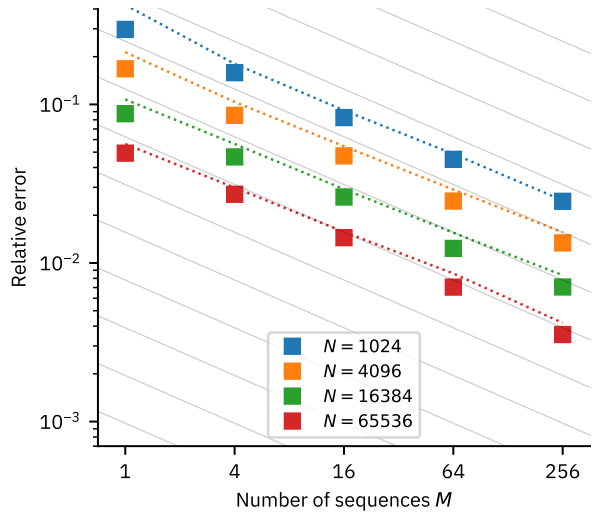

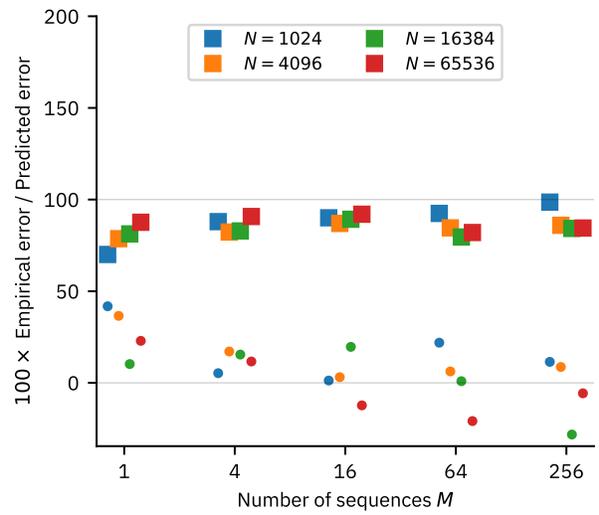

**(d) Validation of the Maximum a Posteriori.**
**See Fig. 8 in the main text for details.**

**(e) Sensitivity of the autocorrelation integral to the cutoff frequency. See Fig. 9 in the main text for details.**

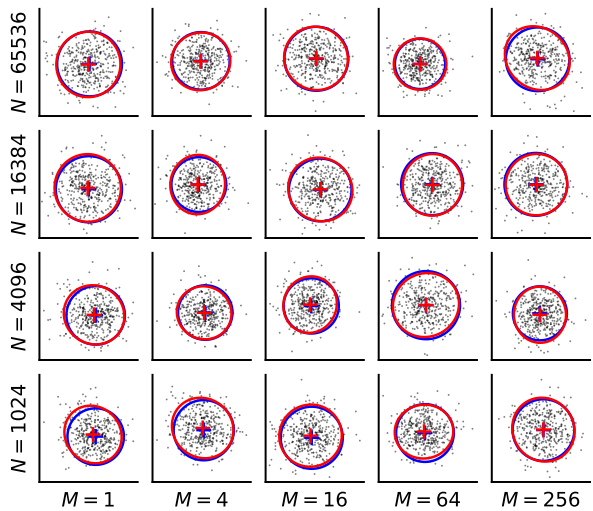

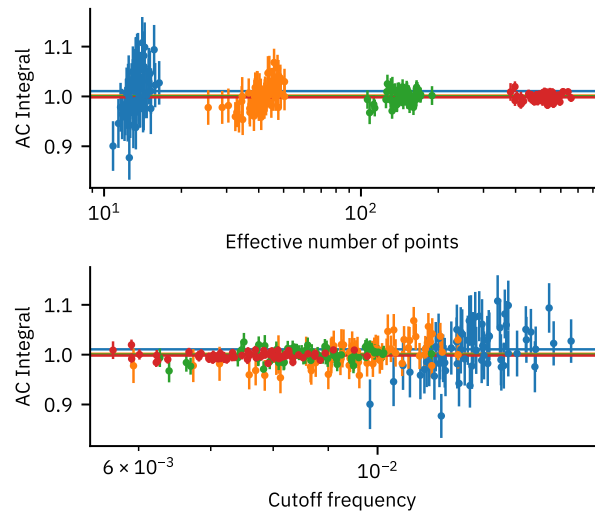

**(f) Sanity check counts for the effective number of points**

|            | $M = 1$ | $M = 4$ | $M = 16$ | $M = 64$ | $M = 256$ |
|------------|---------|---------|----------|----------|-----------|
| $N = 1024$  | 64 | 64 | 64 | 64 | 64 |
| $N = 4096$  | 1  | 0  | 7  | 23 | 45 |
| $N = 16384$ | 0  | 0  | 0  | 0  | 0  |
| $N = 65536$ | 0  | 0  | 0  | 0  | 0  |

**(g) Sanity check counts for the regression cost z-score**

|            | $M = 1$ | $M = 4$ | $M = 16$ | $M = 64$ | $M = 256$ |
|------------|---------|---------|----------|----------|-----------|
| $N = 1024$  | 0 | 0 | 1 | 2 | 0 |
| $N = 4096$  | 0 | 0 | 1 | 1 | 1 |
| $N = 16384$ | 0 | 0 | 0 | 4 | 1 |
| $N = 65536$ | 0 | 0 | 1 | 3 | 4 |

**(h) Sanity check counts for the cutoff criterion z-score**

|            | $M = 1$ | $M = 4$ | $M = 16$ | $M = 64$ | $M = 256$ |
|------------|---------|---------|----------|----------|-----------|
| $N = 1024$  | 2 | 2 | 3 | 3 | 7 |
| $N = 4096$  | 2 | 0 | 0 | 0 | 1 |
| $N = 16384$ | 0 | 0 | 0 | 0 | 0 |
| $N = 65536$ | 0 | 0 | 0 | 1 | 0 |

## S4.10. Kernel sho2crit

**(a) Illustration of input sequences. See Fig. 5 in the main text for details.**

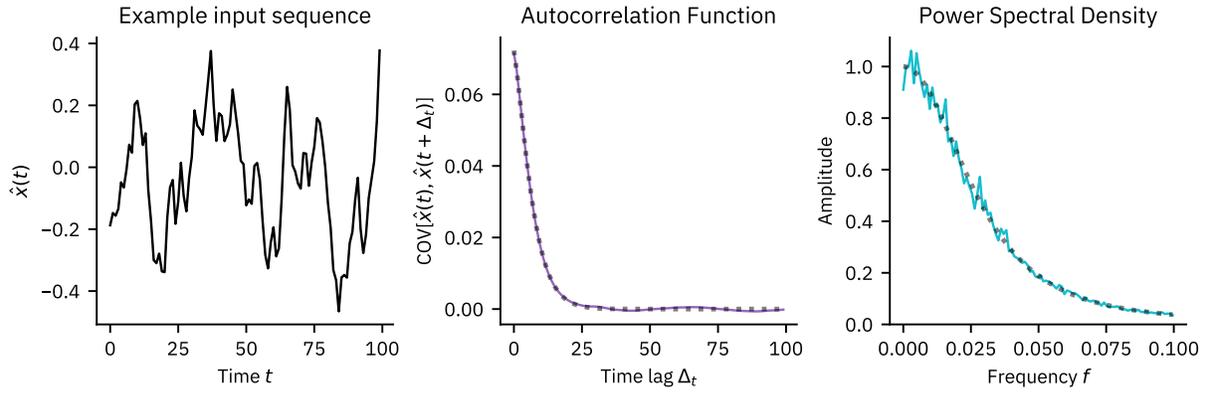

**(b) Scaling of errors with input data.**
**See Fig. 6 in the main text for details.**

**(c) Assessment of the error estimate.**
**See Fig. 7 in the main text for details.**

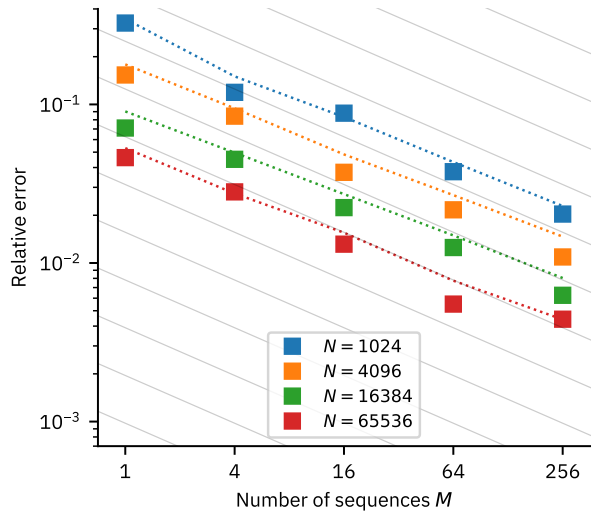

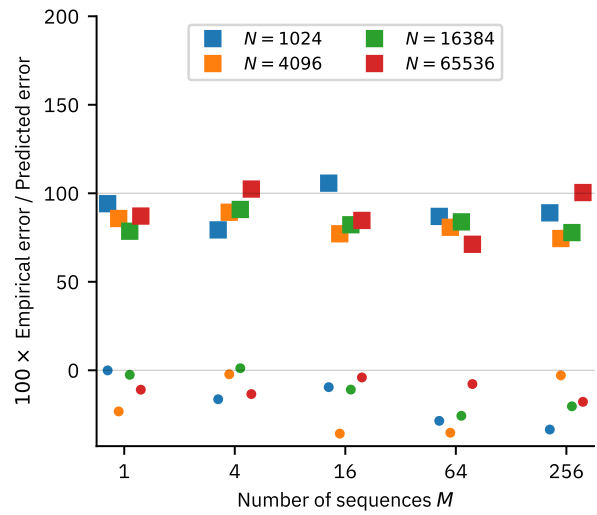

**(d) Validation of the Maximum a Posteriori.**
**See Fig. 8 in the main text for details.**

**(e) Sensitivity of the autocorrelation integral to the cutoff**
**frequency. See Fig. 9 in the main text for details.**

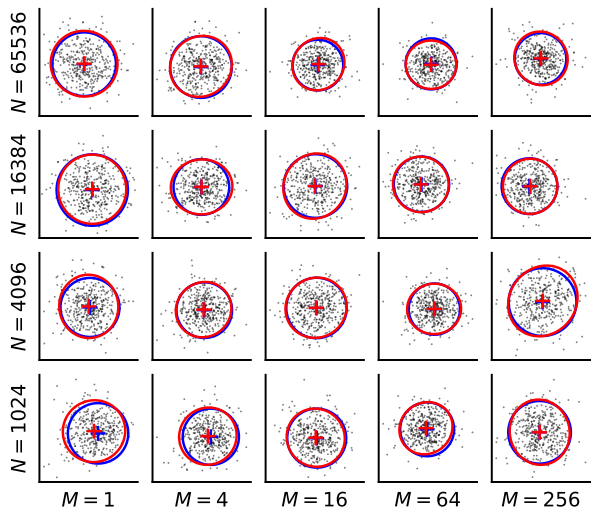

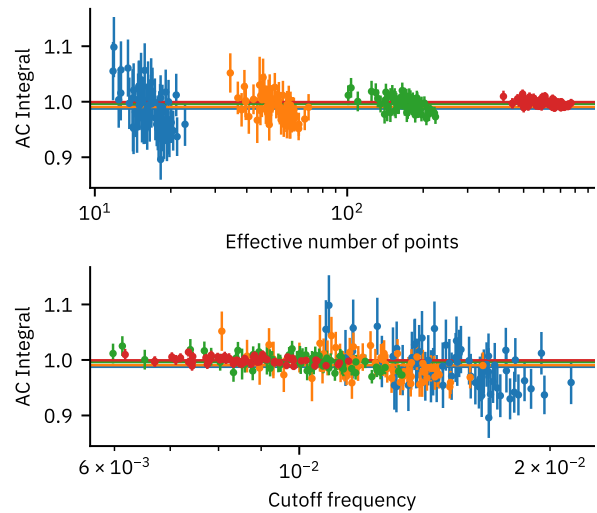

**(f) Sanity check counts for the effective number of points**

|  | $M = 1$ | $M = 4$ | $M = 16$ | $M = 64$ | $M = 256$ |
|---|---|---|---|---|---|
| $N = 1024$ | 62 | 64 | 64 | 64 | 64 |
| $N = 4096$ | 0 | 1 | 1 | 6 | 18 |
| $N = 16384$ | 0 | 0 | 0 | 0 | 0 |
| $N = 65536$ | 0 | 0 | 0 | 0 | 0 |

**(g) Sanity check counts for the regression cost z-score**

|  | $M = 1$ | $M = 4$ | $M = 16$ | $M = 64$ | $M = 256$ |
|---|---|---|---|---|---|
| $N = 1024$ | 0 | 2 | 1 | 1 | 1 |
| $N = 4096$ | 0 | 1 | 2 | 2 | 0 |
| $N = 16384$ | 0 | 0 | 2 | 3 | 1 |
| $N = 65536$ | 0 | 0 | 0 | 0 | 1 |

**(h) Sanity check counts for the cutoff criterion z-score**

|  | $M = 1$ | $M = 4$ | $M = 16$ | $M = 64$ | $M = 256$ |
|---|---|---|---|---|---|
| $N = 1024$ | 1 | 2 | 3 | 2 | 3 |
| $N = 4096$ | 1 | 0 | 1 | 1 | 1 |
| $N = 16384$ | 0 | 0 | 1 | 2 | 0 |
| $N = 65536$ | 0 | 0 | 1 | 0 | 0 |

# S4.11. Kernel sho2over

**(a) Illustration of input sequences. See Fig. 5 in the main text for details.**

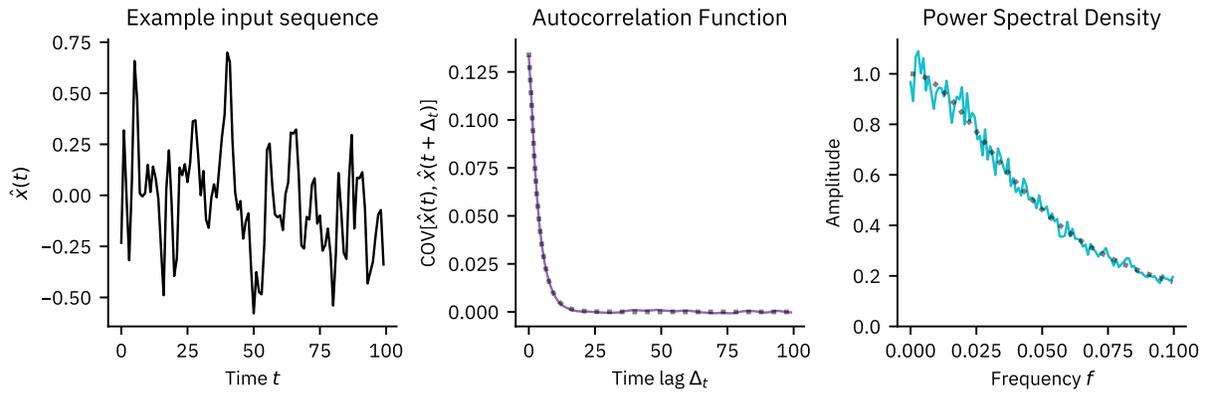

**(b) Scaling of errors with input data.**
**See Fig. 6 in the main text for details.**

**(c) Assessment of the error estimate.**
**See Fig. 7 in the main text for details.**

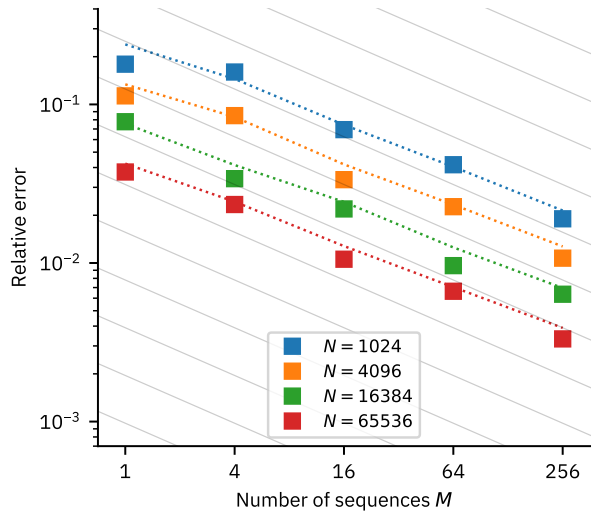

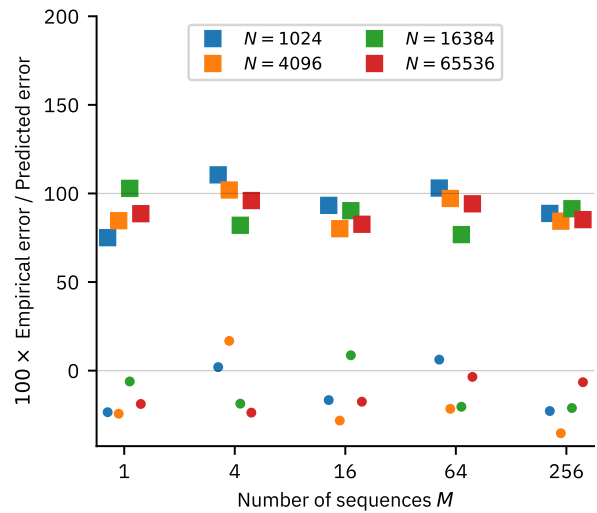

**(d) Validation of the Maximum a Posteriori.**
**See Fig. 8 in the main text for details.**

**(e) Sensitivity of the autocorrelation integral to the cutoff frequency. See Fig. 9 in the main text for details.**

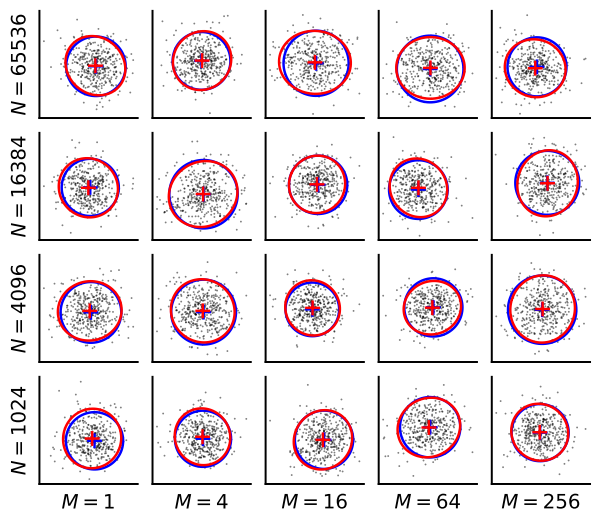

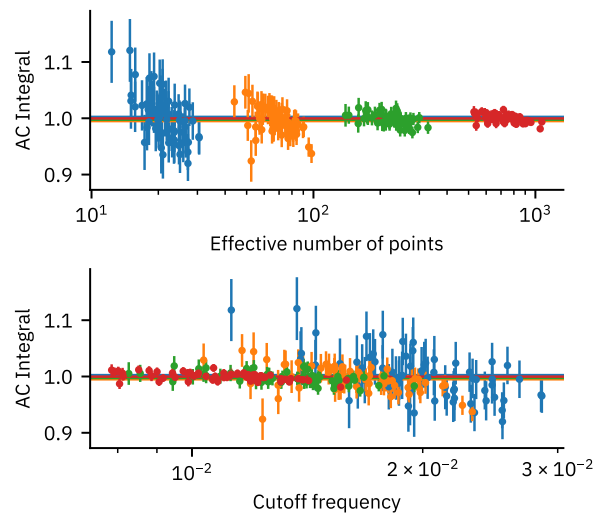

**(f) Sanity check counts for the effective number of points**

|  | $M = 1$ | $M = 4$ | $M = 16$ | $M = 64$ | $M = 256$ |
|---|---|---|---|---|---|
| $N = 1024$ | 29 | 55 | 64 | 64 | 64 |
| $N = 4096$ | 0 | 0 | 0 | 0 | 7 |
| $N = 16384$ | 0 | 0 | 0 | 0 | 0 |
| $N = 65536$ | 0 | 0 | 0 | 0 | 0 |

**(g) Sanity check counts for the regression cost z-score**

|  | $M = 1$ | $M = 4$ | $M = 16$ | $M = 64$ | $M = 256$ |
|---|---|---|---|---|---|
| $N = 1024$ | 0 | 2 | 2 | 0 | 3 |
| $N = 4096$ | 0 | 0 | 0 | 0 | 1 |
| $N = 16384$ | 0 | 1 | 1 | 2 | 0 |
| $N = 65536$ | 0 | 1 | 0 | 1 | 0 |

**(h) Sanity check counts for the cutoff criterion z-score**

|  | $M = 1$ | $M = 4$ | $M = 16$ | $M = 64$ | $M = 256$ |
|---|---|---|---|---|---|
| $N = 1024$ | 0 | 1 | 3 | 3 | 2 |
| $N = 4096$ | 1 | 3 | 0 | 0 | 0 |
| $N = 16384$ | 0 | 0 | 1 | 0 | 0 |
| $N = 65536$ | 0 | 1 | 0 | 0 | 0 |

## S4.12. Kernel sho2under

**(a) Illustration of input sequences. See Fig. 5 in the main text for details.**

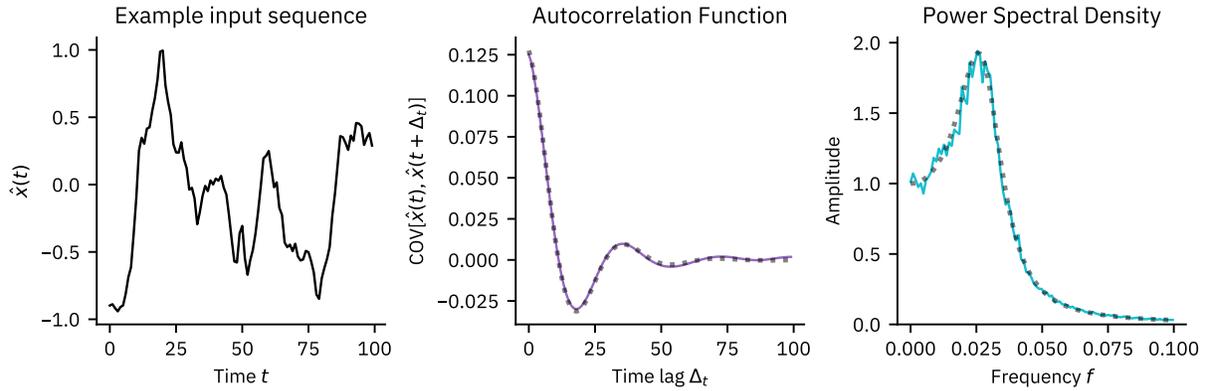

**(b) Scaling of errors with input data.**
**See Fig. 6 in the main text for details.**

**(c) Assessment of the error estimate.**
**See Fig. 7 in the main text for details.**

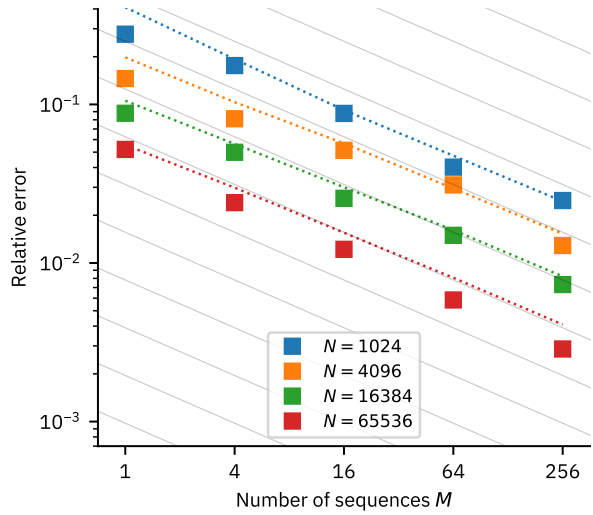

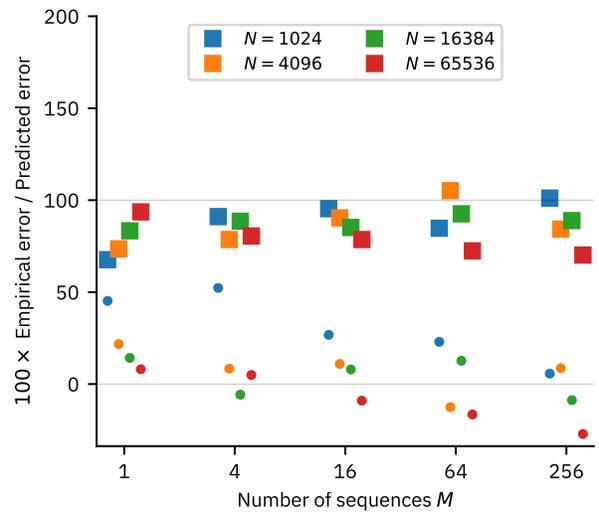

**(d) Validation of the Maximum a Posteriori.**
**See Fig. 8 in the main text for details.**

**(e) Sensitivity of the autocorrelation integral to the cutoff frequency. See Fig. 9 in the main text for details.**

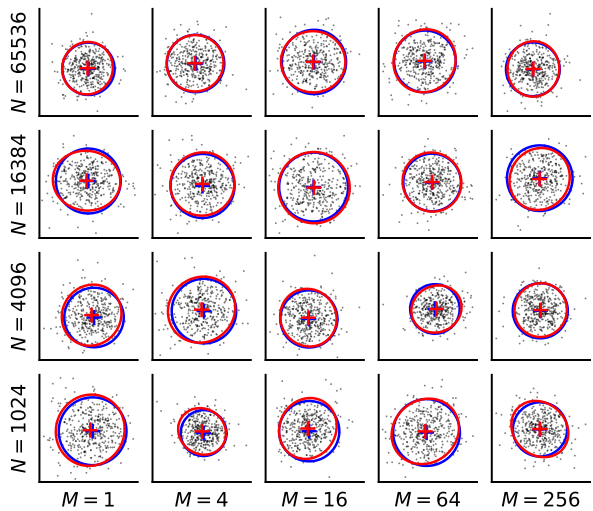

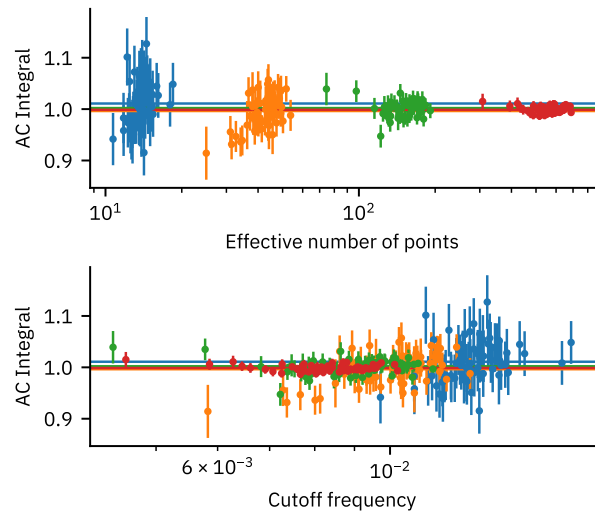

**(f) Sanity check counts for the effective number of points**

|           | $M = 1$ | $M = 4$ | $M = 16$ | $M = 64$ | $M = 256$ |
|-----------|---------|---------|----------|----------|-----------|
| $N = 1024$  | 64 | 64 | 64 | 64 | 64 |
| $N = 4096$  | 0  | 3  | 5  | 17 | 37 |
| $N = 16384$ | 0  | 0  | 0  | 0  | 0  |
| $N = 65536$ | 0  | 0  | 0  | 0  | 0  |

**(g) Sanity check counts for the regression cost z-score**

|           | $M = 1$ | $M = 4$ | $M = 16$ | $M = 64$ | $M = 256$ |
|-----------|---------|---------|----------|----------|-----------|
| $N = 1024$  | 0 | 1 | 3 | 3 | 4 |
| $N = 4096$  | 0 | 0 | 2 | 3 | 1 |
| $N = 16384$ | 0 | 0 | 1 | 1 | 0 |
| $N = 65536$ | 0 | 0 | 1 | 0 | 0 |

**(h) Sanity check counts for the cutoff criterion z-score**

|           | $M = 1$ | $M = 4$ | $M = 16$ | $M = 64$ | $M = 256$ |
|-----------|---------|---------|----------|----------|-----------|
| $N = 1024$  | 1 | 1 | 5 | 3 | 10 |
| $N = 4096$  | 0 | 0 | 1 | 0 | 0  |
| $N = 16384$ | 0 | 0 | 1 | 0 | 0  |
| $N = 65536$ | 1 | 1 | 0 | 0 | 0  |

## TOC Graphic

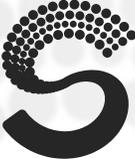



\end{document}